\documentclass[11pt]{article}
\usepackage[preprint]{acl}
\usepackage{times}
\usepackage{latexsym}

\usepackage{booktabs}
\usepackage{multirow}
\usepackage{amsmath}
\usepackage{amsfonts}
\usepackage{tcolorbox}
\usepackage{listings}
\tcbuselibrary{breakable, skins}
\usepackage{caption}
\usepackage{enumitem}
\usepackage{pifont}
\usepackage{makecell}
\usepackage{tabularx}
\usepackage{array}
\newcolumntype{Y}{>{\centering\arraybackslash}X}
\usepackage{url}
\usepackage{algorithm}
\usepackage{algorithmic}
\usepackage{enumitem}

\usepackage{marvosym} 
\usepackage[T1]{fontenc}
\usepackage[utf8]{inputenc}

\usepackage{microtype}
\usepackage{inconsolata}
\usepackage{graphicx}
\usepackage{titletoc}
\usepackage{xspace}

\definecolor{panelbg}{HTML}{F7F7F7}
\definecolor{userbubble}{HTML}{DCF8C6}
\definecolor{botbubble}{HTML}{E3E8EF}
\definecolor{flawred}{HTML}{FDECEC}
\definecolor{fixgreen}{HTML}{E8F5E9}
\definecolor{personapurple}{HTML}{F3E8FD}
\definecolor{problemyellow}{HTML}{FFF3CD}
\definecolor{conceptblue}{HTML}{E8F0FE}
\definecolor{mechgreen}{HTML}{E6F4EA}
\definecolor{intentamber}{HTML}{FEF7E0}
\definecolor{paramgray}{HTML}{F0F0F0}
\usepackage{fontawesome5}
\usepackage{amssymb}

\newcommand{\ourmethod}{{\normalfont\textsc{CoBG}}\xspace}

\newif\ifshowproj
\showprojfalse
\newcommand{\proj}[1]{\ifshowproj{\color{red}#1}\else#1\fi}

\title{\ourmethod: Learning from Player Reviews to Design, Critique, and Revise Board Game Rulebooks}

\author{
\textbf{Zizhen Li\textsuperscript{1,2}},
\textbf{Jianwen Sun\textsuperscript{1,2}},
\textbf{Yukang Feng\textsuperscript{1,2}},
\textbf{Chuanhao Li\textsuperscript{3}},
\textbf{Yibin Wang\textsuperscript{2}},
\\
\textbf{Fanrui Zhang\textsuperscript{2}},
\textbf{Yifei Huang\textsuperscript{4}},
\textbf{Mingzhu Sun\textsuperscript{1}},
\textbf{Kaipeng Zhang\textsuperscript{2,3\,\Letter}}
\\
\textsuperscript{1}Nankai University,
\textsuperscript{2}Shanghai Innovation Institute,
\textsuperscript{3}Alaya Lab,
\textsuperscript{4}The University of Tokyo
\\
\texttt{lizizhen@mail.nankai.edu.cn}, \texttt{kaipeng.zhang@shanda.com}
}

\begin{document}
\maketitle

\begin{abstract}
Board game design alternates between thinking as a designer and experiencing as a player, over repeated cycles of prototyping, critique, and revision. This loop is cognitively demanding: without knowing how players will receive a rule, a designer cannot tell what to revise. We present \ourmethod, a design assistant that supports this workflow end to end with four modules that learn from 2.2K structured rulebooks and 180K player reviews:
\textit{BG-Ideator} turns multi-turn dialogue into a structured design draft;
\textit{BG-Realizer} writes a rulebook and revises it with \textit{BG-Critic}, a review-grounded critic that localizes flaws along the Mechanics--Dynamics--Aesthetics chain and accepts a revision only when it prefers it;
and \textit{BG-Persona} previews feedback from 150 profiled reviewers.
On 207 held-out games, \ourmethod localizes flaws in rulebooks and revises them better than general-purpose LLMs, and iterative revision brings their rating close to that of published rulebooks.
Blind readers prefer \ourmethod’s rulebooks to GPT-5.4’s in 38 of 50 new design intents under the same revision budget.
In a 30-participant user study and paper-prototype playtests, designers using \ourmethod report less blank-page anxiety and catch flaws they would otherwise miss, and the resulting rulebooks hold up at the table with few rule breakdowns.
\end{abstract}

\section{Introduction}
\label{sec:intro}

\begin{figure}[t]
    \centering
    \includegraphics[width=\columnwidth]{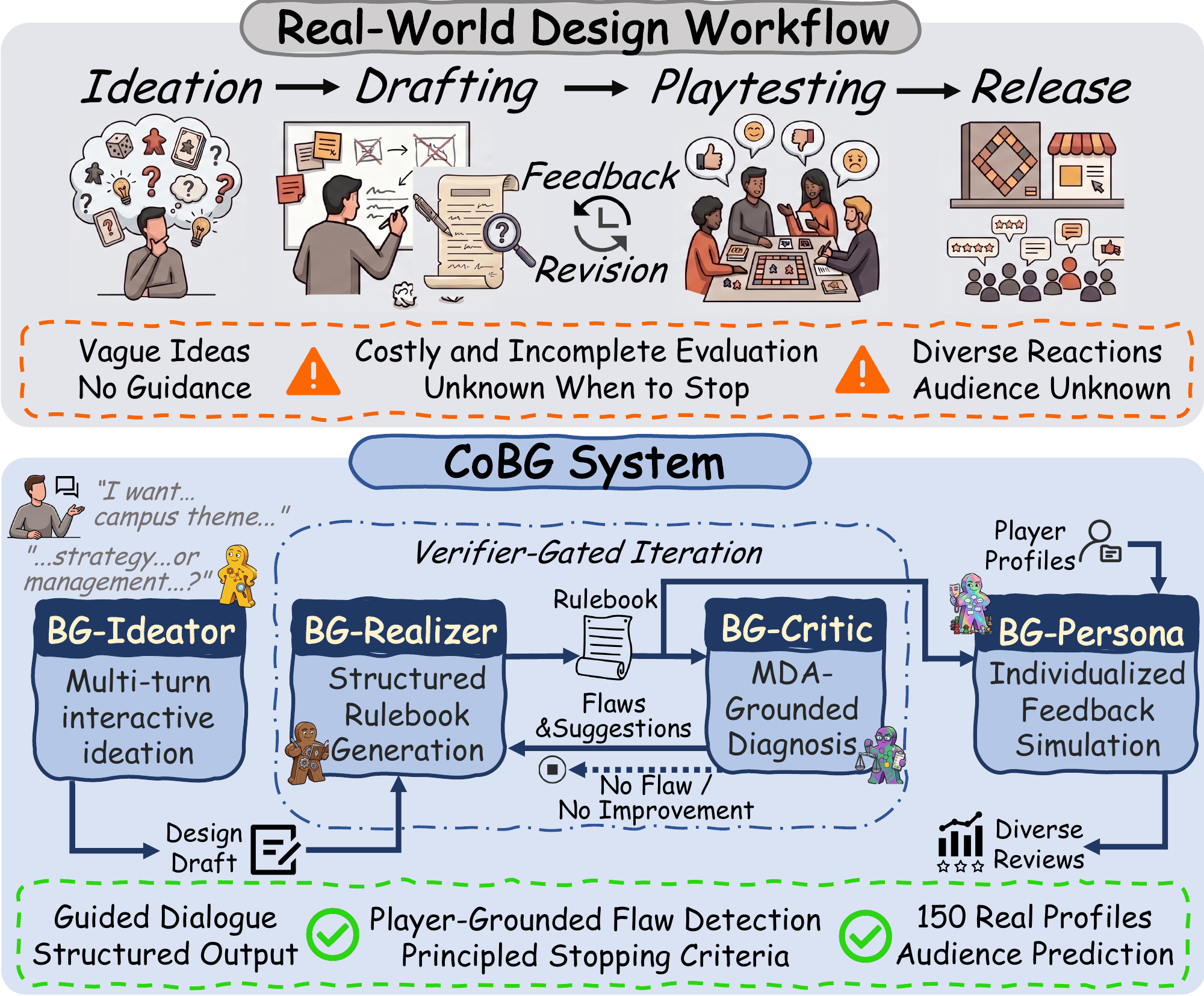}
\caption{\textbf{Overview of \ourmethod.} \textit{Top}: the real-world design workflow poses three challenges. \textit{Bottom}: \ourmethod addresses them in turn. \textbf{BG-Ideator} turns multi-turn dialogue into a structured design draft; \textbf{BG-Realizer} writes the rulebook, and \textbf{BG-Critic}, trained on player reviews, returns localized flaws and suggestions until it finds no flaw or no better revision; \textbf{BG-Persona} previews reactions from 150 real player profiles.}
\label{fig:overview}
\end{figure}

Board games have been widely adopted beyond entertainment, serving important roles in education~\cite{othman2025systematic,conway2026analogue}, therapy~\cite{noda2019effectiveness}, and research on cooperation~\cite{sousa2023playing}. However, designing a board game is cognitively demanding: the designer must constantly switch between writing rules and anticipating how players will experience them, and a finished game emerges only after repeated prototyping, playtesting, and revision~\cite{engelstein2022building,fullerton2024game}. Much of the work within each cycle, such as structuring a vague idea, spotting a flawed rule, or anticipating how players will react, could be shared with an AI assistant, shortening every round of the loop~\cite{Torii_2023,swacha2025supporting}.

Large language models bring such an assistant within reach: they have shown capabilities in brainstorming~\cite{10.1145/3706598.3713146,shahhosseini2026largelanguagemodelsscientific}, copywriting~\cite{wu2025ai,gunawan2026comprehensiveevaluationlargelanguage}, and critique~\cite{xia2026storyalignevaluatingtrainingreward,yang2025matters}, and in the board game domain they have been applied to mechanism ideation~\cite{hu2024game,Torii_2023}, code generation~\cite{Becker_2025,li2025cardiverseharnessingllmsnovel}, and experience evaluation of published games~\cite{li2026meeplelmvirtualplaytestersimulating}. However, each of these addresses an isolated stage, and supporting a designer from a vague idea to a revised, audience-tested rulebook raises three challenges (Figure~\ref{fig:overview}, top).
\textbf{(1)~Idea structuring:} early ideas are often little more than a theme or a few mechanics, so an assistant must ask guiding questions and organize fragments into a coherent draft.
\textbf{(2)~Closed-loop refinement:} revision needs an evaluator that infers dynamics and player experience from static text, localizes flaws, and recognizes when to stop, together with a reviser that can act on its diagnosis.
\textbf{(3)~Audience-specific feedback:} the same design draws different reactions from different players, so feedback should be conditioned on who is playing.

To address these challenges, we present \ourmethod (Figure~\ref{fig:overview}, bottom), a board game design assistant that learns the player perspective from 180K player reviews and 2.2K structured rulebooks through four modules.
\textit{BG-Ideator} guides designers through multi-turn dialogue, transforming scattered ideas into a structured design draft. \textit{BG-Realizer} then converts this draft into a seven-section rulebook. \textit{BG-Critic}, trained on player reviews rewritten along the Mechanics--Dynamics--Aesthetics (MDA) framework~\cite{hunicke2004mda}, learns what players count as a flaw: it rates rulebooks, localizes flaws, and compares versions. Together, BG-Critic and BG-Realizer run Verifier-Gated Iteration (VGI): a revision is kept only when the critic prefers it, and the loop stops when the critic reports no flaw or no better candidate. \textit{BG-Persona}, conditioned on the review histories of 150 individual players, previews how different audiences would react to the same design.

On 207 held-out games, \ourmethod outperforms general-purpose LLMs at diagnosing and repairing design flaws: against GPT-5.4, BG-Critic diagnoses flaws more accurately, including ${\sim}$1K flaws reported in real player reviews (5.71 vs.\ 3.78 out of 10), and blind readers judge that BG-Realizer resolves a given flaw more often (46\% vs.\ \proj{31\%}). Iterating the two brings the rating of its rulebooks close to that of published rulebooks, readers prefer the final version to the initial one in 74\% of sampled games, and BG-Critic improves GPT-5.4's rulebooks as well when used as its gate. BG-Persona, evaluated on held-out comments from real reviewers, ranks each reviewer's games more accurately than prompted models (84.3\% vs.\ 73.4\%).

The system as a whole also holds up with designers and players. On 50 new design intents, blind readers prefer \ourmethod's rulebooks to GPT-5.4's in 38 cases. In a 30-participant user study, most participants report that the dialogue eases the start of a design (19/30) and that BG-Critic surfaces flaws they would otherwise miss (24/30), and in paper-prototype playtests, \ourmethod's rulebooks break down less often at the table than GPT-5.4's (\proj{1.2} vs.\ \proj{3.0} per session).

Our main contributions are as follows:
\begin{itemize}[itemsep=2pt, topsep=2pt, parsep=0pt]
\item We present \ourmethod, an end-to-end board game design assistant from ideation to audience feedback, whose core is a critic trained on player reviews that localizes design flaws and accepts a revision only when it judges the rulebook improved.
\item We construct a dataset of 2.2K structured rulebooks and 180K player reviews, and derive training data for each module from it.
\item We evaluate \ourmethod on 207 held-out games and in three human studies, showing that it diagnoses and repairs flaws better than general-purpose LLMs, produces rulebooks that readers prefer to GPT-5.4’s, and helps designers start and refine their designs.
\end{itemize}

\section{Related Work}
\label{sec:related}

\subsection{Board Game Design and Personas}

LLMs have generated board game rules in formal description languages~\cite{hu2024game,todd2024gavelgeneratinggamesevolution,tanaka2025grammar}, prototyped card games~\cite{li2025cardiverseharnessingllmsnovel}, and implemented games from natural-language rules~\cite{Becker_2025}. Recent systems compose complete tabletop designs with specialized agents~\cite{cheung2026generativeontologystructuredknowledge} or tune rule parameters through LLM self-play~\cite{zeng2026rulesmithmultiagentllmsautomated}, yet strong models still misapply full rulebooks to concrete game states~\cite{peper2026llms}. Design assistants run a design sprint~\cite{Torii_2023} or grow game environments from a prompt~\cite{earle2025dreamgarden}, and MeepleLM simulates archetype-level player feedback on published rulebooks~\cite{li2026meeplelmvirtualplaytestersimulating}. LLM personas can simulate individuals~\cite{aher2023using,park2026llmagentsgroundedselfreports} and underlie synthetic personas~\cite{ge2024scaling}, personalized generation~\cite{salemi2024lamp}, and individualized alignment~\cite{kirk2024prismalignmentdatasetparticipatory,wang2026unified}, but they reproduce real human diversity and subjective judgments poorly~\cite{li2026llm,batzner2025whose,hu2024quantifying,jia2026digitalpersonasreliablyapproximate} and match individual evaluators only under restricted conditions or with person-specific evidence~\cite{dong-etal-2024-llm,zhou2025personaeval,he-etal-2026-personajudge}. None of these systems elicits a draft from a designer, revises a rulebook from a localized, review-grounded diagnosis, or tests its feedback against individual players' held-out comments.

\subsection{Critics for Iterative Refinement}

Self-refinement~\cite{madaan2023self} often fails without an external signal~\cite{huang2024largelanguagemodelsselfcorrect,kamoi-etal-2024-llms}, so many loops separate generation from verification with tools~\cite{gou2024criticlargelanguagemodels,chen2024teaching}, key-condition checks~\cite{wu2024proco}, or trained verifiers~\cite{cobbe2021trainingverifierssolvemath,lightman2023letsverifystepstep}. Algorithm~\ref{alg:vgi} is such a loop, best-of-$K$ selection under a learned preference~\cite{stiennon2020learning}; our focus is the verifier. For open-ended text, critics and reward models have been trained for stories~\cite{xia2026storyalignevaluatingtrainingreward} and writing tasks~\cite{ren-etal-2026-coarse}, and recent critics are trained to produce actionable revisions~\cite{du2026criticgengenerationawareevaluationactionable} or judged by the revision they enable~\cite{kim2026coevolvingactorconditionedcriticsnonverifiable}. Prompted models repair errors well once they are located, and locating them is the hard part~\cite{tyen2024llms}. A rulebook critic must further infer play dynamics and experience from static text, and we find that review-grounded training helps with both localization and repair.

\begin{figure*}[t!]
\centering
\small

\begin{minipage}[t]{0.56\textwidth}
\begin{tcolorbox}[
    colback=panelbg, colframe=gray!60, arc=2pt, boxrule=0.5pt,
    top=-1pt, bottom=2pt, left=3pt, right=3pt,
    title={\small\bfseries (a) $\diamond$ Interactive Ideation},
    coltitle=black, colbacktitle=gray!15,
    height=7.6cm, valign=top,
]
\scriptsize

\colorbox{userbubble}{\faUser\ \textbf{User:}} \textit{``I want to make a board game set on a college campus\ldots classes, clubs, social stuff. Not sure about the mechanics yet\ldots''}

\vspace{-1pt}
\colorbox{botbubble}{\faRobot\ \textbf{Ideator:}} Suggests card scheduling where value depends on campus priorities. Asks: \textit{``placing onto your own schedule, interacting with others, or something else?''}

\vspace{-1pt}
\colorbox{userbubble}{\faUser\ \textbf{User:}} \textit{``Picking sounds cool, but I don't want it to feel like solitaire. Can there be some competition for the same time slots\ldots only one gets in.''} \hfill {\color{blue!70}\scriptsize\textsf{[pushback]}}

\vspace{-1pt}
\colorbox{botbubble}{\faRobot\ \textbf{Ideator:}} Proposes shared campus grid with limited spots per row. When someone takes the last spot, that row disappears --- tension from scarcity.

\vspace{-1pt}
\colorbox{userbubble}{\faUser\ \textbf{User:}} \textit{``Reminds me of course registration\ldots everyone trying to grab electives before they fill up. That panic was real haha.''} \hfill {\color{violet}\scriptsize\textsf{[tangent]}}

\vspace{-1pt}
\colorbox{botbubble}{\faRobot\ \textbf{Ideator:}} Maps ``registration panic'' to design: time is the scarce resource. Proposes streak-scoring with an active window.

\vspace{-1pt}
\colorbox{userbubble}{\faUser\ \textbf{User:}} \textit{``The streak thing sounds clever but kind of fiddly\ldots I'd have to explain `active window' to new players\ldots Is there something in between?''} \hfill {\color{orange!80}\scriptsize\textsf{[hesitation]}}

\vspace{-1pt}
\colorbox{botbubble}{\faRobot\ \textbf{Ideator:}} Simplifies: runs anywhere, no window rule. \hfill {\color{teal}\scriptsize\textsf{$\leftarrow$ adapts to concern}}

\vspace{-1pt}
\colorbox{userbubble}{\faUser\ \textbf{User:}} \textit{``I specifically want actual trading. Not adjacency bonuses\ldots the talking, the deals --- that IS college. Don't water it down.''} \hfill {\color{red!70}\scriptsize\textsf{[opinionated]}}

\vspace{-1pt}
\colorbox{botbubble}{\faRobot\ \textbf{Ideator:}} Accepts. Bounds it: one swap per day. \hfill 
{\color{teal}\scriptsize\textsf{$\leftarrow$ respects \& constrains}}

\vspace{-1pt}
\hfill \textit{\ldots\; 11 turns total $\to$ complete draft} \hfill
\end{tcolorbox}
\end{minipage}%
\hspace{0.005\textwidth}%
\begin{minipage}[t]{0.42\textwidth}
\begin{tcolorbox}[
    colback=panelbg, colframe=gray!60, arc=2pt, boxrule=0.5pt,
    top=2pt, bottom=2pt, left=3pt, right=3pt,
    title={\small\bfseries (b) $\triangleright$ Structured Design Draft},
    coltitle=black, colbacktitle=gray!15,
    height=7.6cm, valign=top,
]
\scriptsize

\begin{tcolorbox}[colback=conceptblue, colframe=conceptblue!60, boxrule=0.3pt, arc=1pt, left=2pt, right=2pt, top=1pt, bottom=1pt, before skip=0pt, after skip=2pt]
\faLightbulb \, \textbf{Concept:} \textit{``Draft classes, clubs, and parties off a shared campus grid, then sequence them into streaks that pay big at day's end.''}
\end{tcolorbox}

\begin{tcolorbox}[colback=mechgreen, colframe=mechgreen!60, boxrule=0.3pt, arc=1pt, left=2pt, right=2pt, top=1pt, bottom=1pt, before skip=0pt, after skip=2pt]
\faCogs \, \textbf{Mechanics}\\
\textit{Core:} Open Drafting $\cdot$ Set Collection $\cdot$ Trading \\
\textit{Supporting:} Take That (denial via drafting) \\
\textit{Structural:} Score-and-Reset Game $\cdot$ Square Grid
\end{tcolorbox}

\begin{tcolorbox}[colback=intentamber, colframe=intentamber!60, boxrule=0.3pt, arc=1pt, left=2pt, right=2pt, top=1pt, bottom=1pt, before skip=0pt, after skip=2pt]
\faBullseye \, \textbf{Core Tension:} \textit{``Lock in a powerful early-slot opportunity now, or wait for the exact connector that makes your run pay this round?''}\\[2pt]
\faPuzzlePiece \, \textbf{Theme--Mechanic Fit:} \textit{``College life is inherently about scheduling conflicts and opportunistic planning, which maps cleanly to open drafting and ordering your own day. Social hangouts justify trading as a natural way to exchange favors without heavy negotiation.''}
\end{tcolorbox}

\begin{tcolorbox}[colback=paramgray, colframe=paramgray!60, boxrule=0.3pt, arc=1pt, left=2pt, right=2pt, top=1pt, bottom=0pt, before skip=0pt, after skip=0pt]
\faSlidersH \, \textbf{Parameters}\\[1pt]
\faChess \, \textit{Type:} Strategy \; \faTags \, \textit{Categories:} College, Party Game\\[1pt]
\faUserFriends \, \textit{Players:} 2--4 \; \faClock \, \textit{Time:} 30--45\,min \; \faBalanceScale \, \textit{Weight:} 2.1/5\\[1pt]
\faLanguage\, \textit{Language:} Some necessary text, easily memorized
\end{tcolorbox}

\end{tcolorbox}
\end{minipage}

\vspace{0.15em}

\begin{minipage}[t]{0.68\textwidth}
\begin{tcolorbox}[
    colback=panelbg, colframe=gray!60, arc=2pt, boxrule=0.5pt,
    top=4pt, bottom=2pt, left=3pt, right=3pt,
    title={\small\bfseries (c) $\circlearrowleft$ Iterative Refinement},
    coltitle=black, colbacktitle=gray!15,
    height=7.8cm, valign=top
]
\scriptsize

\begin{minipage}[t]{0.47\linewidth}
\raggedright
\colorbox{botbubble}{\makebox[\linewidth][s]{\textbf{Rulebook v0} \hfill {\color{gray}\scriptsize generated by BG-Realizer}}}
\faScroll\,\textbf{1. Lore \& Objective}\\
{\color{gray}\textit{In The Campus, players are college students navigating daily life --- balancing classes, clubs, parties, and study\ldots}}\\[1pt]
\faCubes\,\textbf{2. Components}\\
{\color{gray}\textit{108 Activity Cards (4 categories $\times$ 4 time slots, point values 1--4, some with special abilities), 4 Player Boards with ``Today's Plan'' track, Campus Display Board (4$\times$4 grid)\ldots}}\\[1pt]
\faPlayCircle\,\textbf{3. Setup}\\
{\color{gray}\textit{Shuffle cards, fill 4$\times$4 grid face-up, deal 3 cards each, choose first player\ldots}}\\[1pt]
\faRedo\,\textbf{4. Gameplay Flow} {\color{gray}(6 rounds)}\\
{\color{gray}\textit{Draft Activities $\to$ Plan Your Day $\to$ Hang Out (Trading) $\to$ Execute \& Score}}\\
\faCog\,\textbf{5. Core Mechanics}\\
{\color{gray}\textit{Open drafting from shared grid, contiguous same category streak scoring, one-for-one trading\ldots}}\\[1pt]
\faTrophy\,\textbf{6. Scoring \& End Game}\\
{\color{gray}\textit{Run bonuses: 2-in-a-row = 3pt, 3 = 6pt, 4+ = 10pt. 6 rounds, highest VP wins\ldots}}\\[1pt]
\faQuestionCircle\,\textbf{7. FAQ}\\

\end{minipage}%
\hspace{0.02\linewidth}%
\vrule width 0.4pt%
\hspace{0.01\linewidth}%
\begin{minipage}[t]{0.495\linewidth}
\colorbox{botbubble}{\makebox[\linewidth][s]{\textbf{v0} \hfill Rating: 6.17}}

\colorbox{flawred}{\textbf{\scriptsize \texttimes\ Flaw 1}} \texttt{\scriptsize M\_major}: \textit{``Place up to 3 cards into empty slots\ldots may be placed in any order\ldots''}\\
\colorbox{problemyellow}{\textbf{\scriptsize Problem:}}  Time-slot constraint removed; cards go in any slot regardless of printed time.\\
\colorbox{fixgreen}{\textbf{\scriptsize \checkmark\ Fix:}} Cards must be placed in matching time slot.

\vspace{1pt}
\colorbox{botbubble}{\makebox[\linewidth][s]{\textbf{v1} \hfill Rating: 6.27}}

\colorbox{flawred}{\textbf{\scriptsize \texttimes\ Flaw 2}} \texttt{\scriptsize D\_major}: \textit{``Phase 1: Draft $\to$ Phase 2: Plan $\to$ Phase 3: Hang Out $\to$ Phase 4: Score''}\\
\colorbox{problemyellow}{\textbf{\scriptsize Problem:}}  Trading after planning; cannot adjust hands before committing cards.\\
\colorbox{fixgreen}{\textbf{\scriptsize \checkmark\ Fix:}} Move Hang Out before Plan Your Day.

\vspace{1pt}
\colorbox{botbubble}{\makebox[\linewidth][s]{\textbf{v2} \hfill Rating: 6.40}}

\colorbox{flawred}{\textbf{\scriptsize \texttimes\ Flaw 3}} \texttt{\scriptsize M\_major}: \textit{``Players may hold up to 6 cards at a time.''}\\
\colorbox{problemyellow}{\textbf{\scriptsize Problem:}}  Hand limit raised from 5 to 6, reducing drafting pressure.\\
\colorbox{fixgreen}{\textbf{\scriptsize \checkmark\ Fix:}} Restore hand limit to 5.

\hrule
\vspace{3pt}
Rating: \textbf{6.17} $\to$ \textbf{6.27} $\to$ \textbf{6.40} \;{\color{gray}(30$\times$, BG-Critic)}
\end{minipage}

\end{tcolorbox}
\end{minipage}%
\hspace{0.005\textwidth}%
\begin{minipage}[t]{0.30\textwidth}
\begin{tcolorbox}[
    colback=panelbg, colframe=gray!60, arc=2pt, boxrule=0.5pt,
    top=4pt, bottom=2pt, left=3pt, right=3pt,
    title={\small\bfseries (d) $\bigstar$ Persona Feedback},
    coltitle=black, colbacktitle=gray!15,
    height=7.8cm, valign=top,
]
\scriptsize

\begin{tcolorbox}[colback=personapurple, colframe=personapurple!50, boxrule=0.3pt, arc=1pt, left=2pt, right=2pt, top=0pt, bottom=0pt, before skip=0pt, after skip=1.5pt]
\faUserCircle\,\textbf{Persona A} {\color{gray}\scriptsize Deterministic optimization, spatial logic, high agency}\\
\faStar\,Rating: \textbf{9}/10 --- \textit{``A delightful surprise\ldots building runs turns simple set collection into a spatial optimization challenge. \ldots fast, deterministic, and incredibly replayable.''}
\end{tcolorbox}

\begin{tcolorbox}[colback=personapurple, colframe=personapurple!50, boxrule=0.3pt, arc=1pt, left=2pt, right=2pt, top=0pt, bottom=0pt, before skip=0pt, after skip=1.5pt]
\faUserCircle\,\textbf{Persona B} {\color{gray}\scriptsize Tight Euro structures, tension without chaos, clean pacing}\\
\faStar\,Rating: \textbf{7}/10 --- \textit{``Solid engine builder with a drafting twist\ldots scoring is all about the runs, which makes the trading phase surprisingly tense. Clean, low-friction, respects your time.''}
\end{tcolorbox}

\begin{tcolorbox}[colback=personapurple, colframe=personapurple!50, boxrule=0.3pt, arc=1pt, left=2pt, right=2pt, top=0pt, bottom=0pt, before skip=0pt, after skip=0pt]
\faUserCircle\,\textbf{Persona C} {\color{gray}\scriptsize Demands elegance, hates fiddliness, harsh scorer}\\
\faStar\,Rating: \textbf{2}/10 --- \textit{``Theme is cute, but mechanics are a slog\ldots plays like solitaire with four people. Trading feels pointless\ldots needs a discard mechanic.''}
\end{tcolorbox}

\vfill
\vspace{5pt}
\hfill \textit{\scriptsize 150 profiles available; 3 shown}

\end{tcolorbox}
\end{minipage}

\caption{\textbf{Illustrative running example.} \textbf{(a)}~BG-Ideator elicits the design through dialogue, handling pushback, tangents, hesitation, and strong opinions. \textbf{(b)}~The resulting structured draft. \textbf{(c)}~Rulebook $v_0$ and three critic-localized revisions; tags give MDA layer and severity (\texttt{M\_major}: a major mechanics flaw), and ratings average 30 BG-Critic samples. \textbf{(d)}~Three reviewer profiles rate the same rulebook 9, 7, and 2, each for different reasons.}
\label{fig:case_study}
\end{figure*}

\section{Method}
\label{sec:method}

\subsection{System Overview}
\label{sec:overview}

\ourmethod mirrors the real-world board game design pipeline through four modules, as illustrated in Figure~\ref{fig:case_study}.
\textbf{BG-Ideator} (\S\ref{sec:ideation}) guides designers through multi-turn dialogue to convert vague, evolving ideas into a structured design draft (Figure~\ref{fig:case_study}a--b).
\textbf{BG-Realizer} (\S\ref{sec:realization}) converts the draft into a complete seven-section rulebook, and \textbf{BG-Critic} (\S\ref{sec:critic}) evaluates it along the MDA chain~\citep{hunicke2004mda}: \emph{mechanics} are the written rules, \emph{dynamics} are the behaviors those rules produce in play, and \emph{aesthetics} are the experiences players report. Each flaw is localized at one layer and traced along this chain.
The two models form a closed loop: BG-Critic pinpoints specific flaws, BG-Realizer revises, and a revision is kept only when BG-Critic prefers it to the current text (Figure~\ref{fig:case_study}c).
\textbf{BG-Persona} (\S\ref{sec:persona}) previews feedback from 150 real player profiles (Figure~\ref{fig:case_study}d).

\paragraph{Data Foundation.}
We extend the dataset of \citet{li2026meeplelmvirtualplaytestersimulating} to 2.2K verified structured rulebooks covering 192 mechanics and 190 themes, paired with 180K quality-filtered player reviews. Dataset statistics and the processing pipeline are provided in Appendix~\ref{app:data_statistics} and~\ref{app:data_processing}.

\subsection{BG-Ideator: Interactive Ideation}
\label{sec:ideation}

The ideation model, \textbf{BG-Ideator}, helps designers articulate their ideas into a structured design draft through multi-turn dialogue.
The draft schema, developed by domain experts from design literature, community blogs, and design guides, captures the designer's intent through five field groups: \emph{concept}, \emph{classification}, \emph{mechanics}, \emph{design intent}, and \emph{parameters} (Appendix~\ref{app:ideation_schema}).

\paragraph{Draft Construction.}
We select 1,200 seed games ranked in the top-6 per mechanic category from our rulebook collection.
For each seed, GPT-5.4 generates a structured draft from its rulebook, given definitions of the referenced mechanics and themes (Appendix~\ref{app:ideation_draft_prompt}).
Theme migration and mechanic hybridization then produce novel designs (Appendix~\ref{app:ideation_mutation_detail}), and verification by Gemini-3.1-Pro on 11 dimensions (Appendix~\ref{app:ideation_verify}) yields ${\sim}$3.3K mutated drafts with a mean quality of 4.84/5.

\paragraph{Interaction Data.}
GPT-5.4 produces ${\sim}$20K single-turn pairs, each targeting one of 15 aspects in six categories, and ${\sim}$4.7K multi-turn conversations by reverse construction from a target draft.
Conversations are conditioned on six user archetypes and eight dialogue dynamics (Appendix~\ref{app:ideation_aspects},~\ref{app:ideation_dialogue}).

\paragraph{Training.}
We train Qwen3.5-27B with LoRA in two stages: single-turn data builds atomic design capabilities across the six aspect categories, and multi-turn dialogues then teach conversational guidance and information elicitation (Appendix~\ref{app:ideation_hyperparams}).

\subsection{BG-Critic: Community-Grounded Critic}
\label{sec:critic}

\textbf{BG-Critic} produces MDA-structured assessments for three tasks: (1)~\emph{Rating}: given a rulebook, output a 1--10 score with a justification;
(2)~\emph{Diagnostic}: given a rulebook, identify flaw types, severity, affected components, and repair suggestions;
(3)~\emph{Comparison}: given two versions of a design, pick the one with fewer flaws to guide revision.

\paragraph{Data Construction.}
(1)~\emph{Rating.} GPT-5.4 rewrites the ${\sim}$180K reviews into a unified MDA format (Appendix~\ref{app:critic_mda_rewrite}). Three experienced players audited ${\sim}$3.6K chains from 200 games for factual consistency, causal plausibility, and stance; 96.3\% passed ($\kappa{=}0.74$), and invalid chains were removed.
(2)~\emph{Diagnostic.} About \proj{7.7K} examples come from three sources: ${\sim}$5K synthetic flaws injected by GPT-5.4 into real rulebooks with known labels, ${\sim}$\proj{2K} consensus flaws that several players independently report in reviews of the same game, and ${\sim}$800 top-30\% rulebooks as no-flaw examples (Appendix~\ref{app:critic_diagnostic_data}). Synthetic flaws provide exact locations and repair targets, while consensus flaws ground the labels in what players actually report.
(3)~\emph{Comparison.} Each perturbed rulebook is paired with its original (${\sim}$5K pairs).

\paragraph{Training.}
BG-Critic is trained in two stages on Qwen3.5-27B with LoRA.
Stage~1 trains on the ${\sim}$180K rewritten reviews to learn MDA conventions and domain knowledge; Stage~2 performs multi-task SFT on the Diagnostic, Comparison, and a replayed subset of Rating data (${\sim}$\proj{18K} instances; Appendix~\ref{app:critic_hyperparams}).

\subsection{BG-Realizer: Rulebook Generation}
\label{sec:realization}

\textbf{BG-Realizer} turns a draft into a seven-section rulebook and revises it from BG-Critic's diagnosis.

\paragraph{Data Construction.}
Generation training uses 1.2K real draft--rulebook pairs, plus ${\sim}$1.5K rulebooks that GPT-5.4 wrote for high-scoring mutated drafts with the parent rulebook as a one-shot example. A round-trip check converts each generated rulebook back into a draft and keeps the pair only if it matches the source draft (Appendix~\ref{app:real_generation}).
Revision training uses the perturbed rulebooks, with the ground-truth flaw as input and the original rulebook as the target (${\sim}$5K instances; Appendix~\ref{app:real_hyperparams}).

\paragraph{Training.}
BG-Realizer is trained on Qwen3.5-27B with LoRA via multi-task SFT on the combined generation and revision data, with seed pairs replicated $3\times$ to emphasize grounded examples.
Full hyperparameters are listed in Appendix~\ref{app:real_hyperparams}.

\paragraph{Verifier-Gated Iteration (VGI).}
At inference, BG-Realizer and BG-Critic alternate (Algorithm~\ref{alg:vgi}).
\textsc{Diagnose} returns \textsc{No\_Flaw} or a flaw report that conditions revision.
\textsc{Prefer}$(a,b)$ runs the comparison task twice with the two texts swapped and is true only if both orders prefer $a$.
Each of the $K$ candidates is scored by its number of wins against the others; ties are broken by Rating and then by index, so a winner exists even when preferences cycle.
The winner replaces the current rulebook only if it is preferred to it; otherwise the loop stops.

\begin{algorithm}[t]
\caption{Verifier-Gated Iteration}
\label{alg:vgi}
\begin{algorithmic}[1]
\REQUIRE Rulebook $v_0$; rounds $T$; candidates $K$; critic $C$; realizer $R$
\FOR{$t = 1$ \TO $T$}
    \STATE $d \leftarrow C.\textsc{Diagnose}(v_{t-1})$
    \IF{$d$ = \textsc{No\_Flaw}}
        \STATE \textbf{return} $v_{t-1}$
    \ENDIF
    \STATE $\{c_i\}_{i=1}^{K} \leftarrow R.\textsc{Revise}(v_{t-1},\; d)$
    \STATE $w_i \leftarrow |\{j \neq i : \mathrm{Prefer}(c_i, c_j)\}|$
    \STATE $c^* \leftarrow \arg\max_i\, (w_i,\; \mathrm{Rate}(c_i),\; -i)$
    \IF{$\mathrm{Prefer}(c^*,\, v_{t-1})$}
        \STATE $v_t \leftarrow c^*$
    \ELSE
        \STATE \textbf{return} $v_{t-1}$
    \ENDIF
\ENDFOR
\STATE \textbf{return} $v_T$
\end{algorithmic}
\end{algorithm}

\subsection{BG-Persona: Individualized Feedback}
\label{sec:persona}

For designers seeking targeted feedback, \textbf{BG-Persona} simulates responses from 150 real reviewers, each represented by a natural-language profile.
Given the profile and a rulebook, it outputs an MDA analysis from that reviewer's perspective, a comment, and a numeric rating.
For each reviewer with enough high-quality reviews, we draw 20--30 reviews by rating-stratified sampling, group them into three tiers (8--10, 5--7, 1--4), and have GPT-5.4 summarize them into a 200--300 word profile of preferences and evaluative habits. Usernames are removed before profiling (see Ethics Statement).
BG-Persona is fine-tuned from the BG-Critic checkpoint via LoRA on ${\sim}$3K instances (Appendix~\ref{app:persona_hyperparams}).

\section{Experiments}
\label{sec:exp}

Our experiments follow the design loop of \ourmethod. We first test whether BG-Critic recognizes what players count as a flaw (\S\ref{sec:exp_critic}) and whether BG-Realizer can repair a diagnosed flaw (\S\ref{sec:exp_realizer}), and then whether iterating the two under the critic's gate improves rulebooks further (\S\ref{sec:exp_closedloop}). \S\ref{sec:exp_persona} tests whether BG-Persona captures individual reviewers' preferences, and \S\ref{sec:exp_system} evaluates the assembled system with designers and players.

\subsection{Experimental Setup}
\label{sec:exp_setup}

\paragraph{Evaluation Set.}
We use the 207-game test split of MeepleLM~\cite{li2026meeplelmvirtualplaytestersimulating}, which is stratified by complexity and average rating, includes recent titles, and is disjoint from our training data (Appendix~\ref{app:data_statistics}). Each test game is processed through the same pipeline as training to provide a structured rulebook, a design draft, player reviews, and perturbed rulebooks with ground-truth flaw labels.

\paragraph{Baselines.}
We compare against GPT-5.4, Gemini-3.1-Flash, Qwen3.5-397B, and the untuned Qwen3.5-27B backbone~\cite{qwen35blog}, all given the same instruction and input format as our modules (Appendix~\ref{app:repro}). For rating and opinion coverage, we also rerun the public MeepleLM checkpoint, which writes player reviews for a given rulebook (Appendix~\ref{app:meeplelm_proto}).

\paragraph{Statistics.}
Unless noted, we use paired Wilcoxon signed-rank tests, exact McNemar tests for binary outcomes, and 95\% Wilson intervals. 

\subsection{BG-Critic Reliability}
\label{sec:exp_critic}

We test BG-Critic on its three tasks.
\textbf{Rating} takes a rulebook and outputs an MDA-structured review ending in a 1--10 score. We sample 30 reviews per test game and compare their mean score with the game's community rating by \textbf{MAE}, absolute systematic shift ($|\textbf{Bias}|$), and \textbf{Kendall's $\tau$}~\cite{kendall1938new} across games. The same samples give \textbf{Opinion Coverage}: GPT-5.4 lists the distinct opinions in each game's real reviews (36 on average), and Gemini-3.1-Pro judges what fraction of them appear in the sampled reviews.
\textbf{Diagnostic} takes a rulebook and outputs either a flaw report (type, severity, affected component, and repair suggestion) or \textsc{No\_Flaw}. We test on \proj{827} instances drawn from the test games in the same proportions as training (injected flaws, consensus flaws, and flaw-free rulebooks), and a Gemini-3.1-Pro judge scores \textbf{Quality} (0--10) against the reference flaws on six weighted dimensions, penalizing hallucinated flaws.
\textbf{Comparison} takes two versions of a rulebook and names the one with fewer flaws; we report \textbf{Accuracy} on the \proj{531} pairs of an injected-flaw rulebook and its original.
Diagnostic and Comparison use greedy decoding (Appendix~\ref{app:exp_diag_judge}).

\begin{table}[t]
\centering
\footnotesize
\setlength{\tabcolsep}{1.2pt}
\begin{tabular}{lcccccc}
\toprule
 & \multicolumn{3}{c}{\textbf{Rating}} & \textbf{Diag.} & \textbf{Comp.} & \textbf{Opin.} \\
\cmidrule(lr){2-4} \cmidrule(lr){5-5} \cmidrule(lr){6-6} \cmidrule(lr){7-7}
\multirow{-2}{*}{\textbf{Model}} & MAE$\downarrow$ & $|\text{Bias}|$$\downarrow$ & $\tau$$\uparrow$ & Q.$\uparrow$ & Acc.$\uparrow$ & Cov.$\uparrow$ \\
\midrule
GPT-5.4            & 1.30 & 1.29 & 0.336 & 3.92 & 92.1 & 65.2 \\
Gemini-3.1-Flash   & 1.11 & 1.09 & 0.250 & 2.64 & 88.7 & 49.8 \\
Qwen3.5-397B       & 1.05 & 1.04 & 0.232 & 2.39 & 85.5 & 51.7 \\
Qwen3.5-27B        & 0.76 & 0.72 & 0.186 & 1.99 & 90.2 & 23.9 \\
MeepleLM           & 0.68 & 0.24 & 0.282 & -- & -- & 68.9 \\
\midrule
BG-Critic              & 0.50 & 0.18 & 0.360 & \textbf{6.07} & \textbf{96.8} & \textbf{73.5} \\
\quad w/o Stage 2       & \textbf{0.49} & \textbf{0.16} & \textbf{0.368} & 1.76 & 73.1 & 71.8 \\
\bottomrule
\end{tabular}
\caption{\textbf{BG-Critic compared with general-purpose LLMs and MeepleLM on the test games.} Diag.\ Q.: diagnostic quality (0--10); Comp.\ Acc.: comparison accuracy; Opin.\ Cov.: opinion coverage. BG-Critic leads on diagnosis, comparison, and coverage, and Stage~2 is what enables the first two. MeepleLM does not produce our flaw schema.}
\label{tab:critic_all}
\end{table}

As Table~\ref{tab:critic_all} shows, BG-Critic leads every baseline in diagnosis and comparison, with 6.07 and 96.8\% against 3.92 and 92.1\% for GPT-5.4 ($p{<}0.001$ and $p{<}0.01$). Its rating error is 0.18 below that of MeepleLM, the strongest external rater under our protocol. Its sampled reviews also cover more of the opinions in real reviews (73.5\%, against 65.2\% for GPT-5.4 and 23.9\% for the untuned backbone), so review grounding shows up in what the critic says as well as in its scores.

\paragraph{Effect of Stage~2.}
After Stage~1 alone, rating alignment is already strong (MAE 0.49), but diagnostic quality is only 1.76 and comparison accuracy 73.1\%. Stage~2 raises them to 6.07 and 96.8\%, turning a rater into a verifier.

\paragraph{Real Community Flaws.}
We further build a set of ${\sim}$1K flaws mined from player reviews and posed on the unperturbed published rulebooks, ${\sim}$600 of them from 170 additional games disjoint from training and test data (Appendix~\ref{app:real_flaws}).
BG-Critic scores 5.71, against 3.78 for GPT-5.4 and 2.52 for Gemini-3.1-Flash ($p{<}0.001$). Real flaws are harder than the in-domain test set (5.71 vs.\ 6.07), and the ranking is unchanged.
Three experienced players re-judged 200 of these instances; their majority agrees with the judge on 86.7\% (Cohen's $\kappa{=}0.71$). BG-Critic's lead thus holds on flaws that players report.

\subsection{Rulebook Generation and Repair}
\label{sec:exp_realizer}

For generation, each model writes a rulebook from each test draft, and BG-Critic rates it (mean of 30 samples).
For revision, each model receives one perturbed rulebook per test game together with its ground-truth flaw and returns a revised rulebook, so the task isolates repair from diagnosis. \textbf{Impr.\%} is the fraction of repairs that BG-Critic prefers to the perturbed input under position swapping, and \textbf{NoFlaw\%} the fraction for which BG-Critic reports no remaining flaw.

\begin{table}[t]
\centering
\footnotesize
\setlength{\tabcolsep}{3pt}
\begin{tabular}{lccc}
\toprule
 & \textbf{Gen.} & \multicolumn{2}{c}{\textbf{Revision}} \\
\cmidrule(lr){2-2} \cmidrule(lr){3-4}
\multirow{-2}{*}{\textbf{Model}} & Rating$\uparrow$ & Impr.\%$\uparrow$ & NoFlaw\%$\uparrow$ \\
\midrule
GPT-5.4            & \textbf{7.01} & \proj{93.7} & \proj{13.0} {\scriptsize(\phantom{0}\proj{9.1--18.3})} \\
Gemini-3.1-Flash   & 6.69 & \proj{91.3} & \proj{17.9} {\scriptsize(\proj{13.3--23.7})} \\
Qwen3.5-397B       & 6.82 & \proj{90.8} & 12.6 {\scriptsize(\phantom{0}8.7--17.8)} \\
Qwen3.5-27B        & 6.80 & \proj{89.9} & 17.4 {\scriptsize(12.8--23.1)} \\
\midrule
BG-Realizer         & 6.95 & \textbf{\proj{96.1}} & \textbf{32.9} {\scriptsize(26.8--39.5)} \\
\bottomrule
\end{tabular}
\caption{\textbf{Single-pass generation and revision} (given the ground-truth flaw) on the test games, scored by BG-Critic; parentheses give 95\% Wilson intervals. BG-Realizer matches GPT-5.4 in generation and clears nearly twice as many repairs as any baseline.}
\label{tab:realization_all}
\end{table}

As Table~\ref{tab:realization_all} shows, BG-Realizer generates rulebooks on par with GPT-5.4 (6.95 vs.\ 7.01, $p{=}0.31$); the difference lies in revision, where its repairs are cleared nearly twice as often as those of the best baseline (32.9\% vs.\ \proj{17.9}\%, $p{<}0.001$).

\paragraph{Human Repair Check.}
We further test whether revisions resolve complaints that players actually raised, as judged by human readers. We sample 165 review-derived flaws (one per game, excluding games used for revision training). BG-Realizer, GPT-5.4, and Gemini-3.1-Flash each revise the published rulebook once, given the same flaw record and no reference fix. Three blind readers (designers or veteran players) judge whether each revision resolves the complaint without introducing a new contradiction; 15 items that could not be decided without playing are dropped.
A majority judges the complaint cleared for 46\% of BG-Realizer's revisions, against \proj{31\%} for GPT-5.4 and 28\% for Gemini ($\proj{+15}$ and $+18$ points, $p{<}\proj{0.01}$ and $p{<}0.001$; Fleiss' $\kappa{=}0.55$). The gain is largest on mechanics flaws and smallest on aesthetics (Appendix~\ref{app:human_repair}), so learned repair transfers to real complaints, most clearly where the text shows the fix.

\subsection{Verifier-Gated Iteration}
\label{sec:exp_closedloop}

We run VGI from BG-Realizer's initial rulebooks ($v_0$) with $K{=}5$ candidates and at most $T{=}3$ rounds, and compare it with two settings under the same budget: \emph{ungated} revision, which revises from a generic instruction without a flaw report and keeps the first candidate of each round, so neither diagnosis nor selection is used, and a \emph{GPT-5.4 gate}, which replaces BG-Critic in \textsc{Diagnose} and \textsc{Prefer}. We repeat all settings with GPT-5.4 as the generator (Appendix~\ref{app:loop_proto}). Rating and NoFlaw\% follow Table~\ref{tab:realization_all}, now on the generated rulebooks.

\begin{table}[t]
\centering
\footnotesize
\setlength{\tabcolsep}{4.5pt}
\begin{tabular}{lcccc}
\toprule
 & \multicolumn{2}{c}{\textbf{BG-Realizer}} & \multicolumn{2}{c}{\textbf{GPT-5.4}} \\
\cmidrule(lr){2-3} \cmidrule(lr){4-5}
\textbf{Setting} & Rating & NoFlaw\% & Rating & NoFlaw\% \\
\midrule
Initial         & 6.95 & 30.4 & 7.01 & \proj{13.5} \\
Ungated         & 6.99 & \proj{32.4} & \proj{7.03} & \proj{15.5} \\
GPT-5.4 gate    & 7.01 & \proj{33.3} & 7.04 & \proj{15.9} \\
BG-Critic gate  & \textbf{7.08} & \textbf{36.7} & \textbf{7.07} & \textbf{\proj{19.3}} \\
\bottomrule
\end{tabular}
\caption{\textbf{Iteration from each generator's initial rulebooks} ($K{=}5$, $T{\le}3$), scored by BG-Critic; the BG-Critic gate with BG-Realizer is VGI. Ratings average 30 samples and thus span a narrow range (published rulebooks: 7.12 and 33.3\%). The BG-Critic gate gives the largest gain for both generators.}
\label{tab:loop}
\end{table}

\begin{figure}[t]
\centering
\includegraphics[width=\columnwidth]{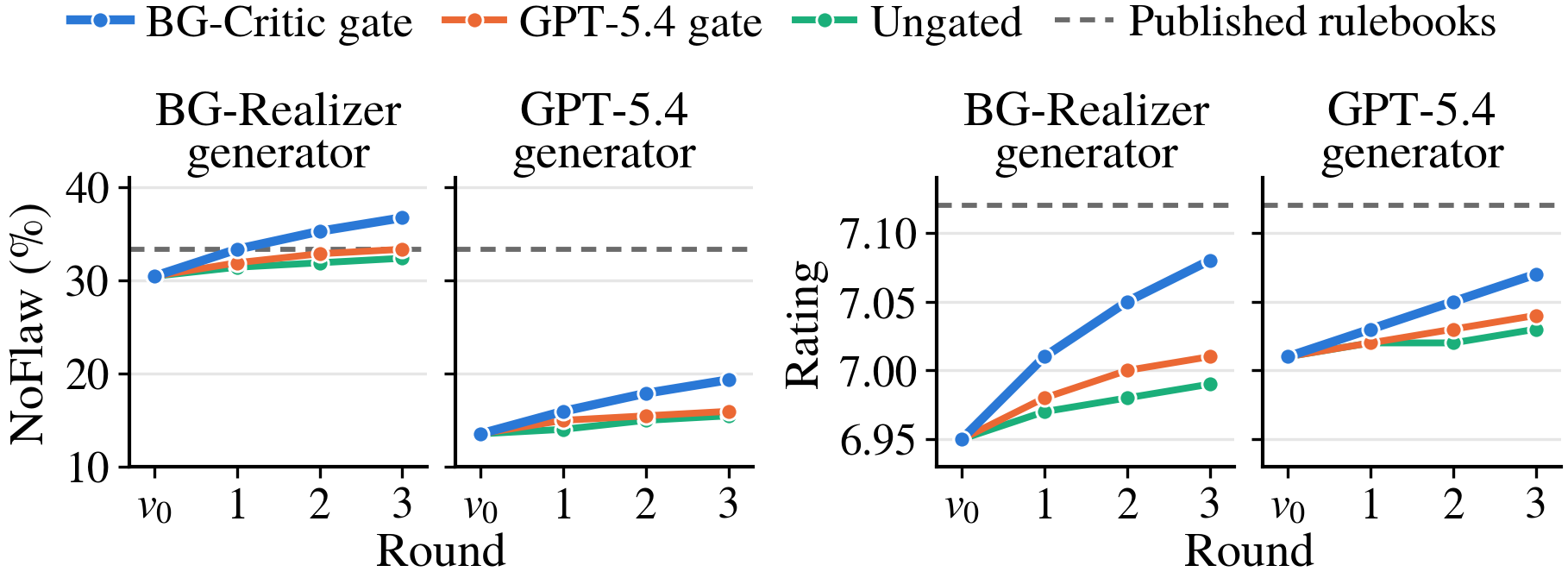}
\caption{\textbf{NoFlaw\% and BG-Critic rating after each round of iteration}; panel titles give the generator, colors the gate, and the dashed line the published rulebooks of the same games (per-round scoring in Appendix~\ref{app:loop_proto}). The BG-Critic gate improves both generators in every round, while the other two settings level off.}
\label{fig:vgi_rounds}
\end{figure}

As Table~\ref{tab:loop} shows, VGI raises the rating from 6.95 to 7.08 and NoFlaw\% from 30.4\% to 36.7\% (both $p{<}\proj{0.001}$); 13 games gain \textsc{No\_Flaw} and, by construction, none lose it. Among the ${\sim}$30K games ranked on BoardGameGeek, ordered by average rating, a rise of this size moves a game up about 1,570 places (Appendix~\ref{app:loop_proto}). As a reference, BG-Critic scores the published rulebooks of the same games at 7.12 and 33.3\%; VGI brings generated rulebooks close to them in rating and past them in NoFlaw\%. On the 144 games flagged at $v_0$, the loop averages \proj{2.44} rounds, and the gate rejects the winning candidate in 17.4\% of rounds.

\paragraph{Choice of Gate.}
Under the same budget, ungated revision adds \proj{2.0} points of NoFlaw\% and a GPT-5.4 gate \proj{2.9}, against 6.3 for BG-Critic. Figure~\ref{fig:vgi_rounds} shows that the gap widens with each round for both generators: the other two settings level off after the first round, while the BG-Critic gate keeps improving NoFlaw\% and rating and alone lifts BG-Realizer past published NoFlaw\%. The gain of iteration thus comes mainly from the review-grounded critic that diagnoses flaws and selects revisions.

\paragraph{Choice of Generator.}
With GPT-5.4 as the generator, the BG-Critic gate adds \proj{5.8} points, more than twice the gain of GPT-5.4's own gate (\proj{2.4}). GPT-5.4's initial rulebooks rate high (7.01) but rarely receive \textsc{No\_Flaw} (\proj{13.5}\%), since \textsc{No\_Flaw} requires the absence of any localizable flaw. BG-Realizer starts \proj{16.9} points ahead and stays ahead under every gate. Most of the advantage therefore comes from learned generation and repair, and the BG-Critic gate adds a consistent increment on top of either generator.

\paragraph{Human Evaluation.}
In a version check, we sample 50 of the \proj{122} games in which VGI accepted at least one revision and show three blind readers the initial and final rulebooks in random order (Appendix~\ref{app:loop_proto}). The majority prefers the final rulebook in 74\% of games, and its choice agrees with BG-Critic's own comparison on 82\% of pairs.
With GPT-5.4 as the generator, six blind evaluators compared its BG-Critic-gated and self-gated rulebooks on 50 new design drafts (Appendix~\ref{app:blind_eval}) and preferred the former in \proj{29} pairs and the latter in \proj{11}, with \proj{10} ties ($p{<}\proj{0.01}$). The improvements accepted by BG-Critic are thus also visible to human readers.

\subsection{Individualized Feedback}
\label{sec:exp_persona}

We evaluate BG-Persona on 210 held-out reviews from 50 profiled players (3--5 games each); their test comments are excluded when profiles are written. Given a player's profile and a rulebook, each model outputs an MDA analysis, a comment, and a 1--10 rating, which we compare with that player's held-out review.
We report rating \textbf{MAE}, two ordering metrics, and Gemini-3.1-Pro comment scores for preference, reasoning, and style (Appendix~\ref{app:persona_judge}). For each player, \textbf{Pair} is the share of their test-game pairs (with different true ratings) that the predicted ratings order correctly, and \textbf{Extreme} checks only the pair of the player's highest- and lowest-rated games. Ordering rewards knowing which game a player prefers even when ratings are off (e.g., 7.4 vs.\ 6.8 for true 9 vs.\ 5), which MAE does not.

\begin{table}[t]
\centering
\footnotesize
\setlength{\tabcolsep}{2.5pt}
\begin{tabular}{lcccccc}
\toprule
 & & & & \multicolumn{3}{c}{\textbf{Comment}$\uparrow$} \\
\cmidrule(lr){5-7}
\multirow{-2}{*}{\textbf{Model}} & \multirow{-2}{*}{MAE$\downarrow$} & \multirow{-2}{*}{Pair$\uparrow$} & \multirow{-2}{*}{Extr.$\uparrow$} & P. & R. & S. \\
\midrule
GPT-5.4          & 1.24 & 73.4 & 68.0 & 5.4 & 4.7 & 3.9 \\
Gemini-3.1-Flash & 1.49 & 70.7 & 70.0 & 5.3 & 4.5 & 3.6 \\
Qwen3.5-397B     & 1.45 & 64.6 & 60.0 & 5.0 & 4.2 & 3.4 \\
Qwen3.5-27B      & 1.90 & 65.7 & 52.0 & 5.0 & 3.8 & 3.2 \\
\midrule
Community mean   & 1.36 & 69.2 & 64.0 & -- & -- & -- \\
\midrule
BG-Persona       & \textbf{1.19} & \textbf{84.3} & \textbf{76.0} & \textbf{5.7} & \textbf{5.0} & \textbf{4.3} \\
\quad w/o profile & 1.33 & 71.8 & 64.0 & 5.1 & 4.3 & 3.4 \\
\quad shuffled profile & 1.47 & 70.4 & 62.0 & 4.9 & 4.1 & 3.6 \\
\bottomrule
\end{tabular}
\caption{\textbf{Individualized feedback on 210 held-out reviews from 50 profiled players.} Pair/Extr.: within-player ordering accuracy (\%); Comment: preference, reasoning, and style (1--10). Community mean, w/o profile, and shuffled profile are controls without the individual; only BG-Persona with the matched profile orders games well above them.}
\label{tab:persona}
\end{table}

As Table~\ref{tab:persona} shows, BG-Persona orders a reviewer's games more accurately than every baseline, reaching 84.3\% pairwise accuracy against 73.4\% for GPT-5.4 with the same profile ($p{<}0.01$). Rating MAE improves only modestly (1.19 vs.\ 1.24) and comment scores stay moderate (reasoning 5.0, style 4.3), so BG-Persona mainly captures \emph{which} games a reviewer prefers, while reproducing the reviewer's voice remains harder.

\paragraph{Profile Controls.}
To isolate the individual, we keep the rulebook and remove the player: \emph{community mean} predicts each game's average community rating for every player, \emph{w/o profile} removes the profile, and \emph{shuffled profile} gives it another player's profile (Appendix~\ref{app:persona_controls}). All three reach only 69.2--71.8\% pairwise accuracy ($p{<}\proj{0.001}$ each), so the ordering gain comes from the matched profile.

\subsection{Evaluation with Designers and Players}
\label{sec:exp_system}

Finally, we compare the assembled system with GPT-5.4 in three human studies.

\paragraph{Blind Comparison.}
Ten participants not involved in development (three designers, seven veteran players) each wrote five design intents, which a separate expert turned into drafts (Appendix~\ref{app:blind_eval}). On each draft, \ourmethod ran BG-Realizer with VGI, and GPT-5.4 ran the same loop and budget with itself as the gate. Six evaluators who did not write the intents, blind to the source, preferred \ourmethod in 38 of 50 pairs and GPT-5.4 in 6, with 6 ties (sign test, $p{<}0.001$; Fleiss' $\kappa{=}0.64$).

\paragraph{User Study.}
Thirty participants (10 designers, 15 hobbyists, 5 novices) each built a rulebook with \ourmethod, and the 22 with prior LLM experience repeated the task with stage-prompted GPT-5.4 in counterbalanced order (Appendix~\ref{app:user_study}). Most reported reduced blank-page anxiety (19/30) and flaws they would otherwise have missed (24/30). \ourmethod was rated higher on all six dimensions (Table~\ref{tab:user_study_retro}), most on feedback helpfulness (6.0 vs.\ 4.3) and iteration improvement (6.0 vs.\ 3.7; both $p{<}\proj{0.001}$), the two stages that BG-Critic drives, and also on intent capture for BG-Ideator (5.6 vs.\ 4.8); novices found critic feedback less actionable (4.8).

\paragraph{Paper-Prototype Playtest.}
In a playtest styled on a designer's internal alpha test (Appendix~\ref{app:playtest}), groups of three players new to the designs played the two rulebooks of \proj{12} participants as paper prototypes, blind to the system and in counterbalanced order, while a facilitator logged every \emph{rule breakdown}, a point where players had to invent or vote on a ruling. \ourmethod's rulebooks broke down less often (\proj{1.2} vs.\ \proj{3.0} per session, $p{<}\proj{0.01}$) and were rated clearer (\proj{5.6} vs.\ \proj{4.2}); the largest share of GPT-5.4's breakdowns (\proj{15} of \proj{36}) were missing edge cases such as tie-breaks and end-game triggers. Completion (\proj{11} vs.\ \proj{8} of 12 games) and willingness to play again (\proj{4.9} vs.\ \proj{4.4}) differed less, so the playtest supports playability, while balance and long-term enjoyment remain untested.

\section{Conclusion}
\label{sec:conclusion}

We presented \ourmethod, a board game design assistant that learns from player reviews to support ideation, rulebook generation, verifier-gated revision, and individualized feedback. Review-grounded training improves flaw localization and repair; VGI adds a consistent increment on top of either generator; individualized ordering depends on the matched profile; and a paper-prototype playtest indicates that the rules hold up at the table, though balance and long-term enjoyment remain untested. Grounding a critic in real audience feedback, and validating it outside the loop it drives, is a recipe we expect to carry over to other feedback-rich creative domains.

\section*{Limitations}
(1)~\textbf{Physical play.} Our playtest uses paper prototypes of short games (at most 60 minutes) in single sessions. It shows that the rules can be followed at the table, while balance, strategic depth, and long-term enjoyment call for repeated play with physical components.
(2)~\textbf{Text-only modality.} \ourmethod reads and writes rulebooks as text, whereas real rulebooks also rely on board layouts, card designs, and illustrations. Extending the modules to multimodal input and output is a natural next step.
(3)~\textbf{Persona profiles.} Each profile summarizes a reviewer's written reviews. Richer signals, such as interaction history and complete play logs, could capture preferences that reviews leave implicit and extend BG-Persona to players without a review history.

\section*{Ethics Statement}
\paragraph{Data Privacy.}
The dataset is built from publicly posted rulebooks and reviews. Usernames and review identifiers are removed before training. We will release the \proj{LoRA weights of all four modules}, prompts, evaluation scripts, and the processed supervision we are permitted to redistribute; raw rulebook PDFs are not ours to redistribute (Appendix~\ref{app:repro}).
\paragraph{Reviewer Profiles.}
The 150 profiles are model-written summaries of public reviews, used to preview how a preference pattern might react to a draft. Because the source posts are public, reviewers were not contacted individually. Profiles are not voice clones, are never presented as coming from a named person, and will not be released in a form that re-identifies an account; we do not endorse using them to impersonate reviewers.
\paragraph{Human Studies.}
All human studies, namely the MDA audit, the human evaluations, the user study, and the playtest, were approved by the authors' institutional ethics board. Participants were recruited from board game communities, gave informed consent, could withdraw at any time, and were paid USD 10 per hour regardless of completion. Playtest sessions were recorded with the players' consent and used only to code rule breakdowns. Logs, recordings, and survey responses are stored anonymized, with access limited to the research team.

\bibliography{custom}

\begin{thebibliography}{61}
\providecommand{\natexlab}[1]{#1}

\bibitem[{Aher et~al.(2023)Aher, Arriaga, and Kalai}]{aher2023using}
Gati~V Aher, Rosa~I Arriaga, and Adam~Tauman Kalai. 2023.
\newblock Using large language models to simulate multiple humans and replicate
  human subject studies.
\newblock In \emph{International conference on machine learning}, pages
  337--371. PMLR.

\bibitem[{Batzner et~al.(2025)Batzner, Stocker, Tang, Natarajan, Chen, Schmid,
  and Kasneci}]{batzner2025whose}
Jan Batzner, Volker Stocker, Bingjun Tang, Anusha Natarajan, Qinhao Chen,
  Stefan Schmid, and Gjergji Kasneci. 2025.
\newblock Whose personae? synthetic persona experiments in llm research and
  pathways to transparency.
\newblock In \emph{Proceedings of the AAAI/ACM Conference on AI, Ethics, and
  Society}, volume~8, pages 343--354.

\bibitem[{Becker et~al.(2025)Becker, Oliveira, Rossato, and
  Tavares}]{Becker_2025}
Álvaro~Guglielmin Becker, Gabriel Bauer~de Oliveira, Lana~Bertoldo Rossato,
  and Anderson~Rocha Tavares. 2025.
\newblock \href {https://doi.org/10.5753/sbgames.2025.10222} {Boardwalk:
  Towards a framework for creating board games with llms}.
\newblock In \emph{Anais do XXIV Simpósio Brasileiro de Jogos e Entretenimento
  Digital (SBGames 2025)}, SBGames 2025, pages 655--667. Sociedade Brasileira
  de Computação.

\bibitem[{Chen et~al.(2024)Chen, Lin, Sch{\"a}rli, and Zhou}]{chen2024teaching}
Xinyun Chen, Maxwell Lin, Nathanael Sch{\"a}rli, and Denny Zhou. 2024.
\newblock \href {https://openreview.net/forum?id=KuPixIqPiq} {Teaching large
  language models to self-debug}.
\newblock In \emph{The Twelfth International Conference on Learning
  Representations}.

\bibitem[{Cheung(2026)}]{cheung2026generativeontologystructuredknowledge}
Benny Cheung. 2026.
\newblock \href {https://arxiv.org/abs/2602.05636} {Generative ontology: When
  structured knowledge learns to create}.
\newblock \emph{Preprint}, arXiv:2602.05636.

\bibitem[{Cobbe et~al.(2021)Cobbe, Kosaraju, Bavarian, Chen, Jun, Kaiser,
  Plappert, Tworek, Hilton, Nakano, Hesse, and
  Schulman}]{cobbe2021trainingverifierssolvemath}
Karl Cobbe, Vineet Kosaraju, Mohammad Bavarian, Mark Chen, Heewoo Jun, Lukasz
  Kaiser, Matthias Plappert, Jerry Tworek, Jacob Hilton, Reiichiro Nakano,
  Christopher Hesse, and John Schulman. 2021.
\newblock \href {https://arxiv.org/abs/2110.14168} {Training verifiers to solve
  math word problems}.
\newblock \emph{Preprint}, arXiv:2110.14168.

\bibitem[{Conway and Smith(2026)}]{conway2026analogue}
Elaine Conway and Ruth Smith. 2026.
\newblock Analogue play in the age of ai: A scoping review of non-digital games
  as active learning strategies in higher education.
\newblock \emph{Behavioral Sciences}, 16(1):133.

\bibitem[{Dong et~al.(2024)Dong, Hu, and Collier}]{dong-etal-2024-llm}
Yijiang~River Dong, Tiancheng Hu, and Nigel Collier. 2024.
\newblock \href {https://doi.org/10.18653/v1/2024.findings-emnlp.592} {Can
  {LLM} be a personalized judge?}
\newblock In \emph{Findings of the Association for Computational Linguistics:
  EMNLP 2024}, pages 10126--10141, Miami, Florida, USA. Association for
  Computational Linguistics.

\bibitem[{Du et~al.(2026)Du, Zuo, Wang, Fan, Wang, and
  Yang}]{du2026criticgengenerationawareevaluationactionable}
Huifang Du, Zecheng Zuo, Sen Wang, Chenghao Fan, Haofen Wang, and Yehui Yang.
  2026.
\newblock \href {https://arxiv.org/abs/2609.05439} {Criticgen: Generation-aware
  evaluation as actionable feedback}.
\newblock \emph{Preprint}, arXiv:2609.05439.

\bibitem[{Earle et~al.(2025)Earle, Parajuli, and
  Banburski-Fahey}]{earle2025dreamgarden}
Sam Earle, Samyak Parajuli, and Andrzej Banburski-Fahey. 2025.
\newblock Dreamgarden: A designer assistant for growing games from a single
  prompt.
\newblock In \emph{Proceedings of the 2025 CHI Conference on Human Factors in
  Computing Systems}, pages 1--19.

\bibitem[{Elias et~al.(2020)Elias, Garfield, and
  Gutschera}]{elias2020characteristics}
George~Skaff Elias, Richard Garfield, and K~Robert Gutschera. 2020.
\newblock \emph{Characteristics of games}.
\newblock MIT Press.

\bibitem[{Engelstein and Shalev(2022)}]{engelstein2022building}
Geoffrey Engelstein and Isaac Shalev. 2022.
\newblock \emph{Building Blocks of tabletop game design: An encyclopedia of
  mechanisms}.
\newblock Crc Press.

\bibitem[{Fullerton(2024)}]{fullerton2024game}
Tracy Fullerton. 2024.
\newblock \emph{Game design workshop: a playcentric approach to creating
  innovative games}.
\newblock AK Peters/CrC Press.

\bibitem[{Ge et~al.(2024)Ge, Chan, Wang, Yu, Mi, and Yu}]{ge2024scaling}
Tao Ge, Xin Chan, Xiaoyang Wang, Dian Yu, Haitao Mi, and Dong Yu. 2024.
\newblock Scaling synthetic data creation with 1,000,000,000 personas.
\newblock \emph{arXiv preprint arXiv:2406.20094}.

\bibitem[{Gou et~al.(2024)Gou, Shao, Gong, Shen, Yang, Duan, and
  Chen}]{gou2024criticlargelanguagemodels}
Zhibin Gou, Zhihong Shao, Yeyun Gong, Yelong Shen, Yujiu Yang, Nan Duan, and
  Weizhu Chen. 2024.
\newblock \href {https://openreview.net/forum?id=Sx038qxjek} {{CRITIC}: Large
  language models can self-correct with tool-interactive critiquing}.
\newblock In \emph{International Conference on Learning Representations}.

\bibitem[{Gunawan and
  Amien(2026)}]{gunawan2026comprehensiveevaluationlargelanguage}
Go~Frendi Gunawan and Mukhlis Amien. 2026.
\newblock \href {https://arxiv.org/abs/2602.07079} {Comprehensive evaluation of
  large language models on software engineering tasks: A multi-task benchmark}.
\newblock \emph{Preprint}, arXiv:2602.07079.

\bibitem[{He et~al.(2026)He, Qi, Chidambaram, Xu, Arannil, Chilton, and
  Williams}]{he-etal-2026-personajudge}
Zeyu He, Xuan Qi, Subramanian Chidambaram, Zhichao Xu, Vinayak Arannil, Lydia
  Chilton, and Alex~C. Williams. 2026.
\newblock \href {https://doi.org/10.18653/v1/2026.sigdial-1.59}
  {{PERSONAJUDGE}: Simulating individual human preference judgments with
  evaluator-specific demonstration data}.
\newblock In \emph{Proceedings of the 27th Annual Meeting of the Special
  Interest Group on Discourse and Dialogue}, pages 841--862, Atlanta, Georgia,
  USA. Association for Computational Linguistics.

\bibitem[{Hu et~al.(2024)Hu, Zhao, and Liu}]{hu2024game}
Chengpeng Hu, Yunlong Zhao, and Jialin Liu. 2024.
\newblock Game generation via large language models.
\newblock In \emph{2024 IEEE Conference on Games (CoG)}, pages 1--4. IEEE.

\bibitem[{Hu and Collier(2024)}]{hu2024quantifying}
Tiancheng Hu and Nigel Collier. 2024.
\newblock Quantifying the persona effect in llm simulations.
\newblock In \emph{Proceedings of the 62nd Annual Meeting of the Association
  for Computational Linguistics (Volume 1: Long Papers)}, pages 10289--10307.

\bibitem[{Huang et~al.(2024)Huang, Chen, Mishra, Zheng, Yu, Song, and
  Zhou}]{huang2024largelanguagemodelsselfcorrect}
Jie Huang, Xinyun Chen, Swaroop Mishra, Huaixiu~Steven Zheng, Adams~Wei Yu,
  Xinying Song, and Denny Zhou. 2024.
\newblock \href {https://arxiv.org/abs/2310.01798} {Large language models
  cannot self-correct reasoning yet}.
\newblock \emph{Preprint}, arXiv:2310.01798.

\bibitem[{Hunicke et~al.(2004)Hunicke, LeBlanc, Zubek et~al.}]{hunicke2004mda}
Robin Hunicke, Marc LeBlanc, Robert Zubek, and 1 others. 2004.
\newblock Mda: A formal approach to game design and game research.
\newblock In \emph{Proceedings of the AAAI Workshop on Challenges in Game AI},
  volume~4, page 1722. San Jose, CA.

\bibitem[{Jia et~al.(2026)Jia, Chen, Sharma, and
  Diaz-Rodriguez}]{jia2026digitalpersonasreliablyapproximate}
Mumin Jia, Yilin Chen, Divya Sharma, and Jairo Diaz-Rodriguez. 2026.
\newblock \href {https://arxiv.org/abs/2605.10659} {When can digital personas
  reliably approximate human survey findings?}
\newblock \emph{Preprint}, arXiv:2605.10659.

\bibitem[{Kamoi et~al.(2024)Kamoi, Zhang, Zhang, Han, and
  Zhang}]{kamoi-etal-2024-llms}
Ryo Kamoi, Yusen Zhang, Nan Zhang, Jiawei Han, and Rui Zhang. 2024.
\newblock \href {https://doi.org/10.1162/tacl_a_00713} {When can {LLM}s
  actually correct their own mistakes? a critical survey of self-correction of
  {LLM}s}.
\newblock \emph{Transactions of the Association for Computational Linguistics},
  12:1417--1440.

\bibitem[{Kendall(1938)}]{kendall1938new}
Maurice~G Kendall. 1938.
\newblock A new measure of rank correlation.
\newblock \emph{Biometrika}, 30(1-2):81--93.

\bibitem[{Kim et~al.(2026)Kim, Khalifa, Logeswaran, Kim, Lee, Lee, and
  Wang}]{kim2026coevolvingactorconditionedcriticsnonverifiable}
Jinyoung Kim, Muhammad Khalifa, Lajanugen Logeswaran, Jaekyeom Kim, Moontae
  Lee, Honglak Lee, and Lu~Wang. 2026.
\newblock \href {https://arxiv.org/abs/2608.30397} {Co-evolving
  actor-conditioned critics for non-verifiable generation}.
\newblock \emph{Preprint}, arXiv:2608.30397.

\bibitem[{Kirk et~al.(2024)Kirk, Whitefield, Röttger, Bean, Margatina, Ciro,
  Mosquera, Bartolo, Williams, He, Vidgen, and
  Hale}]{kirk2024prismalignmentdatasetparticipatory}
Hannah~Rose Kirk, Alexander Whitefield, Paul Röttger, Andrew Bean, Katerina
  Margatina, Juan Ciro, Rafael Mosquera, Max Bartolo, Adina Williams, He~He,
  Bertie Vidgen, and Scott~A. Hale. 2024.
\newblock \href {https://arxiv.org/abs/2404.16019} {The prism alignment
  dataset: What participatory, representative and individualised human feedback
  reveals about the subjective and multicultural alignment of large language
  models}.
\newblock \emph{Preprint}, arXiv:2404.16019.

\bibitem[{Koster(2013)}]{koster2013theory}
Raph Koster. 2013.
\newblock \emph{Theory of fun for game design}.
\newblock " O'Reilly Media, Inc.".

\bibitem[{Li et~al.(2025)Li, Zhang, Sohn, Hu, Usman, and
  Kapadia}]{li2025cardiverseharnessingllmsnovel}
Danrui Li, Sen Zhang, Sam~S. Sohn, Kaidong Hu, Muhammad Usman, and Mubbasir
  Kapadia. 2025.
\newblock \href {https://arxiv.org/abs/2502.07128} {Cardiverse: Harnessing llms
  for novel card game prototyping}.
\newblock \emph{Preprint}, arXiv:2502.07128.

\bibitem[{Li et~al.(2026{\natexlab{a}})Li, Chen, Namkoong, and
  Peng}]{li2026llm}
Leon Li, Haozhe Chen, Hongseok Namkoong, and Tianyi Peng. 2026{\natexlab{a}}.
\newblock Llm generated persona is a promise with a catch.
\newblock \emph{Advances in Neural Information Processing Systems}, 38.

\bibitem[{Li et~al.(2026{\natexlab{b}})Li, Li, Wang, Feng, Sun, Ai, Zhang, Sun,
  Huang, and Zhang}]{li2026meeplelmvirtualplaytestersimulating}
Zizhen Li, Chuanhao Li, Yibin Wang, Yukang Feng, Jianwen Sun, Jiaxin Ai, Fanrui
  Zhang, Mingzhu Sun, Yifei Huang, and Kaipeng Zhang. 2026{\natexlab{b}}.
\newblock \href {https://arxiv.org/abs/2601.07251} {Meeplelm: A virtual
  playtester simulating diverse subjective experiences}.
\newblock \emph{Preprint}, arXiv:2601.07251.

\bibitem[{Lightman et~al.(2023)Lightman, Kosaraju, Burda, Edwards, Baker, Lee,
  Leike, Schulman, Sutskever, and Cobbe}]{lightman2023letsverifystepstep}
Hunter Lightman, Vineet Kosaraju, Yura Burda, Harri Edwards, Bowen Baker, Teddy
  Lee, Jan Leike, John Schulman, Ilya Sutskever, and Karl Cobbe. 2023.
\newblock \href {https://arxiv.org/abs/2305.20050} {Let's verify step by step}.
\newblock \emph{Preprint}, arXiv:2305.20050.

\bibitem[{Madaan et~al.(2023)Madaan, Tandon, Gupta, Hallinan, Gao, Wiegreffe,
  Alon, Dziri, Prabhumoye, Yang et~al.}]{madaan2023self}
Aman Madaan, Niket Tandon, Prakhar Gupta, Skyler Hallinan, Luyu Gao, Sarah
  Wiegreffe, Uri Alon, Nouha Dziri, Shrimai Prabhumoye, Yiming Yang, and 1
  others. 2023.
\newblock Self-refine: Iterative refinement with self-feedback.
\newblock \emph{Advances in neural information processing systems},
  36:46534--46594.

\bibitem[{Noda et~al.(2019)Noda, Shirotsuki, and Nakao}]{noda2019effectiveness}
Shota Noda, Kentaro Shirotsuki, and Mutsuhiro Nakao. 2019.
\newblock The effectiveness of intervention with board games: a systematic
  review.
\newblock \emph{BioPsychoSocial medicine}, 13(1):22.

\bibitem[{Othman et~al.(2025)Othman, Mat, and Sim}]{othman2025systematic}
Mohd~Kamal Othman, Rahimah Mat, and Kah~Ching Sim. 2025.
\newblock A systematic review of paper-based and digital board games for
  collaborative science learning.
\newblock \emph{Review of Education}, 13(3):e70107.

\bibitem[{Park et~al.(2026)Park, Zou, Kamphorst, Egan, Shaw, Hill, Cai, Morris,
  Liang, Willer, and Bernstein}]{park2026llmagentsgroundedselfreports}
Joon~Sung Park, Carolyn~Q. Zou, Jonne Kamphorst, Niles Egan, Aaron Shaw,
  Benjamin~Mako Hill, Carrie Cai, Meredith~Ringel Morris, Percy Liang, Robb
  Willer, and Michael~S. Bernstein. 2026.
\newblock \href {https://arxiv.org/abs/2411.10109} {Llm agents grounded in
  self-reports enable general-purpose simulation of individuals}.
\newblock \emph{Preprint}, arXiv:2411.10109.

\bibitem[{Peper et~al.(2026)Peper, Gandra, Zhang, Chennareddy, Jha, Payani, and
  Wang}]{peper2026llms}
Joseph Peper, Sai~Krishna Gandra, Yunxiang Zhang, Vaibhav Chennareddy, Shloki
  Jha, Ali Payani, and Lu~Wang. 2026.
\newblock \href
  {https://proceedings.iclr.cc/paper_files/paper/2026/file/38324fc35dbd03fadc702510c264c8c9-Paper-Conference.pdf}
  {{LLMs} as rules oracles: Exploring real-world multimodal reasoning in
  tabletop strategy game environments}.
\newblock In \emph{International Conference on Learning Representations},
  volume 2026, pages 33158--33214.

\bibitem[{Qin et~al.(2025)Qin, Yang, Li, Wen, and
  Lee}]{10.1145/3706598.3713146}
Peinuan Qin, Chi-Lan Yang, Jingshu Li, Jing Wen, and Yi-Chieh Lee. 2025.
\newblock \href {https://doi.org/10.1145/3706598.3713146} {Timing matters: How
  using llms at different timings influences writers' perceptions and ideation
  outcomes in ai-assisted ideation}.
\newblock In \emph{Proceedings of the 2025 CHI Conference on Human Factors in
  Computing Systems}, CHI '25, New York, NY, USA. Association for Computing
  Machinery.

\bibitem[{Reimers and Gurevych(2019)}]{reimers2019sentence}
Nils Reimers and Iryna Gurevych. 2019.
\newblock Sentence-bert: Sentence embeddings using siamese bert-networks.
\newblock In \emph{Proceedings of the 2019 conference on empirical methods in
  natural language processing and the 9th international joint conference on
  natural language processing (EMNLP-IJCNLP)}, pages 3982--3992.

\bibitem[{Ren et~al.(2026)Ren, Pan, Chen, and Wang}]{ren-etal-2026-coarse}
Qingyu Ren, Tianjun Pan, Xingzhou Chen, and Xuhong Wang. 2026.
\newblock \href {https://doi.org/10.18653/v1/2026.findings-acl.134} {From
  coarse to fine: Benchmarking and reward modeling for writing-centric
  generation tasks}.
\newblock In \emph{Findings of the {A}ssociation for {C}omputational
  {L}inguistics: {ACL} 2026}, pages 2796--2810, San Diego, California, United
  States. Association for Computational Linguistics.

\bibitem[{Salemi et~al.(2024)Salemi, Mysore, Bendersky, and
  Zamani}]{salemi2024lamp}
Alireza Salemi, Sheshera Mysore, Michael Bendersky, and Hamed Zamani. 2024.
\newblock Lamp: When large language models meet personalization.
\newblock In \emph{Proceedings of the 62nd Annual Meeting of the Association
  for Computational Linguistics (Volume 1: Long Papers)}, pages 7370--7392.

\bibitem[{Schell(2014)}]{schell2014art}
Jesse Schell. 2014.
\newblock The art of game design: A book of lenses.

\bibitem[{Selinker(2011)}]{selinker2011kobold}
Mike Selinker. 2011.
\newblock The kobold guide to board game design.

\bibitem[{Shahhosseini et~al.(2026)Shahhosseini, Marioriyad, Momen, Baghshah,
  Rohban, and Javanmard}]{shahhosseini2026largelanguagemodelsscientific}
Fatemeh Shahhosseini, Arash Marioriyad, Ali Momen, Mahdieh~Soleymani Baghshah,
  Mohammad~Hossein Rohban, and Shaghayegh~Haghjooy Javanmard. 2026.
\newblock \href {https://arxiv.org/abs/2511.07448} {Large language models for
  scientific idea generation: A creativity-centered survey}.
\newblock \emph{Preprint}, arXiv:2511.07448.

\bibitem[{Sousa et~al.(2023)Sousa, Rye, Sousa, Torres, Perim, Mansuklal, and
  Ennami}]{sousa2023playing}
Carla Sousa, Sara Rye, Micael Sousa, Pedro~Juan Torres, Claudilene Perim,
  Shivani~Atul Mansuklal, and Firdaous Ennami. 2023.
\newblock Playing at the school table: Systematic literature review of board,
  tabletop, and other analog game-based learning approaches.
\newblock \emph{Frontiers in Psychology}, 14:1160591.

\bibitem[{Stiennon et~al.(2020)Stiennon, Ouyang, Wu, Ziegler, Lowe, Voss,
  Radford, Amodei, and Christiano}]{stiennon2020learning}
Nisan Stiennon, Long Ouyang, Jeffrey Wu, Daniel Ziegler, Ryan Lowe, Chelsea
  Voss, Alec Radford, Dario Amodei, and Paul~F Christiano. 2020.
\newblock Learning to summarize from human feedback.
\newblock In \emph{Advances in Neural Information Processing Systems},
  volume~33, pages 3008--3021.

\bibitem[{Swacha and Gracel(2025)}]{swacha2025supporting}
Jakub Swacha and Micha{\l} Gracel. 2025.
\newblock Supporting serious game development with generative artificial
  intelligence: Mapping solutions to lifecycle stages.
\newblock \emph{Applied Sciences}, 15(21):11606.

\bibitem[{Tanaka and Simo-Serra(2025)}]{tanaka2025grammar}
Tsunehiko Tanaka and Edgar Simo-Serra. 2025.
\newblock Grammar and gameplay-aligned rl for game description generation with
  llms.
\newblock In \emph{2025 IEEE Conference on Games (CoG)}, pages 1--8. IEEE.

\bibitem[{Team(2026)}]{qwen35blog}
Qwen Team. 2026.
\newblock \href {https://qwen.ai/blog?id=qwen3.5} {Qwen3.5: Accelerating
  productivity with native multimodal agents}.

\bibitem[{Todd et~al.(2024)Todd, Padula, Stephenson, Éric Piette, Soemers, and
  Togelius}]{todd2024gavelgeneratinggamesevolution}
Graham Todd, Alexander Padula, Matthew Stephenson, Éric Piette, Dennis J.
  N.~J. Soemers, and Julian Togelius. 2024.
\newblock \href {https://arxiv.org/abs/2407.09388} {Gavel: Generating games via
  evolution and language models}.
\newblock \emph{Preprint}, arXiv:2407.09388.

\bibitem[{Torii et~al.(2023)Torii, Murakami, and Ochiai}]{Torii_2023}
Maya~Grace Torii, Takahito Murakami, and Yoichi Ochiai. 2023.
\newblock \href {https://doi.org/10.1145/3573382.3623706} {Lottery and sprint:
  Generate a board game with design sprint method on autogpt}.
\newblock In \emph{Companion Proceedings of the Annual Symposium on
  Computer-Human Interaction in Play}, CHI PLAY ’23, pages 259--265. ACM.

\bibitem[{Tyen et~al.(2024)Tyen, Mansoor, C{\u{a}}rbune, Chen, and
  Mak}]{tyen2024llms}
Gladys Tyen, Hassan Mansoor, Victor C{\u{a}}rbune, Yuanzhu~Peter Chen, and Tony
  Mak. 2024.
\newblock Llms cannot find reasoning errors, but can correct them given the
  error location.
\newblock In \emph{Findings of the Association for Computational Linguistics:
  ACL 2024}, pages 13894--13908.

\bibitem[{Wang et~al.(2024)Wang, Xu, Zhao, Ouyang, Wu, Zhao, Xu, Liu, Qu,
  Shang, Zhang, Wei, Sui, Li, Shi, Qiao, Lin, and
  He}]{wang2024mineruopensourcesolutionprecise}
Bin Wang, Chao Xu, Xiaomeng Zhao, Linke Ouyang, Fan Wu, Zhiyuan Zhao, Rui Xu,
  Kaiwen Liu, Yuan Qu, Fukai Shang, Bo~Zhang, Liqun Wei, Zhihao Sui, Wei Li,
  Botian Shi, Yu~Qiao, Dahua Lin, and Conghui He. 2024.
\newblock \href {https://arxiv.org/abs/2409.18839} {Mineru: An open-source
  solution for precise document content extraction}.
\newblock \emph{Preprint}, arXiv:2409.18839.

\bibitem[{Wang et~al.(2026)Wang, Zang, Han, Bu, Zhou, Jin, and
  Wang}]{wang2026unified}
Yibin Wang, Yuhang Zang, Feng Han, Jiazi Bu, Yujie Zhou, Cheng Jin, and Jiaqi
  Wang. 2026.
\newblock Unified personalized reward model for vision generation.
\newblock \emph{arXiv preprint arXiv:2602.02380}.

\bibitem[{Wu et~al.(2025)Wu, Yang, Wu, Mahns, Wang, Zhu, Fang, and
  Xu}]{wu2025ai}
Jibang Wu, Chenghao Yang, Yi~Wu, Simon Mahns, Chaoqi Wang, Hao Zhu, Fei Fang,
  and Haifeng Xu. 2025.
\newblock Ai realtor: Towards grounded persuasive language generation for
  automated copywriting.
\newblock \emph{arXiv preprint arXiv:2502.16810}.

\bibitem[{Wu et~al.(2024)Wu, Zeng, Zhang, Tan, Shen, and Jiang}]{wu2024proco}
Zhenyu Wu, Qingkai Zeng, Zhihan Zhang, Zhaoxuan Tan, Chao Shen, and Meng Jiang.
  2024.
\newblock \href {https://doi.org/10.18653/v1/2024.emnlp-main.714} {Large
  language models can self-correct with key condition verification}.
\newblock In \emph{Proceedings of the 2024 Conference on Empirical Methods in
  Natural Language Processing}, pages 12846--12867, Miami, Florida, USA.
  Association for Computational Linguistics.

\bibitem[{Xia et~al.(2026)Xia, Peng, Qi, Wang, Xu, Hou, and
  Li}]{xia2026storyalignevaluatingtrainingreward}
Haotian Xia, Hao Peng, Yunjia Qi, Xiaozhi Wang, Bin Xu, Lei Hou, and Juanzi Li.
  2026.
\newblock \href {https://arxiv.org/abs/2605.04831} {Storyalign: Evaluating and
  training reward models for story generation}.
\newblock \emph{Preprint}, arXiv:2605.04831.

\bibitem[{Yang et~al.(2025)Yang, Li, Yang, Zhang, Hui, Zheng, Yu, Gao, Huang,
  Lv et~al.}]{yang2025qwen3}
An~Yang, Anfeng Li, Baosong Yang, Beichen Zhang, Binyuan Hui, Bo~Zheng, Bowen
  Yu, Chang Gao, Chengen Huang, Chenxu Lv, and 1 others. 2025.
\newblock Qwen3 technical report.
\newblock \emph{arXiv preprint arXiv:2505.09388}.

\bibitem[{Yang and Jin(2025)}]{yang2025matters}
Dingyi Yang and Qin Jin. 2025.
\newblock What matters in evaluating book-length stories? a systematic study of
  long story evaluation.
\newblock In \emph{Proceedings of the 63rd Annual Meeting of the Association
  for Computational Linguistics (Volume 1: Long Papers)}, pages 16375--16398.

\bibitem[{Zeng et~al.(2026)Zeng, Liu, Liu, Wang, Sun, Yang, Liu, and
  Fan}]{zeng2026rulesmithmultiagentllmsautomated}
Ziyao Zeng, Chen Liu, Tianyu Liu, Hao Wang, Xiatao Sun, Fengyu Yang, Xiaofeng
  Liu, and Zhiwen Fan. 2026.
\newblock \href {https://arxiv.org/abs/2602.06232} {Rulesmith: Multi-agent llms
  for automated game balancing}.
\newblock \emph{Preprint}, arXiv:2602.06232.

\bibitem[{Zheng et~al.(2024)Zheng, Zhang, Zhang, Ye, Luo, Feng, and
  Ma}]{zheng2024llamafactoryunifiedefficientfinetuning}
Yaowei Zheng, Richong Zhang, Junhao Zhang, Yanhan Ye, Zheyan Luo, Zhangchi
  Feng, and Yongqiang Ma. 2024.
\newblock \href {https://arxiv.org/abs/2403.13372} {Llamafactory: Unified
  efficient fine-tuning of 100+ language models}.
\newblock \emph{Preprint}, arXiv:2403.13372.

\bibitem[{Zhou et~al.(2025)Zhou, Zhang, Gao, Jiang, and
  Wang}]{zhou2025personaeval}
Lingfeng Zhou, Jialing Zhang, Jin Gao, Mohan Jiang, and Dequan Wang. 2025.
\newblock Personaeval: Are llm evaluators human enough to judge role-play?
\newblock \emph{arXiv preprint arXiv:2508.10014}.

\end{thebibliography}

\appendix
\clearpage

\section*{Appendix Overview}
\startcontents[appendix]
\printcontents[appendix]{l}{1}{\setcounter{tocdepth}{2}}

\section{Data Foundation}
\label{app:data}

\subsection{Dataset Statistics}
\label{app:data_statistics}

Our dataset comprises 2,233 board games in total, of which 207 are reserved as a held-out evaluation set.
Games are sourced from BoardGameGeek (BGG), covering 192 mechanics\footnote{\url{https://boardgamegeek.com/browse/boardgamemechanic}}, 85 categories\footnote{\url{https://boardgamegeek.com/browse/boardgamecategory}}, and over 190 themes\footnote{\url{https://boardgamegeek.com/wiki/page/Games_by_theme}}.

\begin{figure}[t]
    \centering
    \includegraphics[width=\columnwidth]{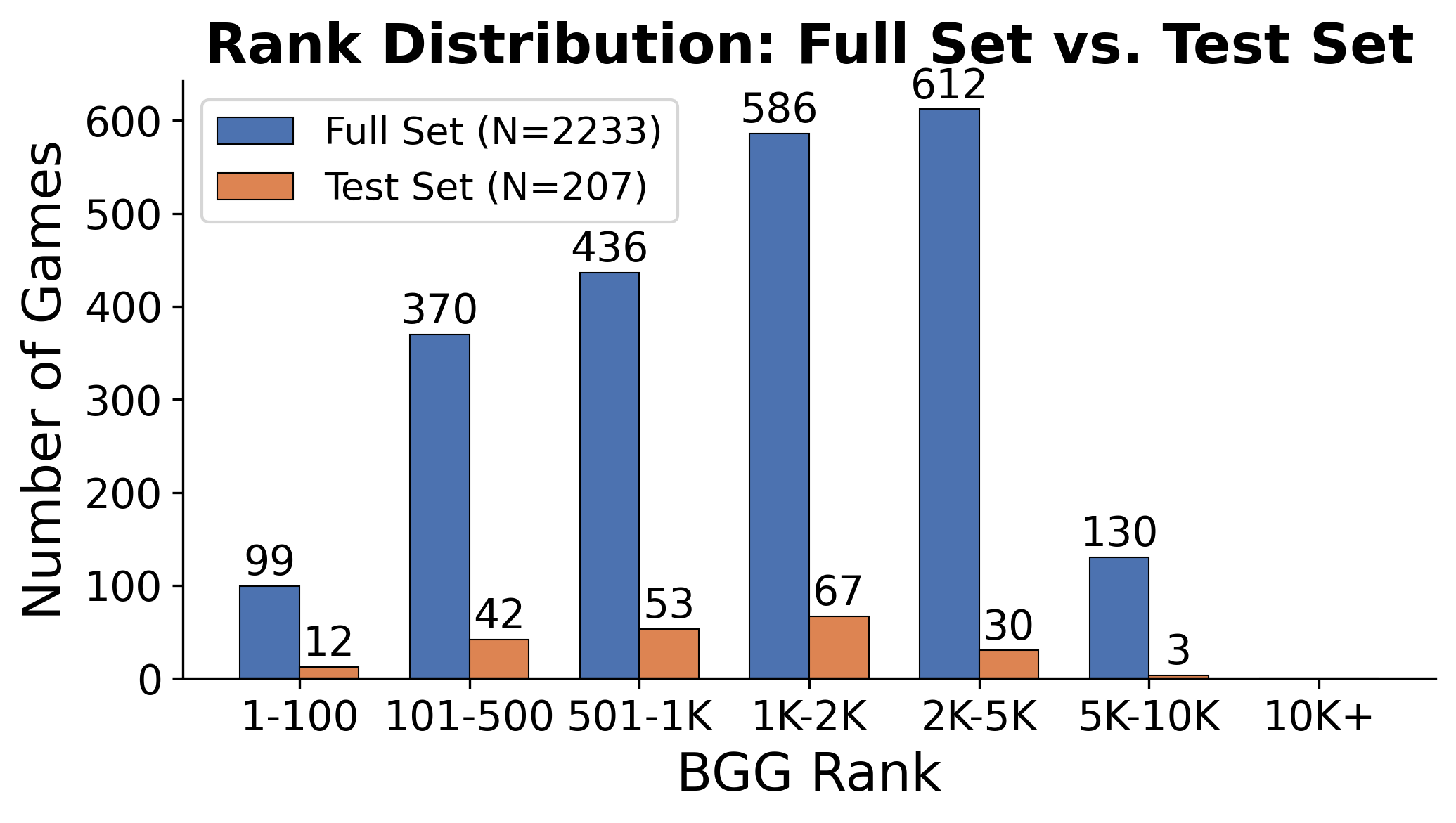}
    \caption{Rank distribution of the full set (N=2,233) and the evaluation set (N=207). The evaluation set has a rank composition close to that of the full set.}
    \label{fig:rank_dist}
\end{figure}

Figure~\ref{fig:rank_dist} shows the rank distribution of both the full set and the evaluation set.
The evaluation set is the 207-game test split of MeepleLM~\cite{li2026meeplelmvirtualplaytestersimulating}, built by stratified sampling over BGG weight (complexity) and average rating, so top-ranked games and long-tail designs appear in proportions close to those of the full set.
It includes 34 titles released in 2024--2025 and shares no game with any of our training data.

\begin{figure}[t]
    \centering
    \includegraphics[width=\columnwidth]{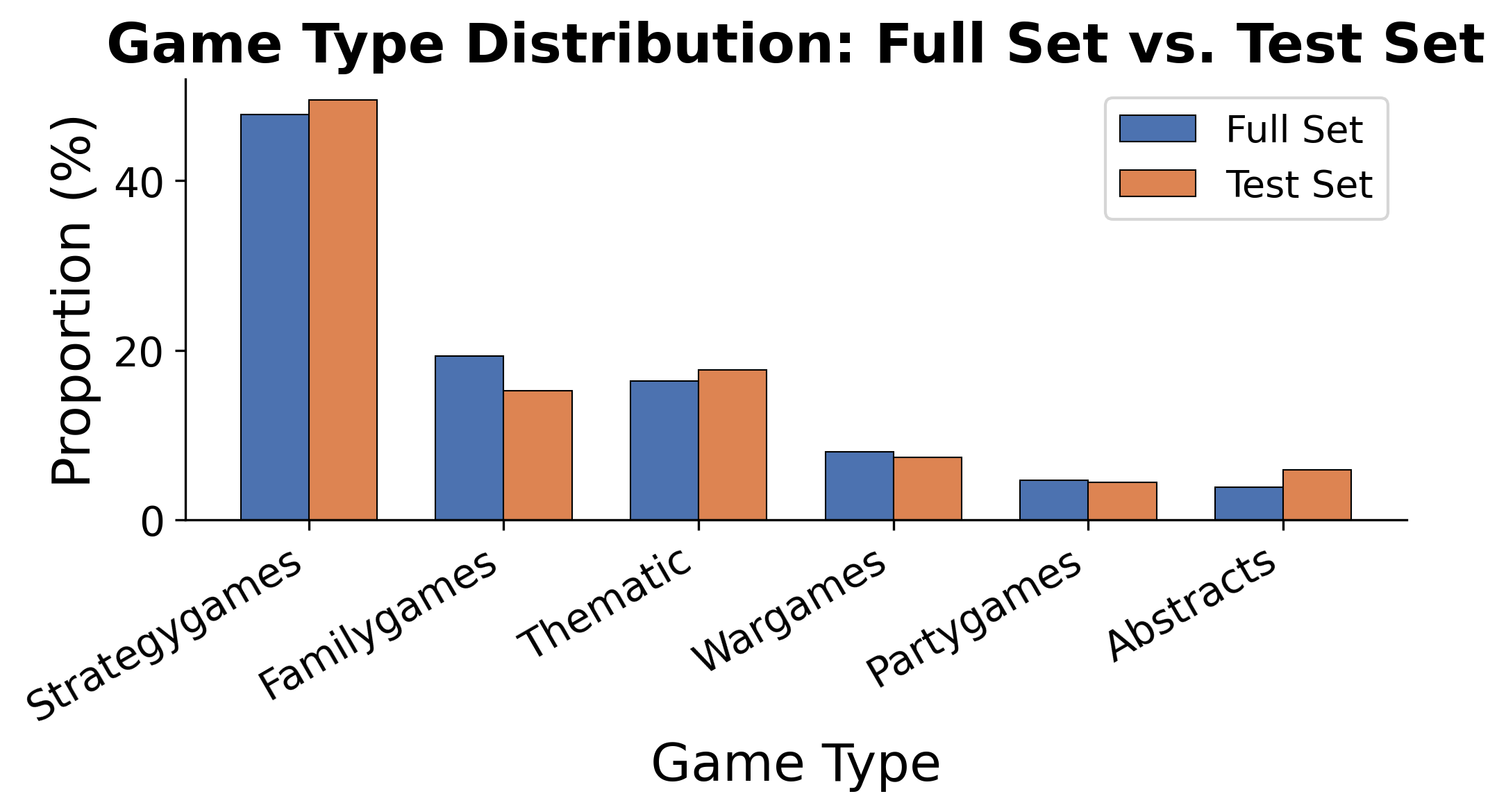}
    \caption{Game type proportion comparison between the full set and the evaluation set. Only the top-6 types present in both sets are shown.}
    \label{fig:type_dist}
\end{figure}

Figure~\ref{fig:type_dist} compares the game type distribution between the two sets.
Game types are derived from BGG's secondary ranking categories (e.g., \textit{strategygames}, \textit{wargames}, \textit{familygames}).
The top-6 types shared by both sets are shown; the proportions remain closely aligned, confirming that the evaluation set is representative of the full collection.

\begin{figure}[t]
    \centering
    \includegraphics[width=\columnwidth]{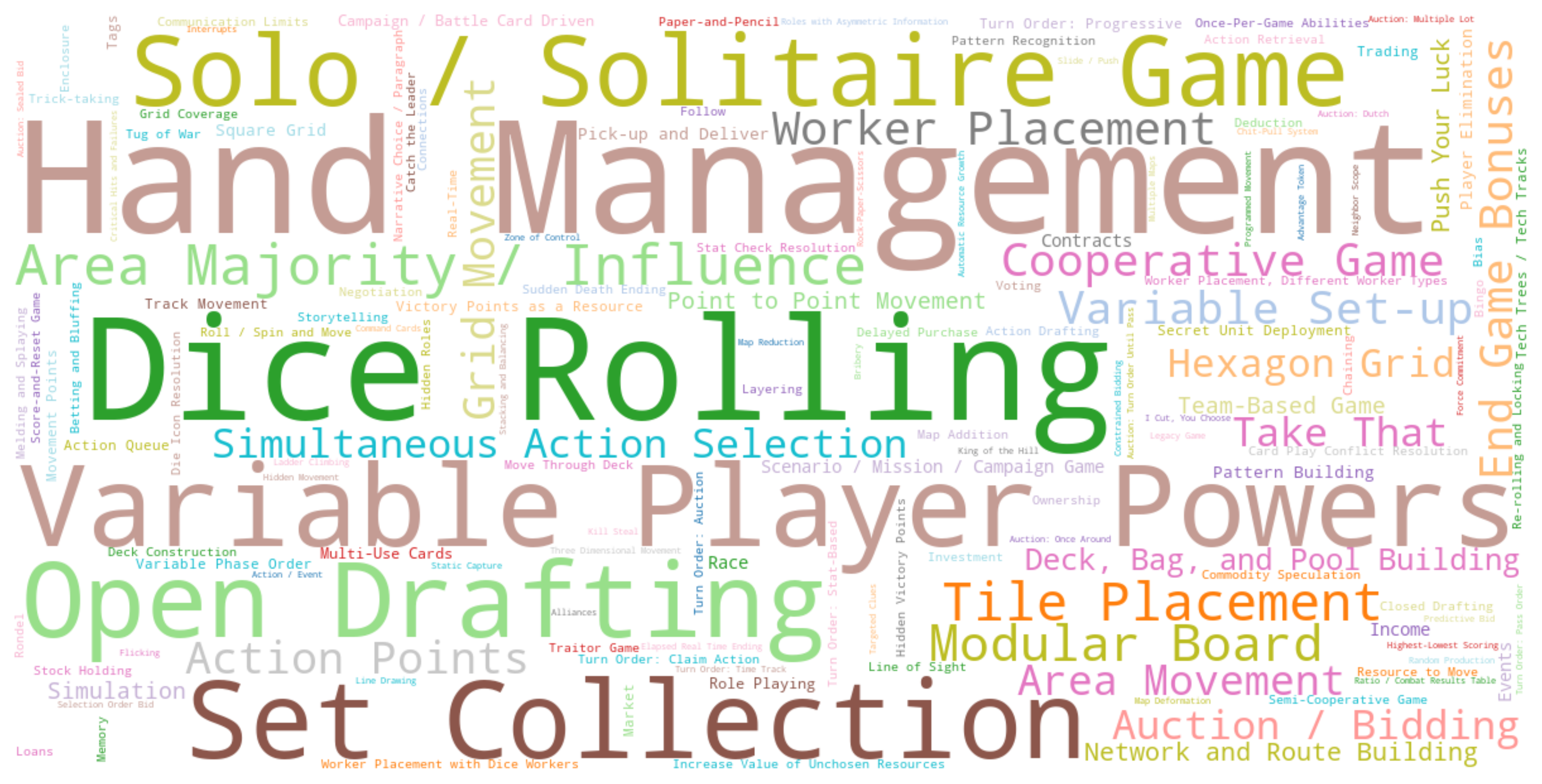}
    \caption{Word cloud of board game mechanics across the full dataset, covering 192 unique mechanics.}
    \label{fig:wc_mech}
\end{figure}

\begin{figure}[t]
    \centering
    \includegraphics[width=\columnwidth]{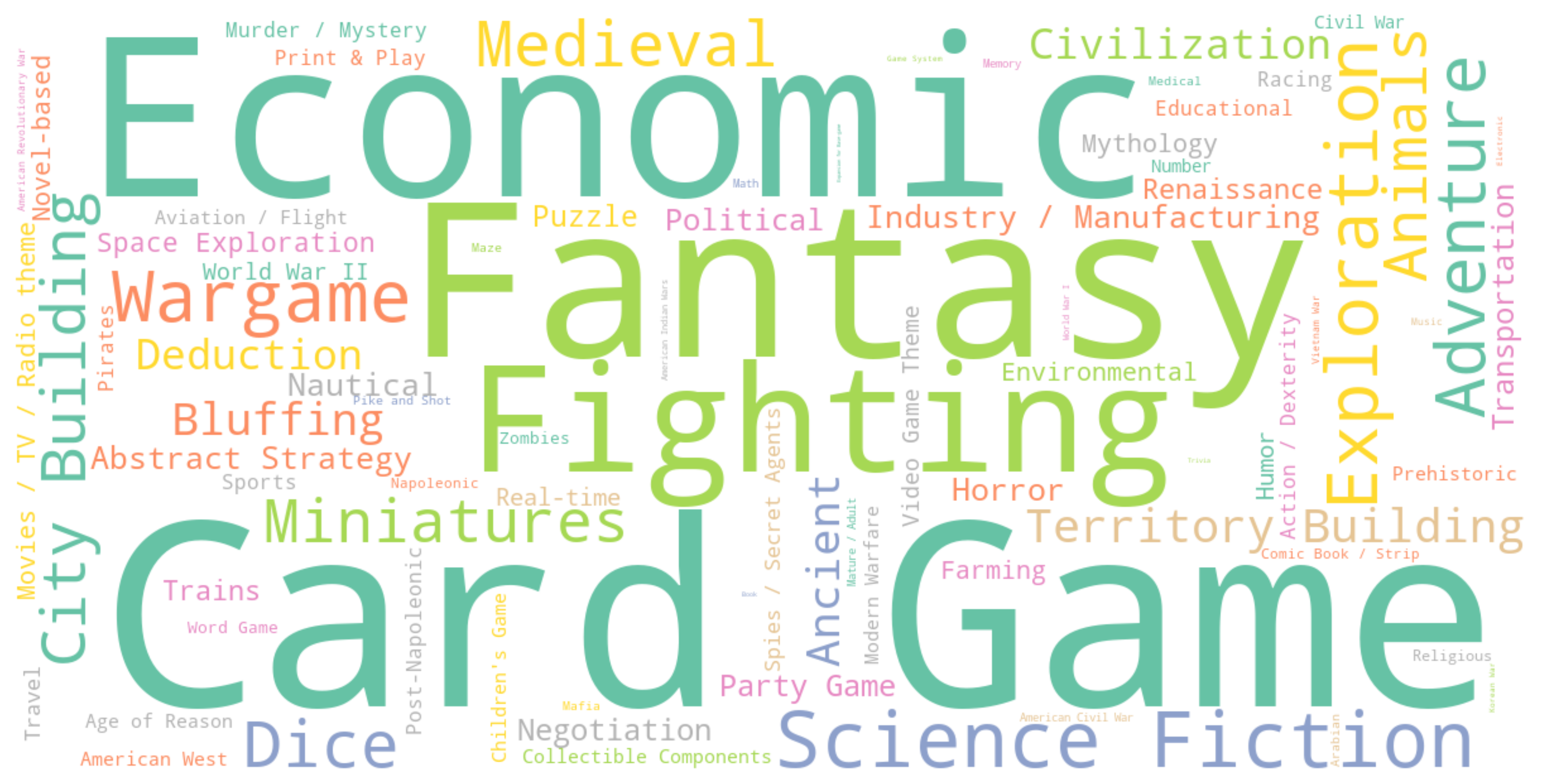}
    \caption{Word cloud of board game categories across the full dataset, covering 85 unique categories.}
    \label{fig:wc_cat}
\end{figure}

To illustrate the diversity of the collected games, Figures~\ref{fig:wc_mech} and~\ref{fig:wc_cat} present word clouds of mechanics and categories across the full dataset.

\subsection{Data Processing}
\label{app:data_processing}

We adopt and extend the data processing pipeline of \citet{li2026meeplelmvirtualplaytestersimulating}: rulebook pdfs are parsed via Mineru~\citep{wang2024mineruopensourcesolutionprecise}, restructured into a seven-section schema by Qwen3-235B~\citep{yang2025qwen3}, and rectified by GPT-5.1; reviews are filtered through MDA-based quality scoring and facet-aware stratified sampling.
The resulting dataset contains 2,233 structured rulebooks and ${\sim}$180K high-quality reviews.

\section{Interactive Ideation (BG-Ideator)}
\label{app:ideation}

\subsection{Draft Schema Definition}
\label{app:ideation_schema}

The design draft schema is developed through a collaborative effort between domain experts and the research team.
Experts reviewed a curated collection of board game design literature, community forums, and practitioner guides (listed below), and distilled the recurring structural elements of game design into a unified JSON schema.
The schema captures designer intent through five field groups: \emph{concept}, \emph{classification}, \emph{mechanics} (with core / supporting / structural layers and rationales), \emph{design\_intent}, and \emph{parameters}.
The complete field definitions are shown in Figure~\ref{fig:draft_schema}, and a filled example is shown in Figure~\ref{fig:draft_example}.

\paragraph{Expert Reference Sources.}
Key sources for the schema design:

\begin{itemize}[leftmargin=1.5em, itemsep=2pt]
    \item \textit{Building Blocks of Tabletop Game Design}~\cite{engelstein2022building} --- a taxonomy of board game mechanics.
    \item \textit{The Art of Game Design: A Book of Lenses}~\cite{schell2014art} --- principles of experience-driven design.
    \item \textit{Theory of Fun for Game Design}~\cite{koster2013theory} --- player engagement and fun.
    \item \textit{Characteristics of Games}~\cite{elias2020characteristics} --- formal properties of game systems.
    \item \textit{The Kobold Guide to Board Game Design}~\cite{selinker2011kobold} --- practitioner perspectives on the design-to-publication pipeline.
    \item BGG Designer Forum\footnote{\url{https://boardgamegeek.com/forums}} and the \textit{Brief Crash Course on Game Design}\footnote{\url{https://boardgamegeek.com/thread/1018707}} --- community discussions and design workflows.
    \item Gcores columns by practitioners\footnote{\url{https://www.gcores.com/articles}} --- thematic design perspectives and cultural context.
\end{itemize}

\subsection{Draft Generation Prompt}
\label{app:ideation_draft_prompt}
 
We prompt GPT-5.4 with the seed game's structured rulebook, BGG metadata (mechanics, categories, complexity, player count), and official BGG definitions for each tagged mechanic and category.
The system prompt instructs the model to act as a design analyst and produce a JSON draft following the schema in Figure~\ref{fig:draft_schema}.
Key instructions include: classifying every metadata mechanic into exactly one tier (core / supporting / structural), citing specific rules in rationales, and writing from a designer's perspective rather than a reviewer's.
The full prompt is shown in Figure~\ref{fig:draft_gen_prompt}.

\subsection{Mutation Strategies and Prompts}
\label{app:ideation_mutation_detail}
 
We apply three mutation strategies to the seed drafts to generate novel designs.
 
\paragraph{Theme Migration.}
Core mechanic names are preserved; categories, description, elevator pitch, design intent, and theme--mechanic fit are rewritten for the new theme.
Supporting and structural mechanics may be replaced (60--80\% retention) if the new theme demands it.
For each seed, 20 candidate themes are randomly sampled from a curated BGG theme list (excluding themes similar to the original), and the model selects the most interesting one.
This produces ${\sim}$1,200 mutated drafts.
The prompt is shown in Figure~\ref{fig:theme_prompt}.
 
\paragraph{2-Core Hybridization.}
Two seed games of the same type are paired.
From their combined core mechanics, one candidate pair is selected such that at least three distinct mechanics appear and the new pair differs from both parents.
The new design inherits one parent's theme.
Supporting and structural mechanics are drawn from both parents or introduced fresh.
Games with exactly 2 core mechanics (488 seeds) are paired via within-type shuffling across 6 rounds, producing ${\sim}$1,200 hybrids.
 
\paragraph{3-Core Hybridization.}
Two seeds of the same type are paired.
The new design inherits 2 core mechanics from parent A and 1 from parent B (which must not already appear in A's cores).
Theme is inherited from parent A.
Games with exactly 3 core mechanics (532 seeds) are paired via within-type shuffling across 5 rounds, producing ${\sim}$1,200 hybrids.
 
The hybridization prompt (shared by both variants with minor parameter differences) is shown in Figure~\ref{fig:hybrid_prompt}.

\subsection{Mutation Statistics}
\label{app:ideation_mutation_stats}
 
Table~\ref{tab:mutation_stats} shows the distribution of hybridized drafts across game types.
Theme migration operates on all seeds regardless of core count, while hybridization is stratified by type to preserve genre coherence.
The total yield before quality filtering is ${\sim}$3,600 drafts (1,200 per strategy); after verification (\S\ref{app:ideation_verify}), ${\sim}$4,500 pass the quality threshold (including the 1,200 original seeds).

\begin{table}[t]
\centering
\small
\begin{tabular}{lrrr}
\toprule
\textbf{Game Type} & \textbf{2-Core} & \textbf{3-Core} & \textbf{Total} \\
\midrule
Strategy Games      & 450 & 489 & 939 \\
Thematic Games      & 177 & 301 & 478 \\
Family Games        & 290 & 159 & 449 \\
Wargames            & 106 & 160 & 266 \\
Party Games         &  81 &  52 & 133 \\
Abstract Strategy   &  75 &  31 & 106 \\
Children's Games    &  17 &   0 &  17 \\
Customizable Games  &   4 &   8 &  12 \\
\midrule
Total               & 1,200 & 1,200 & 2,400 \\
\bottomrule
\end{tabular}
\caption{Distribution of hybridized drafts by game type. Theme migration (not shown) produces an additional ${\sim}$1,200 drafts across all types.}
\label{tab:mutation_stats}
\end{table}

\subsection{Quality Verification}
\label{app:ideation_verify}
 
All mutated drafts are evaluated by Gemini-3.1-Pro using a structured rubric covering 11 dimensions in four blocks:
 
\begin{itemize}[leftmargin=1.5em, itemsep=1pt]
    \item \textbf{Design Vision} (A1--A3): pitch clarity, experience authenticity, core tension clarity.
    \item \textbf{Mechanical Integrity} (B1--B3): core coherence, layer necessity, rationale specificity.
    \item \textbf{Theme--Mechanic Integration} (C1--C2): bidirectional mapping, thematic language consistency.
    \item \textbf{Parameter Coherence} (D1--D3): audience alignment, type fit, rationale grounding.
\end{itemize}
 
Each dimension is scored 1--5 with anchored examples.
Drafts with an average score $\geq 3.5$ and no major issues are accepted.
The mean score across accepted drafts is 4.84/5.
The full rubric is shown in Figure~\ref{fig:verify_rubric}.

\subsection{Single-Turn Interaction: Aspect Definitions and Prompts}
\label{app:ideation_aspects}
 
Each single-turn training instance is a (question, answer) pair generated by GPT-5.4 from a design draft.
The model receives a shared system prompt and a task-specific instruction derived from one of 15 predefined aspects.
The output follows a three-tag format: \texttt{<user\_message>}, \texttt{<reasoning>}, and \texttt{<response>}.
 
Table~\ref{tab:aspects} lists all 15 aspects grouped into six categories, with target counts and brief descriptions.
 
\begin{table*}[t]
\centering
\small
\begin{tabularx}{\textwidth}{clXr}
\toprule
\textbf{Cat.} & \textbf{Aspect} & \textbf{Description} & \textbf{Count} \\
\midrule
\multirow{3}{*}{\rotatebox{90}{\footnotesize Mech.\ Reason.}}
& A1: Constraint Recommend & Given one core mechanic + theme + experience goal, recommend a second core and explain synergy. & 2,000 \\
& A2: Mechanic Comparison & Compare two candidate mechanics for a design slot; analyze trade-offs and recommend one. & 2,000 \\
& A3: Experience Mapping & From a desired player experience, reverse-engineer the mechanic combination that produces it. & 2,000 \\
\midrule
\multirow{2}{*}{\rotatebox{90}{\footnotesize Transl.}}
& B1: Theme $\to$ Mechanic & User describes a theme/setting; model explains which mechanics embody it and why. & 1,500 \\
& B2: Reference $\to$ Design & User cites existing games as reference; model analyzes what to borrow, change, and build around. & 1,500 \\
\midrule
\multirow{3}{*}{\rotatebox{90}{\footnotesize Draft Anal.}}
& C1: Mechanic Tiering & Given a shuffled mechanic list + concept, classify each into core / supporting / structural with rationale. & 1,500 \\
& C2: Coherence Check & A flaw is injected (missing mechanic, contradictory addition, or wrong tier); model must identify it. & 1,500 \\
& C3: Gap Diagnosis & Given only core mechanics, diagnose what the design needs next and prioritize gaps. & 1,000 \\
\midrule
\multirow{3}{*}{\rotatebox{90}{\footnotesize Synth.}}
& D1: Elevator Pitch & Write a one-sentence design hook from complete mechanics + theme. & 1,200 \\
& D2: Concept Description & Write 2--4 paragraphs explaining what players do, how mechanics connect, and why the theme fits. & 1,600 \\
& D3: Design Intent & Articulate target experience, core tension, and theme--mechanic fit. & 1,200 \\
\midrule
\multirow{2}{*}{\rotatebox{90}{\footnotesize Param.}}
& E1: Parameter Estimation & Estimate complexity weight, player count, and play time from the mechanic set with reasoning. & 1,200 \\
& E2: Classification & Determine BGG type and category labels from a concept description. & 800 \\
\midrule
\multirow{3}{*}{\rotatebox{90}{\footnotesize Edge Cases}}
& F1: Gentle Correction & User has a mechanic misconception (e.g., ``engine building = literal factory''); model corrects naturally. & 400 \\
& F2: Scope Management & User wants too many mechanics; model helps focus on core and defers extras to expansions. & 300 \\
& F3: Infeasible Combination & User proposes conflicting mechanics (e.g., cooperative + player elimination); model explains the friction. & 300 \\
\bottomrule
\end{tabularx}
\caption{The 15 single-turn aspects across six categories, totaling ${\sim}$20K training instances.}
\label{tab:aspects}
\end{table*}
 
\paragraph{Mechanic Description Injection.}
Depending on the aspect, BGG mechanic definitions are injected at different depths to teach the model to explain mechanics through their consequences rather than dictionary definitions:
\emph{deep} injection (full descriptions) for A3, B1, B2, and F1;
\emph{brief} injection (first sentence only) for A1, A2, and C3;
no injection for all remaining aspects.
The response is constrained to prose style---bullet points and numbered lists are explicitly prohibited.
 
The shared system prompt and all per-aspect task instructions are shown in Figure~\ref{fig:s2_full_prompt}.

\subsection{Multi-Turn Dialogue: User Archetypes and Dynamics}
\label{app:ideation_dialogue}
 
Each multi-turn training instance is generated via reverse construction: given a target draft, GPT-5.4 simulates a natural conversation that converges to it.
Diversity is controlled along two dimensions: user archetypes (Table~\ref{tab:user_types}) and conversation dynamics (Table~\ref{tab:quirks}).
The final turn's \texttt{draft\_update} is programmatically replaced with the complete target draft JSON.
 
\begin{table}[t]
\centering
\small
\begin{tabular}{lcrp{3.2cm}}
\toprule
\textbf{Type} & \textbf{Wt.} & \textbf{Turns} & \textbf{Key Trait} \\
\midrule
mechanic & 10\% & 3--4 & Uses BGG terms; concise, direct \\
theme    & 20\% & 4--5 & Describes setting/story; no jargon \\
experience & 20\% & 5--6 & ``I want players to feel...'' \\
reference & 15\% & 4--5 & ``Like Game X but with Y''; must name $\geq$2 real games \\
explorer         & 20\% & 6--7 & Almost no direction; needs guidance \\
casual     & 15\% & 5--7 & Loose concept; very short messages \\
\bottomrule
\end{tabular}
\caption{Six user archetypes controlling information-reveal pace, message length, and dialogue style. Each maps to a specific assistant tone.}
\label{tab:user_types}
\end{table}
 
\begin{table}[t]
\centering
\small
\begin{tabular}{lcp{3.5cm}}
\toprule
\textbf{Quirk} & \textbf{Wt.} & \textbf{Effect} \\
\midrule
smooth          & 35\% & Normal flow \\
early\_pivot    & 12\% & User changes theme/mechanic mid-conversation \\
pushback        & 12\% & User rejects a suggestion with reasons \\
confused\_user  & 10\% & Subtle mechanic misconception \\
scope\_creep    &  8\% & User keeps adding features \\
sparse\_user    &  8\% & Extremely short replies \\
opinionated     &  8\% & Strong preferences; no alternatives wanted \\
tangent\_return &  7\% & Brief anecdote, then back on track \\
\bottomrule
\end{tabular}
\caption{Eight conversation dynamics injected per conversation for behavioral diversity.}
\label{tab:quirks}
\end{table}
 
Each user type has an associated assistant tone (e.g., \emph{efficient\_peer} for mechanic-driven, \emph{patient\_teacher} for explorers) and enforced message-length variance rules ensuring realistic variation within a single conversation.
Turn count is dynamically adjusted based on the draft's mechanic count to avoid rushing complex designs.
 
To prevent repetitive assistant behavior, we enforce five response shapes (validate+question, propose+justify, push-back, build-on, clarify) with a constraint that no two consecutive turns use the same shape.
The final turn uses one of four summary formats (bullet recap, narrative walkthrough, tension-centered, comparison frame), randomly assigned per conversation.
The complete generation prompt is shown in Figure~\ref{fig:s3_full_prompt}.

\subsection{Training Hyperparameters}
\label{app:ideation_hyperparams}

BG-Ideator is trained in two stages using LLaMA-Factory~\citep{zheng2024llamafactoryunifiedefficientfinetuning} (used for all module training hereafter).
Stage~1 (atomic SFT) trains on ${\sim}$20K single-turn data with a higher learning rate to acquire foundational design capabilities.
Stage~2 (dialogue SFT) continues from the Stage~1 checkpoint on ${\sim}$4.7K multi-turn conversations with a lower learning rate and longer context window.
Both stages enable the thinking mode to train the model's internal reasoning process.
The full configurations are listed in Table~\ref{tab:ideator_hparams}.

\begin{table}[t]
\centering
\small
\begin{tabular}{lcc}
\toprule
\textbf{Hyperparameter} & \textbf{Stage 1} & \textbf{Stage 2} \\
\midrule
Backbone           & \multicolumn{2}{c}{Qwen3.5-27B} \\
Initialization     & Base model & Stage 1 ckpt \\
\midrule
LoRA Rank ($r$)    & 64 & 64 \\
LoRA Alpha ($\alpha$) & 128 & 128 \\
Target Modules     & All linear & All linear \\
\midrule
Context Window     & 4,096 & 8,192 \\
Per-Device Batch   & 4 & 2 \\
Gradient Accum.    & 2 & 2 \\
Effective Batch    & 64 & 32 \\
\midrule
Learning Rate      & $1 \times 10^{-4}$ & $2 \times 10^{-5}$ \\
LR Scheduler       & Cosine & Cosine \\
Warmup Ratio       & 0.05 & 0.05 \\
Epochs             & 3 & 3 \\
Precision          & bf16 & bf16 \\
\midrule
Thinking Mode      & Enabled & Enabled \\
Attention          & Flash Attention 2 & Flash Attention 2 \\
\bottomrule
\end{tabular}
\caption{Training hyperparameters for BG-Ideator. Stage~1 acquires atomic design capabilities; Stage~2 learns dialogue-level guidance from the Stage~1 checkpoint.}
\label{tab:ideator_hparams}
\end{table}

\section{Community-Grounded Critic (BG-Critic)}
\label{app:critic}

\subsection{MDA Rewriting Procedure}
\label{app:critic_mda_rewrite}
 
We rewrite each entry into a unified three-section format using GPT-5.4 (temperature 0.3).
 
The rewriter receives both the reasoning chain and the original review, and produces a structured evaluation with three paragraph-based sections:
(1)~\emph{Mechanics \& Design}---concrete game systems and how they work;
(2)~\emph{Gameplay Dynamics}---emergent patterns and tensions during play;
(3)~\emph{Player Experience}---emotional and evaluative response grounded in the dynamics above, written in first person and concluding with the original rating.
The output is validated for section completeness, rating consistency, and prose style (no bullet points).
The full rewriting prompt is shown in Figure~\ref{fig:mda_rewrite_prompt}.

\paragraph{Human Audit.}
Three experienced players checked ${\sim}$3.6K rewritten chains from 200 games.
A chain was accepted only if it was factually consistent with that game, the Mechanics$\to$Dynamics$\to$Aesthetics links were causally plausible, and the stance matched the source review.
96.3\% were accepted (inter-rater $\kappa{=}0.74$). Invalid chains were removed from training.

\subsection{Diagnostic Training Data}
\label{app:critic_diagnostic_data}
 
The Diagnostic task trains BG-Critic to identify, classify, and localize design flaws in rulebooks.
Training data comes from three complementary sources.
 
\paragraph{Source A: Synthetic Perturbations.}
We systematically inject flaws into real rulebooks using GPT-5.4.
Flaws span eight types across three MDA layers and three severity levels, as defined in Table~\ref{tab:flaw_types}.
Source rulebooks are drawn from a rank-stratified pool: high-tier games (top 50\% by BGG rank) receive Critical and Major perturbations, where the contrast with their polished original design is most informative; mid-tier games (next 30\%) receive Minor perturbations, which require finer-grained judgment.
Each perturbation is specified as a list of edit operations (delete, replace, replace\_all, or rewrite\_section) applied locally to the rulebook, preserving all unrelated content.
The perturbed rulebook is validated for length ratio (0.5--1.3$\times$ original) and non-identity.
The total target is ${\sim}$5,000 perturbations with the allocation ratios shown in Table~\ref{tab:flaw_types}.
The perturbation prompt and edit schema are shown in Figure~\ref{fig:perturbation_prompt}.
 
\begin{table}[t]
\centering
\small
\setlength{\tabcolsep}{3pt}
\begin{tabular}{llp{4.0cm}r}
\toprule
\textbf{Type} & \textbf{Sev.} & \textbf{Target} & \textbf{\%} \\
\midrule
M\_critical & Crit. & Win/loss condition, mandatory gating rule, core resolution steps, phase transition & 14 \\
M\_major    & Maj.  & Numerical cap, cost/penalty, frequency restriction, prerequisite & 18 \\
M\_minor    & Min.  & Small numerical error, ambiguous wording, edge-case inconsistency & 8 \\
\midrule
D\_critical & Crit. & Inter-mechanic linkage severed; individual mechanics intact but system incoherent & 12 \\
D\_major    & Maj.  & Feedback loop broken; game still loops but central dynamic tension missing & 18 \\
D\_minor    & Min.  & Timing/sequencing detail shifted between phases & 8 \\
\midrule
A\_major    & Maj.  & Design intent misaligned with mechanics; stated experience contradicts actual gameplay & 14 \\
A\_minor    & Min.  & Thematic presentation disconnected; flavor text generic or off-theme & 8 \\
\bottomrule
\end{tabular}
\caption{Eight flaw types across three MDA layers. Allocation ratios sum to 100\% of the ${\sim}$5,000 target.}
\label{tab:flaw_types}
\end{table}
 
\paragraph{Source B: Community-Consensus Flaws.}
Player reviews frequently articulate concrete design problems.
For every training game outside the top 30\% of our collection (the no-flaw pool of Source~C), we group review complaints that target the same rule or component and keep a complaint as a consensus flaw when at least three reviewers raise it independently, with at most three flaws per game.
GPT-5.4 maps each consensus flaw to our flaw taxonomy (type, severity, affected component), and the flaw is posed on the published, unperturbed rulebook.
This yields \proj{1,986} flaws from \proj{1,127} games, a diagnostic signal grounded in authentic community feedback.
Games in the real-flaw evaluation set (Appendix~\ref{app:real_flaws}) are excluded.

\paragraph{Source C: No-Flaw Calibration.}
We include games ranked in the top 30\% of our collection as no-flaw samples.
These well-established rulebooks have undergone extensive community validation; BG-Critic should learn to report \textsc{No\_Flaw} for such designs rather than over-diagnosing.

\subsection{Training Hyperparameters}
\label{app:critic_hyperparams}

BG-Critic is trained in two stages using LLaMA-Factory.
Stage~1 trains on ${\sim}$180K MDA-rewritten community reviews to learn the evaluation format and domain knowledge.
Stage~2 performs multi-task SFT on three task-specific datasets, replaying a subset of rating data to maintain calibration.
Table~\ref{tab:critic_s2_data} shows the Stage~2 data composition; Table~\ref{tab:critic_hparams} lists the full training configurations.
Both stages disable the thinking mode.

\begin{table}[t]
\centering
\small
\begin{tabular}{lr}
\toprule
\textbf{Task} & \textbf{Count} \\
\midrule
Diagnostic (Source A: synthetic)   & 4,972 \\
Diagnostic (Source B: community)   & \proj{1,986} \\
Diagnostic (Source C: no-flaw)     & 781 \\
Comparison                         & 4,972 \\
Rating replay                      & 5,000 \\
\midrule
Total                              & \proj{17,711} \\
\bottomrule
\end{tabular}
\caption{Stage~2 multi-task data composition for BG-Critic.}
\label{tab:critic_s2_data}
\end{table}

\begin{table}[t]
\centering
\small
\begin{tabular}{lcc}
\toprule
\textbf{Hyperparameter} & \textbf{Stage 1} & \textbf{Stage 2} \\
\midrule
Backbone           & Qwen3.5-27B & Stage 1 merged \\
\midrule
LoRA Rank ($r$)    & 64 & 64 \\
LoRA Alpha ($\alpha$) & 128 & 128 \\
Target Modules     & All linear & All linear \\
\midrule
Context Window     & 8,192 & 16,384 \\
Per-Device Batch   & 2 & 2 \\
Gradient Accum.    & 2 & 2 \\
Effective Batch    & 32 & 32 \\
\midrule
Learning Rate      & $5 \times 10^{-5}$ & $2 \times 10^{-5}$ \\
LR Scheduler       & Cosine & Cosine \\
Warmup             & 100 steps & 50 steps \\
Epochs             & 2 & 3 \\
Precision          & bf16 & bf16 \\
\midrule
Thinking Mode      & Disabled & Disabled \\
Attention          & Flash Attention 2 & Flash Attention 2 \\
\bottomrule
\end{tabular}
\caption{Training hyperparameters for BG-Critic. Stage~1 learns MDA format on ${\sim}$180K reviews; Stage~2 fine-tunes on ${\sim}$\proj{18K} multi-task data from the Stage~1 merged checkpoint.}
\label{tab:critic_hparams}
\end{table}

\section{Rulebook Generation (BG-Realizer)}
\label{app:real}
 
\subsection{Rulebook Generation and Verification}
\label{app:real_generation}
 
For ${\sim}$1,500 high-scoring mutated drafts, we generate rulebooks using GPT-5.4 in a one-shot setting.
Each mutation type uses a tailored prompt that provides the parent game's real rulebook as a structural reference:
theme migrations receive one parent rulebook;
2-core and 3-core hybridizations receive both parents' rulebooks along with lineage metadata specifying which core mechanics are inherited from which parent.
The model is instructed to write a complete seven-section rulebook in prose paragraphs, with component counts and numeric values independently derived from the draft's parameters rather than copied from the reference.
The generation prompt is shown in Figure~\ref{fig:rulebook_gen_prompt}.
 
\paragraph{Round-Trip Verification.}
To ensure the generated rulebooks faithfully implement their source drafts, we apply a round-trip verification pipeline.
Each generated rulebook is fed back to GPT-5.4 to produce a reverse-engineered draft (using the same draft generation prompt from Appendix~\ref{app:ideation_draft_prompt}).
The reverse draft is then compared against the original mutation draft across ten dimensions with a weighted scoring scheme:
 
\begin{itemize}[leftmargin=1.5em, itemsep=1pt]
    \item \textbf{Structural} (40\%): core mechanic tier match (20\%), supporting mechanic Jaccard (10\%), type match (5\%), category Jaccard (5\%).
    \item \textbf{Semantic} (55\%): cosine similarity of core tension (20\%), target experience (15\%), elevator pitch (10\%), theme--mechanic fit (5\%), description (5\%).
    \item \textbf{Parametric} (5\%): consistency of complexity, player count, and play time.
\end{itemize}
 
Semantic similarity is computed using sentence embeddings (all-MiniLM-L6-v2 ~\cite{reimers2019sentence}).
Pairs whose total weighted score falls below \proj{0.70} are discarded.

\subsection{Training Hyperparameters}
\label{app:real_hyperparams}
 
BG-Realizer is trained via multi-task SFT on the combined generation and revision data using LLaMA-Factory.
The ${\sim}$1,200 seed draft--rulebook pairs are weighted $3\times$ during training to emphasize grounded examples; together with the ${\sim}$\proj{1,200} generated pairs that pass round-trip verification, this gives an effective generation dataset of ${\sim}$4,800 instances.
Combined with ${\sim}$5K revision instances, the total training set is ${\sim}$9,800 instances.
The full configuration is listed in Table~\ref{tab:realizer_hparams}.
 
\begin{table}[t]
\centering
\small
\begin{tabular}{lc}
\toprule
\textbf{Hyperparameter} & \textbf{Value} \\
\midrule
Backbone           & Qwen3.5-27B \\
\midrule
LoRA Rank ($r$)    & 64 \\
LoRA Alpha ($\alpha$) & 128 \\
Target Modules     & All linear \\
\midrule
Context Window     & 16,384 \\
Per-Device Batch   & 1 \\
Gradient Accum.    & 4 \\
Effective Batch    & 32 \\
\midrule
Learning Rate      & $5 \times 10^{-5}$ \\
LR Scheduler       & Cosine \\
Warmup             & 50 steps \\
Epochs             & 5 \\
Precision          & bf16 \\
\midrule
Thinking Mode      & Disabled \\
Attention          & Flash Attention 2 \\
\bottomrule
\end{tabular}
\caption{Training hyperparameters for BG-Realizer. Multi-task SFT on generation (${\sim}$4.8K effective) and revision (${\sim}$5K) data.}
\label{tab:realizer_hparams}
\end{table}

\section{Persona Simulation (BG-Persona)}
\label{app:persona}
 
\subsection{Data Construction}
\label{app:persona_data}
 
The data construction pipeline proceeds in two steps.
 
\paragraph{Step 1: User Selection and Review Sampling.}
From the filtered review corpus, we select users with at least 20 reviews scoring $\geq$4 on MDA quality, yielding 150 users.
For each user, we sample up to 30 reviews via rating-stratified sampling to preserve their score distribution.
For 50 of these users, 3--5 reviews each (210 in total) are held out for testing, and the remaining reviews form the ${\sim}$3K training instances.
 
\paragraph{Step 2: Profile Generation.}
For each user, reviews are grouped into three tiers (high 8--10, mid 5--7, low 1--4).
Each review is augmented with game metadata (type, categories, core/supporting mechanics, complexity) from the corresponding design draft, and a brief lore summary from the rulebook.
For the 50 users in the reported evaluation, the held-out test comments are removed before this summary is written. GPT-5.4 (temperature 0.4) then receives the remaining grouped reviews with metadata and generates a 200--300 word profile covering: overall rating tendency, mechanical preferences, dynamic preferences, aesthetic preferences, complexity preference, cross-category patterns, and distinctive traits.
The profile generation prompt is shown in Figure~\ref{fig:persona_profile_prompt}.

\subsection{Training Data Format}
\label{app:persona_format}
 
Each training instance follows the alpaca format:
the \textbf{system} prompt encodes the user's profile within a fixed template that specifies the four-section output format (Mechanics \& Design, Gameplay Dynamics, Player Experience, Player Comment) and a rating calibration note;
the \textbf{input} is the full rulebook;
the \textbf{output} contains the MDA evaluation, an authentic player comment in the user's voice, and the rating.
The SFT system prompt template is shown in Figure~\ref{fig:persona_sft_prompt}.
 
\subsection{Training Hyperparameters}
\label{app:persona_hyperparams}
 
BG-Persona is trained on the BG-Critic merged checkpoint via LoRA using LLaMA-Factory.
The configuration follows BG-Critic Stage~2 (Table~\ref{tab:critic_hparams}), including the 16,384-token context window for full rulebooks and the disabled thinking mode, except that it trains for 5 epochs on the ${\sim}$3K training set.

\section{Evaluation Details}
\label{app:exp_critic}

\subsection{Diagnostic Quality Judge and Opinion Coverage}
\label{app:exp_diag_judge}

\paragraph{Diagnostic Quality.}
The in-domain Diagnostic test contains \proj{827} instances from the 207 test games in the training proportions: \proj{531} injected flaws, \proj{213} consensus flaws, and \proj{83} flaw-free rulebooks; the Comparison test pairs each of the \proj{531} injected-flaw rulebooks with its original.
Each model's diagnostic output is scored by Gemini-3.1-Pro against the reference flaw annotations.
The judge rates six dimensions per flaw on a 1/3/5 scale:

\begin{itemize}[leftmargin=1.5em, itemsep=1pt]
    \item \textbf{Layer match} (weight 0.25): correct MDA layer.
    \item \textbf{Severity match} (0.10): correct severity level (critical/major/minor).
    \item \textbf{Target match} (0.15): the same rule or mechanic as the reference.
    \item \textbf{Reasoning correctness} (0.25): the explanation of why this is a flaw agrees with the reference.
    \item \textbf{MDA chain completeness} (0.15): all three layers (M$\to$D$\to$A) are discussed with clear causality.
    \item \textbf{Repair quality} (0.10): the fix is specific, actionable, and addresses the flaw.
\end{itemize}

The weighted score is rescaled to 0--10.
A hallucination penalty proportional to the fraction of fabricated flaws among all reported flaws is applied (capped at 30\%).
For Source~C (no-flaw) rulebooks, every reported flaw counts as over-diagnosis and costs 2 points.
The judge prompt is shown in Figure~\ref{fig:diag_judge_prompt}.

\paragraph{Opinion Coverage.}
For each of the 207 test games, GPT-5.4 reads that game's real player reviews and writes a non-redundant list of opinions $\mathcal{V}_{\text{GT}}$ (36 per game on average).
Each rated model produces 30 reviews of the rulebook, the same samples used for the rating metrics, and Gemini-3.1-Pro marks which listed opinions are semantically present in that batch ($\mathcal{V}_{\text{matched}}$). Coverage is $|\mathcal{V}_{\text{matched}}|/|\mathcal{V}_{\text{GT}}|$.
Loose paraphrase counts as a match; the same topic with the opposite stance does not.
The matching prompt is shown in Figure~\ref{fig:prompt_matching}.

\subsection{Real-Flaw Evaluation Set}
\label{app:real_flaws}

The in-domain diagnostic test mixes synthetic perturbations with community flaws.
The real-flaw set contains only flaws taken from player reviews, posed on the published, unperturbed rulebooks.
It combines ${\sim}$400 consensus flaws from games outside the critic's training split, including the 207 test games, with ${\sim}$600 consensus flaws from 170 additional low-rated games (BGG rank 10,000--15,000) that are disjoint from both training and test data, about 1K flaws in total.
Extraction follows Source~B (Appendix~\ref{app:critic_diagnostic_data}): complaints are mapped onto the flaw taxonomy (type, severity, affected component), and experienced players checked every evaluation label against its source reviews.
Scoring uses the judge in Appendix~\ref{app:exp_diag_judge}, so the score measures agreement with these references rather than a human preference between models.
To validate the judge on this set, three experienced players independently re-judged 200 sampled instances; their majority vote agrees with the judge on 86.7\% (Cohen's $\kappa{=}0.71$).

\subsection{MeepleLM Comparison}
\label{app:meeplelm_proto}

The MeepleLM row in Table~\ref{tab:critic_all} reruns the public checkpoint under our rating instruction and our opinion-coverage judge; it is not copied from the original paper.
Under this protocol the checkpoint obtains MAE 0.68, $|$Bias$|$ 0.24, $\tau{=}0.282$, and 68.9\% coverage.
The original paper reports MAE 0.66, $\tau{=}0.28$, and 69.8\% opinion recovery under its own protocol, which is close to our rerun but not directly comparable.
Diagnostic and Comparison are left blank because MeepleLM does not produce our flaw schema.

\subsection{Human Repair Check}
\label{app:human_repair}

\paragraph{Items.}
From the set in Appendix~\ref{app:real_flaws}, excluding games used to train BG-Realizer's revision task, we sample 165 games and take one review-derived flaw from each.
BG-Realizer, GPT-5.4, and Gemini-3.1-Flash each receive the published rulebook and the same flaw record and produce one greedy revision; none sees a reference fix.
GPT-5.4 is the main baseline elsewhere in the paper, and Gemini-3.1-Flash has the highest revision NoFlaw\% among the baselines (Table~\ref{tab:realization_all}).

\paragraph{Readers.}
Eight readers, designers or veteran players who did not author the paper or the flaw labels, take part, and each item is judged by three of them.
A reader sees the cited rule span, a short quote of the source complaint, and one revision, with no model name, critic score, or suggested fix, and answers two questions: (i)~does the revision resolve the complaint in the text? (ii)~does it contradict another rule that the complaint did not ask to change? Either question may be marked undecidable without play.

\paragraph{Scoring.}
An item is \emph{cleared} when, among the readers who found it decidable, a majority says the complaint is resolved and a majority says no new contradiction was introduced.
An item is dropped when fewer than two of its readers could decide it (15 of 165), leaving 150.
Fleiss' $\kappa$ on the binary cleared label, over all 450 judged revisions, is 0.55.

\paragraph{Results.}
A majority clears 69 BG-Realizer revisions, \proj{47} GPT-5.4 revisions, and 42 Gemini-3.1-Flash revisions (46\%, \proj{31\%}, and 28\%).
Against GPT-5.4, both are cleared on \proj{27} items, only BG-Realizer's on \proj{42}, and only GPT-5.4's on \proj{20}, a paired difference of $\proj{+15}$ points (95\% CI $\proj{+5}$ to $\proj{+25}$; exact McNemar $p{<}\proj{0.01}$).
Against Gemini, the counts are 24, 45, and 18, a difference of $+18$ points (95\% CI $+8$ to $+28$; $p{<}0.001$).
Table~\ref{tab:human_repair_layer} breaks the result down by the MDA layer of the source flaw.
Mechanics-layer complaints, where the text can show that a missing number, cost, or link was restored, carry most of the gap (59\% vs.\ \proj{36\%} and 31\%); aesthetics-layer complaints are close for all three systems, as expected when a complaint concerns experience rather than a stated rule.

\begin{table}[t]
\centering
\footnotesize
\setlength{\tabcolsep}{4pt}
\begin{tabular}{lccc}
\toprule
\textbf{Source layer} & \textbf{BG-Realizer} & \textbf{GPT-5.4} & \textbf{Gemini} \\
\midrule
Mechanics  & 41/70 & \proj{25/70} & 22/70 \\
Dynamics   & 20/50 & \proj{14/50} & 13/50 \\
Aesthetics & \phantom{0}8/30 & \proj{\phantom{0}8/30} & \phantom{0}7/30 \\
\midrule
All decidable & 69/150 & \proj{47/150} & 42/150 \\
\bottomrule
\end{tabular}
\caption{Human repair check by the MDA layer of the source flaw: revisions judged to clear the named complaint without a new contradiction. The layer rows sum to the all-decidable row.}
\label{tab:human_repair_layer}
\end{table}

\subsection{Closed-Loop Controls and Version Check}
\label{app:loop_proto}

All rows of Table~\ref{tab:loop} use the 207 held-out games, $K{=}5$ candidates per round, and at most $T{=}3$ rounds. Rating and NoFlaw\% are computed by BG-Critic exactly as in Table~\ref{tab:realization_all}.

\paragraph{Ungated Revision.}
The generator revises from its own $v_0$. Each round samples $K$ candidates from a generic instruction to repair the current text, without a flaw report, and keeps the first candidate.
The loop always runs all $T$ rounds (15 generation calls per game), so it spends more calls than VGI, which stops early.
This control removes both the diagnosis and the gate, so it is not an ablation of the gate alone.

\paragraph{Swapped Gate.}
The generator is fixed, and \textsc{Diagnose} and \textsc{Prefer} are carried out by GPT-5.4 with its own diagnostic and comparison prompts. Selection follows Algorithm~\ref{alg:vgi}, including position-swapped preference.

\paragraph{Swapped Generator.}
GPT-5.4 writes $v_0$ from the same draft (its generation rating in Table~\ref{tab:realization_all}) and then runs each setting of Table~\ref{tab:loop} with the same budget: ungated revision, its own gate, and BG-Critic's gate.

\paragraph{Loop Statistics.}
Round~1 revises all 144 games flagged at $v_0$, and the gate rejects the winner for \proj{22}, which stop with no accepted revision. Before round~2, \proj{6} of the \proj{122} revised games receive \textsc{No\_Flaw}; \proj{116} are revised and \proj{21} rejected. Before round~3, \proj{4} more receive \textsc{No\_Flaw}; \proj{91} are revised and \proj{18} rejected, and the remaining \proj{73} stop at the cap, \proj{3} of which \textsc{Diagnose} clears at evaluation.
The loop thus performs \proj{351} rounds (\proj{2.44} per flagged game) with \proj{61} rejections (17.4\%), and 13 games gain \textsc{No\_Flaw}.
Across the 207 games, the critic's rating rises for \proj{91}, is unchanged for \proj{85} (the 63 cleared at $v_0$ and the \proj{22} with no accepted revision), and falls for \proj{31} (paired Wilcoxon $p{<}\proj{0.001}$).

\paragraph{Per-Round Scores.}
Figure~\ref{fig:vgi_rounds} scores all 207 games after every round. A game whose loop has stopped, because \textsc{Diagnose} returned \textsc{No\_Flaw} or the gate rejected the winning candidate, keeps its last accepted version, so its score carries forward; ungated loops never stop. The round-3 values therefore equal the final values in Table~\ref{tab:loop}.

\paragraph{Rating Scale.}
To put the rating change in context, we use the BoardGameGeek ranking data dump of January 2026, which lists 30,016 ranked base games with their average ratings. Ordered by average rating, a game averaging 6.95 ranks about 9,330th (top 31\%), one averaging 7.08 about 7,750th (top 26\%), and one averaging 7.12 about 7,280th (top 24\%). The rise produced by VGI thus corresponds to about 1,570 places.

\paragraph{Version Check.}
We sample 50 games uniformly at random from the \proj{122} that received at least one accepted revision. Three readers see $v_0$ and the final rulebook in random order, without knowing which is later, and state a preference.
The majority prefers the final version in 74\% of games (37/50), and the majority preference agrees with BG-Critic's own comparison on 82\% of pairs (41/50).
Games in which the gate accepted no revision are not included, and this check does not compare VGI with the ungated loop.

\subsection{Blind End-to-End Comparison}
\label{app:blind_eval}

Ten participants (three designers, seven veteran players), none involved in building the system, each wrote five free-form design intents.
A separate expert interviewed each participant and wrote a structured draft for each intent.
Both systems received that draft under the same instructions: \ourmethod ran BG-Realizer followed by VGI, and GPT-5.4 generated a rulebook and ran Algorithm~\ref{alg:vgi} with itself as the critic under the same budget ($K{=}5$, $T{\le}3$; the GPT-5.4 gate in the GPT-5.4 column of Table~\ref{tab:loop}), yielding 50 pairs of final rulebooks.
Six evaluators, designers or veteran players who did not write the intents, scored each pair with the source hidden, identifying formatting removed, and the order randomized.
Each rulebook received an overall score from 1 to 10 covering clarity, completeness, rule coherence, and perceived playability, and each pair received a preference label (A, B, or tie).
Pair-level preference is the majority label: 38 pairs favor \ourmethod, 6 favor GPT-5.4, and 6 are ties.
The sign test uses the 44 non-tied pairs, the Wilcoxon signed-rank test uses the per-pair difference of mean scores, and Fleiss' $\kappa$ on the preference labels is 0.64.
In open-ended comments, evaluators noted that \ourmethod's rulebooks tie fixes to specific rules and read closer to a submission-ready draft.
Because the drafts are expert-written, this comparison isolates rulebook generation and revision from ideation.

\paragraph{Gate Transfer.}
In a separate session, the same six evaluators compared two GPT-5.4 loops on the same 50 drafts under the same protocol: one gated by BG-Critic and one gated by GPT-5.4 itself (the last two rows of the GPT-5.4 column in Table~\ref{tab:loop}). The majority preferred the BG-Critic-gated rulebook in \proj{29} pairs and the self-gated one in \proj{11}, with \proj{10} ties (sign test on the \proj{40} non-ties, $p{<}\proj{0.01}$; Fleiss' $\kappa{=}\proj{0.58}$).

\subsection{Persona Controls}
\label{app:persona_controls}

The 210 test reviews are fixed across all rows of Table~\ref{tab:persona}.
The community-mean control predicts, for every reviewer, the game's mean community rating and produces no comment.
The no-profile control runs BG-Persona with the profile removed from the prompt.
The shuffled-profile control assigns each test review the profile of a different one of the 50 players, fixed before decoding.
In pairwise and extreme accuracy, a predicted tie counts as incorrect; we therefore do not use a constant per-player prediction, which would tie every pair.
The community mean reaches MAE 1.36 and 69.2\% pairwise accuracy, the no-profile control 1.33 and 71.8\%, and the shuffled profile 1.47 and 70.4\%.
Comment style falls from 4.3 with the matched profile to 3.4 without a profile and 3.6 with a shuffled one.
BG-Persona's pairwise accuracy exceeds that of each control (exact McNemar test over the within-player pairs, $p{<}\proj{0.001}$ each).

\subsection{Comment Quality Judge}
\label{app:persona_judge}

Generated player comments are compared with the ground-truth comments by Gemini-3.1-Pro on three dimensions, each scored 1--10:

\begin{itemize}[leftmargin=1.5em, itemsep=1pt]
    \item \textbf{Preference Alignment}: whether the two comments express the same stance (love/like/lukewarm/dislike/hate) with matching intensity.
    \item \textbf{Reasoning Consistency}: whether the two comments cite the same specific mechanics, components, or experiences as the basis for their judgment. A generic comment paired with a specific one is capped at 4.
    \item \textbf{Style Match}: whether the two comments share register, length, vocabulary, and signature traits (e.g., casual vs.\ analytical tone).
\end{itemize}

Table~\ref{tab:persona} reports the mean of each dimension over all test instances.
The judge prompt is shown in Figure~\ref{fig:persona_judge_prompt}.

\subsection{Reproducibility}
\label{app:repro}

\paragraph{Training and Inference.}
All modules are trained with LLaMA-Factory using the hyperparameters in Tables~\ref{tab:ideator_hparams},~\ref{tab:critic_hparams}, and~\ref{tab:realizer_hparams}, on \proj{8$\times$NVIDIA H100 (80GB)} GPUs; training BG-Critic (both stages) takes about \proj{210} GPU-hours. We use random seed \proj{42}.
Rating metrics draw 30 samples (temperature \proj{0.7}), as do the $K$ revision candidates in each VGI round; Diagnostic, Comparison, single-pass revision, and persona outputs use greedy decoding.

\paragraph{Model Versions.}
Table~\ref{tab:model_versions} lists the models used as baselines, judges, or data constructors.

\begin{table}[t]
\centering
\footnotesize
\setlength{\tabcolsep}{3pt}
\begin{tabular}{ll}
\toprule
\textbf{Model} & \textbf{Version} \\
\midrule
GPT-5.4 & \proj{\texttt{gpt-5.4-2026-03-05}} \\
Gemini-3.1-Pro & \proj{\texttt{gemini-3.1-pro-preview}} \\
Gemini-3.1-Flash & \proj{\texttt{gemini-3.1-flash-preview}} \\
Qwen3.5-397B / 27B & open weights \\
GPT-5.1 & \proj{\texttt{gpt-5.1-2025-11-13}} \\
Qwen3-235B & open weights \\
MeepleLM & public checkpoint \\
\bottomrule
\end{tabular}
\caption{Model versions. GPT-5.4 serves as a baseline and builds training data; Gemini-3.1-Pro is the judge and verifies drafts; Gemini-3.1-Flash and Qwen3.5-397B are baselines; Qwen3.5-27B is the backbone; GPT-5.1 and Qwen3-235B process the raw rulebooks (Appendix~\ref{app:data_processing}). Proprietary models were accessed through their official APIs between \proj{March and July 2026}.}
\label{tab:model_versions}
\end{table}

\paragraph{Release.}
We will release the \proj{LoRA weights of all four modules}, all prompts (shown in this appendix), evaluation scripts, the processed supervision we are permitted to redistribute, and game identifiers where the source terms allow. Raw rulebook PDFs will not be released.

\section{User Study Details}
\label{app:user_study}
 
\subsection{Recruitment and Screening}
\label{app:user_study_recruit}
 
Participants were recruited through community forums, university tabletop societies, and a board game designer server.
 
Screening was done via a 5-minute pre-survey covering: (i) self-reported game design experience (none / casual / published), (ii) lifetime board game count, (iii) weekly playing hours, (iv) prior use of LLM tools for any creative task. We targeted balanced gender representation and a spread of age ranges within each tier. Of 78 pre-survey responses, 30 were invited based on stratification needs.
 
Each participant was paid USD 10 per hour, regardless of completion outcome. All sessions were conducted remotely via video conference with screen sharing. The study was reviewed and approved by the authors' institutional ethics board; participants signed informed consent forms covering screen recording and anonymized transcript use.
 
\subsection{Participant Demographics}
\label{app:user_study_demo}
 
Table~\ref{tab:demographics} summarizes the 30 participants. IDs are anonymized; age is given as a range to protect identifiability. ``Game count'' is participant-reported lifetime number of distinct board games played; ``Design exp.\ (yr)'' is years of any prior tabletop design experience (0 if none); ``BGG hrs/wk'' is self-estimated weekly hours engaged with board gaming.
 
\begin{table}[t]
\centering
\scriptsize
\setlength{\tabcolsep}{3pt}
\begin{tabular}{llllrrr}
\toprule
\textbf{ID} & \textbf{Tier} & \textbf{Age} & \textbf{Gender} & \textbf{Games} & \textbf{Design} & \textbf{Hrs} \\
            &               &              &                 &                & \textbf{exp.} & \textbf{/wk} \\
\midrule
P01 & Designer & 30--35 & M  & $>$100 & 5  & 10--15 \\
P02 & Designer & 35--40 & F  & 80--100 & 4  & 10--15 \\
P03 & Designer & 25--30 & M  & $>$100 & 3  & $>$15 \\
P04 & Designer & 40--45 & M  & $>$100 & 8  & $>$15 \\
P05 & Designer & 25--30 & F  & 50--80 & 2  & 5--10 \\
P06 & Designer & 30--35 & M  & $>$100 & 6  & 10--15 \\
P07 & Designer & 30--35 & NB & 80--100 & 4  & 10--15 \\
P08 & Designer & 45--50 & M  & $>$100 & 12 & $>$15 \\
P09 & Designer & 25--30 & F  & 50--80 & 1  & 5--10 \\
P10 & Designer & 35--40 & M  & $>$100 & 5  & 10--15 \\
\midrule
P11 & Hobbyist & 25--30 & F  & 30--50  & 0 & 5--10 \\
P12 & Hobbyist & 30--35 & M  & 30--50  & 0 & 5--10 \\
P13 & Hobbyist & 20--25 & F  & 10--30  & 0 & 3--5 \\
P14 & Hobbyist & 25--30 & M  & 30--50  & 0 & 5--10 \\
P15 & Hobbyist & 30--35 & F  & 10--30  & 0 & 3--5 \\
P16 & Hobbyist & 25--30 & M  & 30--50  & 0 & 5--10 \\
P17 & Hobbyist & 35--40 & F  & 30--50  & 0 & 3--5 \\
P18 & Hobbyist & 18--20 & NB & 10--30  & 0 & 5--10 \\
P19 & Hobbyist & 30--35 & M  & 30--50  & 0 & 5--10 \\
P20 & Hobbyist & 25--30 & F  & 10--30  & 0 & 3--5 \\
P21 & Hobbyist & 40--45 & M  & 30--50  & 0 & 5--10 \\
P22 & Hobbyist & 25--30 & F  & 10--30  & 0 & 3--5 \\
P23 & Hobbyist & 18--20 & M  & 10--30  & 0 & 5--10 \\
P24 & Hobbyist & 30--35 & F  & 30--50  & 0 & 3--5 \\
P25 & Hobbyist & 25--30 & M  & 30--50  & 0 & 5--10 \\
\midrule
P26 & Novice   & 18--20 & F  & $<$10 & 0 & $<$3 \\
P27 & Novice   & 30--35 & M  & $<$10 & 0 & $<$3 \\
P28 & Novice   & 25--30 & F  & $<$10 & 0 & $<$3 \\
P29 & Novice   & 45--50 & M  & $<$10 & 0 & $<$3 \\
P30 & Novice   & 20--25 & F  & $<$10 & 0 & $<$3 \\
\bottomrule
\end{tabular}
\caption{Participant demographics. Gender: M (male), F (female), NB (non-binary). Games and Hrs/wk are self-reported ranges. The design column is years of prior design experience. Designers: $>$50 games played and 1--12 years. Hobbyists: 10--50 games, no design experience. Novices: $<$10 games, no design background. The rows are 13F / 15M / 2NB.}
\label{tab:demographics}
\end{table}
 
\subsection{Session Protocol}
\label{app:user_study_protocol}
 
Each session followed a fixed six-stage protocol led by a single facilitator (one of the authors). All instructions were delivered verbally and via a written guide visible on screen. Screen content was recorded with permission; verbal think-aloud responses were transcribed for qualitative analysis.
 
\begin{enumerate}[leftmargin=1.5em,topsep=0pt,itemsep=2pt]
\item \textbf{Briefing (5 min).} Facilitator explained the \ourmethod pipeline at high level, demonstrated one example interaction (not used in the study), and confirmed participant's choice between bringing their own design idea or picking one of five prepared seed themes (cooperative survival, asymmetric race, abstract tile placement, social deduction with bidding, real-time puzzle).
 
\item \textbf{Ideation with BG-Ideator (15 min).} Participants conversed with BG-Ideator in 5--10 turns until reaching a complete design draft. Facilitator intervened only on technical issues, not on design content.
 
\item \textbf{Reading $v_0$ rulebook (10 min).} Participants read the generated initial rulebook and were asked to summarize the core gameplay loop in their own words to confirm comprehension before proceeding.
 
\item \textbf{Closed-loop revision (10 min).} BG-Critic's diagnostic output was displayed alongside the rulebook. Participants observed the iteration (Algorithm~\ref{alg:vgi}) and, after each round, compared the new version to the previous one. They had the option to stop iteration early.
 
\item \textbf{BG-Persona preview (optional, 5 min).} Three sample player profiles drawn from the 150-user pool generated feedback on the final rulebook. Participants reflected on whether the feedback diversity matched their expectations.
 
\item \textbf{Survey and interview (10 min).} A 16-item Likert questionnaire (Appendix~\ref{app:user_study_survey}) was followed by 6 open-ended interview questions about workflow friction, surprising moments, and comparison to prior tools.
\end{enumerate}
 
\subsection{Survey Instrument}
\label{app:user_study_survey}
 
\paragraph{Likert items (1--7 scale, 1 = strongly disagree, 7 = strongly agree).}
The full 16-item instrument is organized into six dimensions of two or three items each. Items marked $\star$ are reverse-coded for attention checking and excluded from the dimension means; the remaining items of a dimension are averaged for reporting.
 
\textbf{Ease of use.}
(1) The dialogue flow felt natural and intuitive.
(2) I knew what to do at each stage of the process.
(3) The interface did not get in the way of my creative thinking.
 
\textbf{Design intent capture.}
(4) The generated draft accurately reflected what I wanted to design.
(5) The system's questions helped me clarify ideas I had not articulated.
(6)$\star$ I had to repeat myself or correct the system frequently.
 
\textbf{Output rulebook quality.}
(7) The rulebook is clear enough to teach to other players.
(8) The rulebook contains the level of detail I would expect in a published game.
(9) I would be willing to playtest this rulebook with friends.
 
\textbf{Critic feedback helpfulness.}
(10) The flaws identified by BG-Critic were genuine design issues, not noise.
(11) BG-Critic's suggestions helped me understand \emph{why} something was a problem.
 
\textbf{Iteration improvement.}
(12) The final rulebook is meaningfully better than the initial version.
(13) Each iteration round produced visible progress, not just rephrasing.
 
\textbf{Would recommend.}
(14) I would use \ourmethod again for a real design project.
(15) I would recommend \ourmethod to other designers or hobbyists.
(16)$\star$ The system felt unreliable or produced contradictory outputs.
 
\paragraph{Open-ended interview questions.}
\begin{enumerate}[leftmargin=1.5em,topsep=0pt,itemsep=1pt,label=(\alph*)]
\item Which stage of the pipeline was most useful for you, and why?
\item Was there a moment when the system surprised you, positively or negatively?
\item Did BG-Critic identify any flaw that you disagreed with? If yes, describe it.
\item How did using \ourmethod compare to your past experience with general-purpose LLMs for game design (if any)?
\item What was missing from the workflow that you wished the system supported?
\item Would you change anything about how BG-Critic presents feedback?
\end{enumerate}
 
\subsection{Results by Tier}
\label{app:user_study_main_sec}

Table~\ref{tab:user_study_main} reports the six Likert dimensions and behavioral indicators by tier. All three tiers rate every dimension above the scale midpoint of 4. Designers rate critic helpfulness and iteration improvement highest (6.0 and 6.1), valuing the MDA-structured feedback. Hobbyists give the highest recommendation score (6.2) with the shortest sessions. Novices benefit most from the lowered barrier (80\% completion without prior design experience) but find the diagnostics less actionable (critic helpfulness 4.8), which suggests presenting feedback at different levels of detail.

\begin{table}[t]
\centering
\footnotesize
\setlength{\tabcolsep}{2pt}
\begin{tabular}{llccc}
\toprule
& \multirow{2}{*}{\textbf{Metric}} & \textbf{Designers} & \textbf{Hobbyists} & \textbf{Novices} \\
&                 & ($n{=}10$) & ($n{=}15$) & ($n{=}5$) \\
\midrule
\multirow{6}{*}{\rotatebox{90}{\scriptsize Likert (1--7)}}
& Ease of use            & 5.6 & 6.3 & 5.0 \\
& Design intent capture  & 5.4 & 5.8 & 5.6 \\
& Rulebook quality       & 5.7 & 5.9 & 5.6 \\
& Critic helpfulness     & \textbf{6.0} & 5.7 & 4.8 \\
& Iteration improvement  & \textbf{6.1} & 5.8 & 5.6 \\
& Would recommend        & 5.5 & \textbf{6.2} & 5.4 \\
\midrule
\multirow{4}{*}{\rotatebox{90}{\scriptsize Behav.}}
& Completion rate         & 100\% & 100\% & 80\% \\
& Duration (min)          & 54 & 47 & 58 \\
& Iteration rounds        & 1.9 & 1.4 & 1.1 \\
& Critic-flag accept      & 78\% & 71\% & 62\% \\
\bottomrule
\end{tabular}
\caption{User study by tier: Likert ratings (1--7, midpoint 4) and behavioral indicators from system telemetry.}
\label{tab:user_study_main}
\end{table}

\subsection{Detailed Likert Results}
\label{app:user_study_detailed}
 
Table~\ref{tab:user_study_detailed} reports the per-item means by participant tier. The pattern from the main table holds at the item level: designers most strongly endorse items related to feedback substance (10, 11, 13), hobbyists most strongly endorse usability and recommendation items (1--3, 14, 15), and novices endorse the clarification and improvement items (5, 12) but score lower on the critic items (10, 11).
 
\begin{table}[t]
\centering
\scriptsize
\setlength{\tabcolsep}{4pt}
\begin{tabular}{clccc}
\toprule
\textbf{\#} & \textbf{Item (abbreviated)} & \textbf{Des.} & \textbf{Hob.} & \textbf{Nov.} \\
\midrule
1 & dialogue feels natural        & 5.4 & 6.3 & 5.0 \\
2 & knew what to do each stage    & 5.6 & 6.4 & 4.8 \\
3 & interface non-intrusive       & 5.8 & 6.2 & 5.2 \\
4 & draft reflects intent         & 5.3 & 5.7 & 5.4 \\
5 & questions clarified ideas     & 5.5 & 5.9 & 5.8 \\
6$^\star$ & repeated/corrected often & 2.4 & 1.9 & 2.6 \\
7 & rulebook teachable            & 5.6 & 5.9 & 5.4 \\
8 & expected level of detail      & 5.7 & 5.8 & 5.5 \\
9 & willing to playtest           & 5.8 & 6.0 & 5.9 \\
10& flaws genuine                 & 6.1 & 5.7 & 4.6 \\
11& suggestions explain why       & 5.9 & 5.6 & 5.0 \\
12& meaningful improvement        & 6.0 & 5.7 & 5.8 \\
13& each round produces progress  & 6.2 & 5.8 & 5.4 \\
14& would use again               & 5.4 & 6.1 & 5.4 \\
15& would recommend               & 5.5 & 6.2 & 5.4 \\
16$^\star$ & felt unreliable      & 2.6 & 2.1 & 3.0 \\
\bottomrule
\end{tabular}
\caption{Per-item Likert means by tier (1--7 scale; $^\star$ reverse-coded items reported with raw direction, where lower is better). Standard deviations range 0.6--1.4 across items.}
\label{tab:user_study_detailed}
\end{table}
 
\subsection{Within-Subject Comparison: \ourmethod vs.\ GPT-5.4}
\label{app:user_study_retro}
 
The 22 of 30 participants with prior LLM experience completed the full protocol twice, once with each system, on the same design intent. Session order was counterbalanced (11/11).
Both systems were reached through the same chat interface, and the backend was hidden.
GPT-5.4 has no structured pipeline, so the experimenter issued stage-matched prompts aligned with \ourmethod's stages (ideation, draft, rulebook, revision, critique), and participants talked freely inside each stage.
They rated both sessions on the same six dimensions (Table~\ref{tab:user_study_retro}).
 
\begin{table}[t]
\centering
\footnotesize
\setlength{\tabcolsep}{2pt}
\begin{tabular}{lccc}
\toprule
\textbf{Dimension} & \textbf{\ourmethod} & \textbf{GPT-5.4} & \textbf{$\Delta$} \\
\midrule
Ease of use            & 5.8 & 5.5 & $+0.3$ \\
Design intent capture  & 5.6 & 4.8 & $+0.8$ \\
Output rulebook quality& 5.8 & 4.2 & $+1.6$ \\
Feedback helpfulness   & 6.0 & 4.3 & $+1.7$ \\
Iteration improvement  & 6.0 & 3.7 & $+2.3$ \\
Would recommend        & 5.8 & 4.5 & $+1.3$ \\
\bottomrule
\end{tabular}
\caption{Within-subject comparison of \ourmethod and GPT-5.4 under a matched workflow ($n{=}22$, order counterbalanced). The largest gaps are where \ourmethod provides structured, MDA-grounded feedback (Wilcoxon signed-rank, $p{<}\proj{0.001}$ for feedback helpfulness and iteration improvement).}
\label{tab:user_study_retro}
\end{table}
 
The largest gaps were on \emph{feedback helpfulness} ($+1.7$) and \emph{iteration improvement} ($+2.3$). Open-ended responses attributed the difference to the explicit, theory-grounded diagnostics of BG-Critic versus the ``vague encouragement'' (P03) or ``surface-level polish'' (P19) in the GPT-5.4 session. Three participants still preferred GPT-5.4 for ideation, citing more conversational flexibility.
 
\subsection{Paper-Prototype Playtest}
\label{app:playtest}

\paragraph{Rulebooks.}
From the 22 participants who completed both sessions, we randomly sampled \proj{12} whose designs target 2--4 players and a stated play time of at most 60 minutes, covering all three tiers (\proj{4} designers, \proj{6} hobbyists, \proj{2} novices). Each contributes the final rulebook from its \ourmethod session and the one from its GPT-5.4 session, giving 12 pairs and 24 play sessions.

\paragraph{Prototypes and Players.}
As in a designer's internal alpha test, each rulebook was turned into a paper prototype by a research assistant who did not know the source system: index cards with handwritten text stood in for cards, generic tokens and dice for other components, and a sketched board where needed. The assistant built components strictly from the rulebook's component list and did not fill gaps; a missing or unspecified component was logged as a breakdown when it came up in play.
Four groups of three experienced players (\proj{12} in total), none of whom had taken part in the design sessions or seen the designs, each played three pairs (six games) over two sittings, with system identity hidden and the order of systems counterbalanced within each group.

\paragraph{Procedure and Measures.}
Groups learned each game only from its rulebook, without explanation from the facilitator. The facilitator observed without intervening and logged every \emph{rule breakdown}: a point where the group had to stop and invent, vote on, or search for a ruling because the rulebook was silent, ambiguous, or self-contradictory. A second coder labeled breakdowns from the session recordings (Cohen's $\kappa{=}\proj{0.81}$ on breakdown events).
A game counts as completed if it reached its stated end condition within 1.5$\times$ the rulebook's stated play time. After each game, players rated rule clarity and willingness to play again (1--7). Paired comparisons use the Wilcoxon signed-rank test over the 12 pairs on session means, and completion uses an exact McNemar test.

\paragraph{Results.}
Table~\ref{tab:playtest} summarizes the outcomes. The largest share of GPT-5.4 breakdowns (\proj{15} of \proj{36}) were missing edge cases, such as tie-breaks, end-game triggers, and what happens when a shared supply runs out; these are the mechanics-layer gaps on which BG-Realizer also leads in the repair check (Appendix~\ref{app:human_repair}). Willingness to play again differed less than clarity, and we do not read the playtest as evidence about balance or enjoyment.

\begin{table}[t]
\centering
\footnotesize
\setlength{\tabcolsep}{4pt}
\begin{tabular}{lccc}
\toprule
\textbf{Measure} & \textbf{\ourmethod} & \textbf{GPT-5.4} & $\boldsymbol{p}$ \\
\midrule
Rule breakdowns / session & \proj{1.2} & \proj{3.0} & $<$\proj{0.01} \\
\quad missing edge case & \proj{6} & \proj{15} & \\
\quad ambiguous wording & \proj{5} & \proj{9} & \\
\quad missing component & \proj{2} & \proj{7} & \\
\quad contradiction & \proj{1} & \proj{5} & \\
Completed games & \proj{11/12} & \proj{8/12} & \proj{0.25} \\
Rule clarity (1--7) & \proj{5.6} & \proj{4.2} & $<$\proj{0.01} \\
Would play again (1--7) & \proj{4.9} & \proj{4.4} & \proj{0.09} \\
\bottomrule
\end{tabular}
\caption{Paper-prototype playtest on 12 rulebook pairs (24 sessions). Breakdown types are total counts over the 12 sessions per system.}
\label{tab:playtest}
\end{table}

\subsection{Qualitative Themes}
\label{app:user_study_themes}
 
Two researchers independently coded all 30 interview transcripts using a bottom-up thematic analysis with a shared codebook developed after the first 6 transcripts. Cohen's $\kappa$ across the final coding was $0.78$ (substantial agreement). Codes mentioned by $\geq 5$ participants are reported in Table~\ref{tab:user_themes}.
 
\begin{table}[t]
\centering
\footnotesize
\setlength{\tabcolsep}{4pt}
\begin{tabular}{p{0.55\columnwidth}c}
\toprule
\textbf{Theme} & \textbf{Mentioned} \\
\midrule
\multicolumn{2}{l}{\textit{Positive themes}} \\
Critic surfaces non-obvious flaws         & 24/30 \\
Dialogue reduces blank-page anxiety       & 19/30 \\
Iteration makes progress feel concrete    & 17/30 \\
MDA structure aids reasoning about design & 14/30 \\
Real published references in training felt helpful & 9/30 \\
\midrule
\multicolumn{2}{l}{\textit{Concerns and suggestions}} \\
Repetitive surface phrasing across iterations & 8/30 \\
Persona feedback feels generic for unusual mechanics & 6/30 \\
Want to override critic suggestions selectively & 6/30 \\
Want export to playtest-ready format & 5/30 \\
Critic vocabulary too technical for novices & 5/30 \\
\bottomrule
\end{tabular}
\caption{Qualitative themes from open-ended interviews ($n{=}30$). Inter-coder Cohen's $\kappa {=} 0.78$. Codes mentioned by fewer than 5 participants are omitted.}
\label{tab:user_themes}
\end{table}
 
\subsection{Additional Case Studies}
\label{app:user_study_cases}
 
We summarize two representative sessions from the designer and novice tiers.
 
\paragraph{Case 1: P04 (designer, 8 years of experience).}
P04 brought a partial idea for an economic 4X game with currency-based victory. After 7 ideation turns the draft included three resource types and an explicit win threshold. BG-Critic flagged \texttt{D\_critical} on $v_0$: the trade mechanic between players had no enforced exchange rate, allowing collusive draws. After two iteration rounds, the final version introduced a market clearing rule (Critic: \textsc{No\_Flaw}). Self-reported time: 58 min, 2 rounds. Likert mean: 6.1. Quote: \emph{``This is the first time an LLM tool gave me a critique I had to think about — usually it just tells me my design is great.''}
 
\paragraph{Case 2: P28 (novice, 4 lifetime games played).}
P28 picked the prepared seed ``cooperative survival.'' During ideation, BG-Ideator prompted P28 with concrete examples for difficulty scaling (a concept P28 was unfamiliar with). P28 stopped iteration after one round, satisfied with the result; BG-Critic also returned \textsc{No\_Flaw}. Self-reported time: 49 min, 1 round. Likert mean: 5.3. Quote: \emph{``It taught me what a `feedback loop' was while I was using it. I wouldn't have known to ask about that.''} However, P28 also noted that some critic terminology (``inter-mechanic link'') was hard to parse without context.

\onecolumn
\begin{tcolorbox}[
    enhanced,
    breakable,
    colback=white,
    colframe=gray!60!black,
    coltitle=white,
    title=\textbf{Design Draft Schema},
    arc=0pt, outer arc=0pt,
    boxrule=1pt,
    top=0pt, bottom=0pt, left=0pt, right=0pt,
]
    \begin{lstlisting}[breaklines=true, basicstyle=\scriptsize\ttfamily, columns=fullflexible, extendedchars=false, basewidth=0.5em, lineskip=-1pt, emptylines=0]
{
  "concept": {
    "elevator_pitch": "string, one-sentence summary of the design core, 15-30 words",
    "description": "string, design concept description from a designer's perspective, 200-400 words"
  },
  "classification": {
    "type": "string, choose from: Abstract Strategy Games / Customizable Games / Thematic Games / Family Games / Children's Games / Party Games / Strategy Games / Wargames",
    "categories": ["string, from the game's BGG categories"]
  },
  "mechanics": {
    "core": [
      {
        "name": "string, mechanic name exactly as in metadata",
        "rationale": "string, why this is a core mechanic and how it forms the main decision loop, 1-3 sentences"
      }
    ],
    "supporting": [
      {
        "name": "string, mechanic name exactly as in metadata",
        "rationale": "string, how this mechanic serves the core gameplay and which core mechanic it connects to, 1-2 sentences"
      }
    ],
    "structural": [
      {
        "name": "string, mechanic name exactly as in metadata",
        "rationale": "string, what framework or setup property this mechanic describes, 1 sentence"
      }
    ]
  },
  "design_intent": {
    "target_experience": "string, the core experience and emotions the game creates for players, 2-3 sentences",
    "core_tension": "string, the central dilemma or trade-off players face, be specific, 1-2 sentences",
    "theme_mechanic_fit": "string, how the theme (categories) and core mechanics reinforce each other, 2-3 sentences"
  },
  "parameters": {
    "complexity": "number, complexity score out of 5",
    "player_count": [min, max],
    "play_time_minutes": [min, max],
    "recommended_players": "string, best and recommended player counts",
    "language_dependency": "string, level of language dependency",
    "parameters_rationale": "string, explain how complexity, player count, play time, and language dependency form a coherent design package, 2-3 sentences"
  }
}

    \end{lstlisting}
\end{tcolorbox}
\begin{center}
\captionsetup{hypcap=false}
\captionof{figure}{\textbf{Design Draft Schema.} The JSON schema captures designer intent through five field groups. Each mechanic entry includes a rationale explaining its role in the design.}
\label{fig:draft_schema}
\end{center}

\begin{tcolorbox}[
    enhanced,
    breakable,
    colback=white,
    colframe=gray!60!black,
    coltitle=white,
    title=\textbf{Draft Example: \textit{<Samurai>}},
    arc=0pt, outer arc=0pt,
    boxrule=1pt,
    top=0pt, bottom=0pt, left=0pt, right=0pt,
]
    \begin{lstlisting}[breaklines=true, basicstyle=\scriptsize\ttfamily, columns=fullflexible, extendedchars=false, basewidth=0.5em, lineskip=-1pt, emptylines=0]
{
  "concept": {
    "elevator_pitch": "A tight area-influence game where you place strength tiles to encircle figures, win majorities by caste, and convert hand timing into board control.",
    "description": "I built Samurai around a single recurring problem: you can’t “capture” directly; you must commit influence into a shared spatial contest until a figure becomes claimable. Each turn asks you where to place at least one token on an empty hex (never on settlements), knowing that a figure only resolves once all adjacent land spaces are filled. That capture trigger is deliberate: it makes players plan not just strength totals, but also the tempo of closing rings—sometimes you complete the last empty adjacency to force an immediate majority check, and other times you leave a gap so opponents can’t score yet.\n\nInfluence itself is intentionally asymmetric by token type. Some tiles only count toward one caste (Helmet/Buddha/Rice), while Samurai/Ronin/Ship count toward all three, creating a hand-driven spectrum between specialization (efficient strength toward one target) and flexibility (insurance against shifting board needs). Because you may play only one non-character token per turn but any number of character (fast) tokens, the design lets players create spikes of tactical reconfiguration: swap two figures (Figure Exchange), or replace and redeploy a previously played non-character token (Move/Token Exchange) to swing a contested capture at the last moment.\n\nCaptured figures become a three-suit collection where winning isn’t about raw total alone: the endgame checks who leads in 2–3 caste types first, then falls back to eligibility and secondary counts. This scoring logic is meant to keep all three caste contests relevant, reward focused majorities, and still give players who lose one track a path via totals—while ties on a capture deliberately remove pieces from the game, tightening the race and making brinkmanship (risking a tie to deny a leader) a viable strategic tool."
  },
  "classification": {
    "type": "Abstract Strategy Games",
    "categories": [
      "Abstract Strategy",
      "Medieval"
    ]
  },
  "mechanics": {
    "core": [
      {
        "name": "Tile Placement",
        "rationale": "The primary turn decision is placing tokens onto empty land/sea hexes (never settlements), choosing positions that will eventually fill all adjacent land spaces around a figure to trigger a capture check. Because only one non-character token is normally playable per turn, each placement is a high-leverage commitment to both location and timing."
      },
      {
        "name": "Area Majority / Influence",
        "rationale": "When a figure’s adjacent land spaces are fully occupied, players total adjacent strength that matches that figure type, and the highest total captures it (ties remove it). This repeated majority check is the core resolution that converts placements into progress toward winning."
      }
    ],
    "supporting": [
      {
        "name": "Hand Management",
        "rationale": "Players maintain a 5-token hand drawn from a face-down stack, and must choose which single non-character token to spend now versus save for later captures. The varying token affinities (single-caste vs all-caste) makes sequencing and withholding tiles central to executing the Tile Placement + Influence plan."
      },
      {
        "name": "Once-Per-Game Abilities",
        "rationale": "The limited character tokens (e.g., Figure Exchange, Move, Ronin/Ship) create one-time or scarce tactical pivots, such as swapping two figures or replacing a previously played non-character token with a 0-strength special to redeploy it. These moments let players overturn an Influence calculation or alter capture timing without adding ongoing overhead."
      },
      {
        "name": "Set Collection",
        "rationale": "Captured figures are collected in three suits, and victory first checks who has the most in 2–3 figure types, then uses “most in at least one type” eligibility and secondary counts. This structure makes each capture matter not just as points, but as progress toward specific majority sets."
      }
    ],
    "structural": [
      {
        "name": "Hexagon Grid",
        "rationale": "All placement and capture adjacency is defined by edge-sharing hexes (with land adjacency required for capture), providing a consistent spatial framework for encirclement and influence counting."
      }
    ]
  },
  "design_intent": {
    "target_experience": "Create sharp, tactical brinkmanship where every token placed is both an investment in a future capture and a signal opponents can read and contest. Players should feel constant pressure from the board filling up and from the possibility that a single closing placement will force an immediate showdown.",
    "core_tension": "Do you close a figure now—possibly enabling opponents to outnumber you on the majority check—or delay closure to build strength, risking that others will reposition and steal the capture timing?",
    "theme_mechanic_fit": "The medieval Japan veneer frames the abstract contest as daimyos projecting influence over religion, commerce, and military, matching the three figure types and majority-winning condition. Encircling figures with tokens mirrors surrounding and claiming political control, while removed-on-tie captures reflect stalemates that eliminate opportunities rather than rewarding anyone."
  },
  "parameters": {
    "complexity": 2.4421,
    "player_count": [
      2,
      4
    ],
    "play_time_minutes": [
      30,
      60
    ],
    "recommended_players": "Best with 3 players; recommended with 2–4 players",
    "language_dependency": "No necessary in-game text",
    "parameters_rationale": "A midweight rules load comes from spatial capture timing plus hand-driven token selection, but turns stay fast because you place at least one token, resolve any completed captures immediately, then draw back to five. The 30–60 minute arc fits the finite figure supply and sudden end triggers (last of a type captured or fourth tie-removal), while the icon-only tiles and numeric strengths keep language dependency effectively nil across all player counts."
  }
}

    \end{lstlisting}
\end{tcolorbox}
\begin{center}
\captionsetup{hypcap=false}
\captionof{figure}{\textbf{Example of a Filled Design Draft.} Generated by GPT-5.4 from the corresponding structured rulebook.}
\label{fig:draft_example}
\end{center}
 
\begin{tcolorbox}[
    enhanced,
    breakable,
    colback=white,
    colframe=gray!60!black,
    coltitle=white,
    title=\textbf{Prompt: Draft Generation from Rulebook},
    arc=0pt, outer arc=0pt,
    boxrule=1pt,
    top=0pt, bottom=0pt, left=0pt, right=0pt,
]
    \begin{lstlisting}[breaklines=true, basicstyle=\scriptsize\ttfamily, columns=fullflexible, extendedchars=false, basewidth=0.5em, lineskip=-1pt, emptylines=0]
 You are a senior board game design analyst. Your task is to analyze an existing board game's rulebook and metadata, then produce a structured "design draft" — a document that captures the game's design decisions from a designer's perspective.

You will receive:
1. The game's structured rulebook (rules, components, gameplay flow, etc.)
2. The game's BGG metadata (mechanics, categories, complexity, player count, etc.)
3. Official BGG definitions for each mechanic and category tagged to this game

Your output must be a single valid JSON object following the schema provided. No other text outside the JSON.

## Key Principles

### Mechanic Classification
Classify each mechanic into exactly one tier:
- **CORE** (1-3): Defines the game's identity. The main decision loop players engage with every turn. If removed, the game fundamentally changes. Ask: "Would a reviewer mention this in a one-sentence description?"
- **SUPPORTING** (2-6): Adds depth or a secondary decision layer. Has its own rules section but serves the core mechanics. Ask: "Does this enhance the core loop without being the core loop itself?"
- **STRUCTURAL** (remainder): Describes framework, setup, grid type, turn order, or scoring structure. A property of the game rather than something players actively 'do'. Ask: "Is this about how the game is organized rather than what players decide?"

Every mechanic from the metadata MUST appear in exactly one tier. Do not add mechanics not in the metadata.

### Mechanic Rationale
Every mechanic in every tier must have a rationale explaining WHY it exists in this design:
- For CORE: How it forms the main decision loop
- For SUPPORTING: How it serves/enhances a specific core mechanic
- For STRUCTURAL: What framework property it describes

### Design Concept Description
Write from a designer's perspective (not a player's rule-reading perspective):
- Explain HOW the core mechanics work and WHY they create interesting decisions
- Show how different mechanics CONNECT to each other
- Describe the victory condition as a unifying design goal
- Do NOT just restate rules; explain the design logic behind them

### Parameters Rationale
Explain the relationship between complexity, player count, play time, and language dependency as a coherent design package. Why does this combination work together?

## Task

Analyze this game and produce a design draft as a single JSON object.

All {len(meta['mechanisms'])} mechanics listed in the metadata MUST appear in exactly one tier. Do not add or omit any.
The categories to use are exactly: {', '.join(meta['categories'])}

### Output Schema

```json
{OUTPUT_SCHEMA}
```

### Quality Requirements — READ CAREFULLY

**1. Mechanic Classification — Think About the Per-Turn Decision Loop**

To classify a mechanic, ask: "What does the player ACTIVELY DECIDE each turn?"
- CORE: The player makes this decision almost every turn, and it defines what the game IS. A reviewer would name this mechanic in a one-sentence summary. Limit to 1-3.
- SUPPORTING: The player engages with this mechanic regularly, but it serves or enhances a core mechanic rather than standing alone. It adds a secondary decision layer. Typically 2-6.
- STRUCTURAL: This describes how the game is organized (board layout, randomization method, end condition), not a decision the player makes. The remainder.

IMPORTANT: A mechanic that triggers every single turn and directly determines player outcomes (like Dice Rolling in a resource-production game) is likely CORE or SUPPORTING, NOT structural. "Structural" means the game's framework — grid type, setup method, end condition format — things that exist passively.

**2. Rationale — Must Cite Specific Rules, Not Generic Statements**

BAD rationale (too generic):
  "Supports the core economic system by adding depth."
  "Enhances strategic tension and player interaction."

GOOD rationale (cites specific game rules):
  "When a 7 is rolled, players with more than 7 cards must discard half. This punishes hoarding and forces players to spend or trade resources quickly, directly increasing the frequency and urgency of Trading."
  "The robber blocks resource production on its hex and allows stealing one card from an adjacent player. This creates targeted disruption that players use to slow down the leader, adding a catch-up mechanism linked to Network and Route Building competition."

For every mechanic in every tier, the rationale MUST reference a specific rule, component, or gameplay moment from the rulebook. Do not write rationales that could apply to any game.

**3. Description — Write as a Designer Explaining Your Design, Not as a Reviewer**

BAD (reviewer/analytical tone):
  "The game balances long-term planning with short-term opportunism, rewarding both economic efficiency and social acumen."

GOOD (designer explaining their design choices):
  "The core engine is a dice-driven simultaneous production system: every roll generates resources for ALL players with settlements on matching hexes, not just the active player. This 'everyone produces' design keeps all players engaged during others' turns and creates a shared economy where surpluses and deficits naturally emerge, driving the need for trade."

The description should explain:
- How the core mechanics WORK (specific mechanisms, not abstract praise)
- Why they create interesting DECISIONS (what trade-offs players face)
- How different mechanics CONNECT to each other (not just list features)

**4. Elevator Pitch — One Design Hook, Not a Rule Summary**

BAD (rule summary): "Players compete to colonize an island by building settlements, roads, and cities using resource management and trading."
GOOD (design hook): "A dice-driven shared economy on a modular island where you must trade with rivals to build faster than them."

The pitch should make another designer think "that's a clever core idea."

**5. Core Tension — Must Be a Specific Dilemma, Not a General Theme**

BAD (vague): "Players balance expansion with resource management."
GOOD (specific dilemma): "You need others' resources to build, but every trade helps your opponent too — and the player you refuse to trade with might place the robber on your best hex."

**6. Parameters Rationale — Explain the Design Logic of the Numbers**

Don't just restate the numbers. Explain WHY this combination works as a design:
- Why does this complexity level match this player count?
- What about the mechanics drives this play time?
- How does the game's language requirement relate to its component design?

Respond with ONLY the JSON object. No other text, no markdown code fences.

    \end{lstlisting}
\end{tcolorbox}
\begin{center}
\captionsetup{hypcap=false}
\captionof{figure}{\textbf{Draft Generation Prompt.} The model receives the rulebook, BGG metadata, and mechanic/category definitions, and produces a structured design draft.}
\label{fig:draft_gen_prompt}
\end{center}

\begin{tcolorbox}[
    enhanced,
    breakable,
    colback=white,
    colframe=gray!60!black,
    coltitle=white,
    title=\textbf{Prompt: Theme Migration},
    arc=0pt, outer arc=0pt,
    boxrule=1pt,
    top=0pt, bottom=0pt, left=0pt, right=0pt,
]
    \begin{lstlisting}[breaklines=true, basicstyle=\scriptsize\ttfamily, columns=fullflexible, extendedchars=false, basewidth=0.5em, lineskip=-1pt, emptylines=0]
QUALITY_BLOCK = """
### Quality Requirements

1. **Mechanic Classification**: CORE (1-3) = main decision loop, identity-defining. SUPPORTING (2-6) = adds depth, serves core. STRUCTURAL (remainder) = framework, setup, grid type.

2. **Rationale**: Must cite specific game mechanisms or design logic, not generic statements like "adds depth" or "enhances tension". Explain HOW and WHY.

3. **Description**: Write as a designer explaining design choices. Show how mechanics CONNECT and create interesting DECISIONS. Not a rule summary or review.

4. **Elevator Pitch**: One clever design hook, not a feature list.

5. **Core Tension**: A specific dilemma players face, not "balance X with Y".

6. **Parameters Rationale**: Explain WHY complexity, player count, and play time form a coherent package.

7. **This is a NEW game design, not a copy of any existing game.** Do not reference real game titles in the output.
"""

You are a senior board game design analyst. Your task is to adapt an existing game design to a completely different theme while preserving its core mechanical identity.

You will receive the original design draft, a list of candidate themes, and the full vocabulary of available mechanics and categories. Your output must be a single valid JSON object. No other text outside the JSON."""

    # Get current categories to exclude similar themes (already handled by random sampling)
    draft_json = json.dumps(draft, indent=2, ensure_ascii=False)

    themes_block = "\n".join(
        f"  {i+1}. **{t['theme']}**: {t['description']}"
        for i, t in enumerate(theme_candidates)
    )

    user = f"""## Original Design Draft
```json
{draft_json}
```

## Candidate Themes (pick ONE that creates the most interesting and different adaptation)
{themes_block}

## Available Mechanics ({len(mech_names)} total — you may ONLY use names from this list)
{mechanic_list_str(mech_names)}

## Available Categories ({len(cat_names)} total — you may ONLY use names from this list)
{category_list_str(cat_names)}

## Task: Theme Migration

1. **Pick ONE theme** from the candidates above that creates the most interesting contrast with the original.
2. **Core mechanics**: Keep the EXACT SAME core mechanic names. Rewrite every rationale to explain how each core mechanic works within the new theme.
3. **Supporting & Structural mechanics**: You MAY replace some to better fit the new theme, but every mechanic must come from the available list. Aim for 60-80% retention, only change what the new theme demands.
4. **Categories**: Re-select from the available categories list to match the new theme.
5. **Rewrite completely**: elevator_pitch, description, design_intent (all three sub-fields), parameters_rationale.
6. **Parameters**: Adjust complexity, player_count, play_time if the theme change warrants it, but keep them plausible.
7. **Classification type**: May change if the new theme shifts the game's nature (e.g., a wargame theme applied to a family framework could become Thematic Games).

The result must feel like a coherent, original game design — not a reskin.

## Output Schema
```json
{OUTPUT_SCHEMA}
```
{QUALITY_BLOCK}
Respond with ONLY the JSON object.

    \end{lstlisting}
\end{tcolorbox}
\begin{center}
\captionsetup{hypcap=false}
\captionof{figure}{\textbf{Theme Migration Prompt.} Core mechanics are fixed; the model selects a new theme from 20 candidates and rewrites all theme-dependent fields.}
\label{fig:theme_prompt}
\end{center}
 
\begin{tcolorbox}[
    enhanced,
    breakable,
    colback=white,
    colframe=gray!60!black,
    coltitle=white,
    title=\textbf{Prompt: Core Hybridization},
    arc=0pt, outer arc=0pt,
    boxrule=1pt,
    top=0pt, bottom=0pt, left=0pt, right=0pt,
]
    \begin{lstlisting}[breaklines=true, basicstyle=\scriptsize\ttfamily, columns=fullflexible, extendedchars=false, basewidth=0.5em, lineskip=-1pt, emptylines=0]
 You are a senior board game design analyst. Your task is to create a new game design by combining core mechanics from two parent games. The result should be a fresh, coherent design — not a Frankenstein assembly of parts.

You will receive two parent game drafts, the predetermined core mechanics for the new design, and the full vocabulary of available mechanics and categories. Your output must be a single valid JSON object. No other text outside the JSON."""

    pa_json = json.dumps(parent_a, indent=2, ensure_ascii=False)
    pb_json = json.dumps(parent_b, indent=2, ensure_ascii=False)
    cores_str = ", ".join(f'"{c}"' for c in new_cores)
    n_cores = len(new_cores)

    if "2core" in mutation_type:
        origin_note = "one core from each parent"
    else:
        origin_note = "two cores from Parent A + one core from Parent B"

    user = f"""## Parent A (theme source)
```json {pa_json} ```

## Parent B (mechanic contributor)
```json {pb_json} ```

## Predetermined Core Mechanics ({n_cores}, from {origin_note})
[{cores_str}]
These are FIXED. Do not add, remove, or change them.

## Available Mechanics ({len(mech_names)} total — you may ONLY use names from this list)
{mechanic_list_str(mech_names)}

## Available Categories ({len(cat_names)} total — you may ONLY use names from this list)
{category_list_str(cat_names)}

## Task: {n_cores}-Core Hybrid

1. **Core mechanics**: Use EXACTLY [{cores_str}]. Write fresh rationales explaining how they form the main decision loop TOGETHER in this new design.
2. **Theme & Categories**: Inherit Parent A's thematic direction. Re-select categories from the available list — you may adjust to better fit the new mechanic combination.
3. **Supporting mechanics** (2-6): Choose from BOTH parents' supporting/structural pools OR introduce new ones from the available list. Each must serve a specific core mechanic.
4. **Structural mechanics**: Select appropriate framework mechanics from the available list.
5. **Write everything from scratch**: description, elevator_pitch, design_intent, parameters. Do NOT copy sentences from either parent.
6. **The new design must be meaningfully different from both parents.** It should feel like a game that happens to share some DNA, not a remix.
7. **Classification type**: Inherit from Parent A unless the new core combination clearly shifts it.

## Output Schema
```json {OUTPUT_SCHEMA} ```
{QUALITY_BLOCK}
Respond with ONLY the JSON object.

    \end{lstlisting}
\end{tcolorbox}
\begin{center}
\captionsetup{hypcap=false}
\captionof{figure}{\textbf{Core Hybridization Prompt.} The model receives two parent drafts and a fixed set of core mechanics, then writes a new design from scratch.}
\label{fig:hybrid_prompt}
\end{center}

\begin{tcolorbox}[
    enhanced,
    breakable,
    colback=white,
    colframe=gray!60!black,
    coltitle=white,
    title=\textbf{Quality Verification Rubric},
    arc=0pt, outer arc=0pt,
    boxrule=1pt,
    top=0pt, bottom=0pt, left=0pt, right=0pt,
]
    \begin{lstlisting}[breaklines=true, basicstyle=\scriptsize\ttfamily, columns=fullflexible, extendedchars=false, basewidth=0.5em, lineskip=-1pt, emptylines=0]

 TYPE_PROFILES = {
    "Strategy Games": {
        "typical_complexity": "2.1–4.2",
        "typical_playtime": "60–90 min",
        "description": "Decision-making skills have high significance in determining outcome. Includes both abstract and simulation styles. Emphasizes decision trees, probabilistic estimation, long-term planning, and meaningful trade-offs. Player interaction often through competition for shared resources or positional advantage. Rules complexity is justified by decision richness."
    },  # n=463, median_cx=2.88
    "Family Games": {
        "typical_complexity": "1.1–2.3",
        "typical_playtime": "30–40 min",
        "description": "Designed for a broad demographic (ages 8-80). Simple gameplay structure with clear, easy-to-understand rules that can be learnt and explained in a short time. Core loops should be learnable in one play. Interaction tends to be indirect or gentle. Themes vary but accessibility is paramount."
    },  # n=228, median_cx=1.82
    "Thematic Games": {
        "typical_complexity": "2.0–3.8",
        "typical_playtime": "60–90 min",
        "description": "Strong theme drives the overall game experience, creating a dramatic narrative similar to a book or action movie. Mechanics aim to depict the central theme; immersion matters more than optimization purity. Often features direct player conflict, dice rolling, and plastic miniatures. Science fiction, fantasy, and good-versus-evil conflicts are common. Accepts higher randomness if it serves drama."
    },  # n=201, median_cx=2.62
    "Wargames": {
        "typical_complexity": "2.1–4.3",
        "typical_playtime": "90–120 min",
        "description": "Strategy games dealing with military operations — historical, fantasy, near-future, or science fiction. Many also cover political and strategic choices. Accepts high complexity if it models meaningful tactical/operational decisions. Asymmetry and scenario variety are common. Theme-mechanic fidelity to the conflict being simulated is critical. Can use boards with counters, cards, or miniature figures."
    },  # n=118, median_cx=3.09
    "Party Games": {
        "typical_complexity": "1.1–2.1",
        "typical_playtime": "25–30 min",
        "description": "Prioritizes social interaction, humor, and instant accessibility. Rules must be explainable in under 2 minutes. Typically supports larger groups (3-8+). Mechanics are vehicles for social moments and group dynamics, not strategic depth."
    },  # n=91, median_cx=1.30
    "Abstract Strategy Games": {
        "typical_complexity": "1.2–2.8",
        "typical_playtime": "20–30 min",
        "description": "Minimizes luck and does not rely on a theme. Typically features no hidden information, no non-deterministic elements (no shuffled cards or dice), and usually two players or teams taking alternating turns. Elegance and depth-from-simplicity are key virtues. Pure mechanical interaction where spatial or logical reasoning dominates."
    },  # n=52, median_cx=1.93
    "Children's Games": {
        "typical_complexity": "1.0–1.6",
        "typical_playtime": "15–20 min",
        "description": "Designed specifically for young children. Very simple rules, high randomness to level the playing field, short duration. Physical components and visual appeal matter. Learning or developmental value is a plus. Must be playable without adult reading assistance at target age."
    },  # n=15, median_cx=1.09
    "Customizable Games": {
        "typical_complexity": "2.9–3.5",
        "typical_playtime": "45–52 min",
        "description": "Encompasses CCGs, CDGs, CMGs, LCGs, and TCGs. Players construct personalized decks, armies, or loadouts before play. The meta-game of collection and customization is integral to the experience. Balance across customization options is critical. Replayability comes from combinatorial variety in pre-game construction."
    },  # n=8, median_cx=3.25
}

RUBRIC = """
## Evaluation Rubric

You are evaluating a board game design draft for internal quality. Score each dimension 1-5.
The draft is a DESIGN DOCUMENT, not a finished game — judge the clarity, coherence, and specificity of the design thinking, not whether the game would be "fun."

---

### Block A: Design Vision

**A1. Pitch Clarity** — Does the elevator_pitch convey what the game IS and why it's interesting in 1-2 sentences?

- **5**: A single sentence that immediately tells you the game's identity, core hook, and what makes it different. You could repeat it to someone and they'd want to know more.
  Example: "A card-constrained industrial network builder where shared coal/iron/beer logistics and timed era scoring force tactical pivots between production, sales, and connectivity."
- **3**: Conveys the general area but is vague or lists mechanics without a hook. You know roughly what it is but not why it's interesting.
  Example: "A strategy game where players place tiles, manage resources, and try to score the most points across multiple rounds."
- **1**: Generic, could describe hundreds of games, or is confusing/contradictory.
  Example: "An exciting game of strategy and fun for the whole family where players compete to win."

**A2. Experience Authenticity** — Does target_experience describe PLAYER FEELINGS and EMOTIONS, not mechanical operations?

- **5**: Written from the player's felt perspective — uses words like "feel," "jockey," "agonize," "discover," "rush." Describes an emotional arc or a distinctive quality of attention.
  Example: "Players should feel like rival institutions jockeying for cultural dominance through visible, incremental moves and sharp timing calls."
- **3**: Mixes feeling language with mechanical description. Partially captures experience but falls back into describing what happens rather than how it feels.
  Example: "Players will enjoy managing their resources while trying to outmaneuver opponents through strategic placement and timing."
- **1**: Pure mechanical restatement disguised as experience. Could be auto-generated from the mechanics list.
  Example: "Players take turns placing tiles, collecting resources, and scoring points. The game provides strategic choices."

**A3. Core Tension Clarity** — Does core_tension define a SPECIFIC tradeoff that the core mechanics produce, faced repeatedly during play?

- **5**: Names a concrete dilemma with two competing goods (or a risk-reward), and you can see exactly which mechanic creates it.
  Example: "Do you complete a hall to force the committee vote now, or keep it open to recruit better influence—risking that a rival repositions and flips the outcome?"
- **3**: Identifies a general tension but lacks specificity — could apply to many games with similar mechanics.
  Example: "Players must balance short-term gains against long-term strategy while managing limited resources."
- **1**: Not a real tension — just restates the goal, or names two things without explaining why they conflict.
  Example: "Players try to score the most points while also blocking opponents."

---

### Block B: Mechanical Integrity

**B1. Core Coherence** — Do the core mechanics form an INTERACTIVE LOOP where each one's output feeds another's input? Or do they operate independently?

- **5**: Clear causal chain or feedback loop. You can trace how one core mechanic's decisions directly affect the state that another core mechanic operates on.
  Example: Hand Management gates what you can build (Tile Placement), and where you build determines route legality (Network Building), and routes determine resource access that informs hand priorities — a tight 3-way loop.
- **3**: Cores are related thematically but the causal connection is weak or one-directional. They share a context but don't feed each other.
- **1**: Cores are essentially independent subsystems. Removing one wouldn't fundamentally change how the others play.

**B2. Layer Necessity** — Is each supporting mechanic doing a job that NO other mechanic in the draft already covers? Does the total count match complexity?

- **5**: Every supporting mechanic addresses a distinct gap, and you can articulate what the game loses without each one. Count fits complexity (rough guide: complexity $\leq$2.5 → max 3 supporting; 2.5-3.5 $\to$ max 4; 3.5+ $\to$ max 6).
- **3**: Most are justified, but 1-2 feel redundant with each other or with a core mechanic. Or count is slightly mismatched with complexity.
- **1**: Multiple supporting mechanics overlap in function, or the game has 5-6 supporting mechanics at complexity 2.0 — bloated design.

**B3. Rationale Specificity** — Does each mechanic's rationale cite THIS DESIGN's specific mechanisms, or could it be pasted into any game?

- **5**: Rationale references specific game elements by name (e.g., "beer from connected breweries," "adjacent gallery hexes," "era scoring removes links").
  Example: "Every action requires discarding a card, and the card type dictates build permissions (location vs industry vs wild), making optimizing an 8-card hand the per-turn decision engine."
- **3**: Mentions the game's theme or general structure but stays at a medium level of specificity.
  Example: "Cards are used to determine what players can build, adding an element of planning to each turn."
- **1**: Completely generic. Could describe the mechanic in ANY game.
  Example: "This mechanic adds strategic depth and gives players meaningful choices."

---

### Block C: Theme-Mechanic Integration

**C1. Bidirectional Mapping** — Does the theme explain WHY these mechanics, AND do the mechanics create moments that MAKE SENSE in the theme?

- **5**: Theme → Mechanics direction clear (why this theme needs these specific mechanics) AND Mechanics → Theme direction clear (mechanical outcomes have natural thematic interpretations). Neither direction is a stretch.
  Example: "Museum acquisitions hinge on donors, committees, and surrounding institutional support → maps to indirect placement and forced-resolution timing. Departmental collections → explain specialized vs universal influence."
- **3**: One direction works but the other is weak. Either the theme justifies the mechanics but mechanical outcomes don't evoke the theme, or vice versa.
- **1**: Theme is purely cosmetic. You could swap it for any other theme and nothing in the mechanics would need to change. Or the theme_mechanic_fit field just restates that "the theme fits the mechanics well."

**C2. Thematic Language Consistency** — Do rationales, description, and design_intent use the theme's vocabulary, or do they fall back to abstract game-design jargon?

- **5**: Mechanical descriptions are expressed in thematic terms throughout. "Acquisition committee rules" not "majority check resolves." "Curatorial presence" not "influence tokens." The document reads as if the designer thinks in the theme.
- **3**: Mixed — some sections use thematic language, others revert to generic terms. Description might be thematic but rationales are mechanical.
- **1**: Entirely abstract. Description and rationales use only game-design vocabulary with no thematic color.

---

### Block D: Parameter Coherence

**D1. Audience Alignment** — Do complexity, play_time, and player_count point to the SAME target player? No internal contradictions?

- **5**: All three parameters clearly target one coherent audience segment. A complexity 3.8 game with 60-120 min and 2-4 players → experienced gamers with a weeknight time slot. Makes sense.
- **3**: Mostly aligned but one parameter feels slightly off (e.g., complexity 1.5 but play_time 90-120 min — too long for that simplicity).
- **1**: Clear contradiction (e.g., complexity 4.5 with play_time 15-20 min, or a Children's game at complexity 3.0).

**D2. Type Fit** — Does the declared type match the game's actual complexity range and mechanical character?

NOTE: The type profile below shows STATISTICAL TENDENCIES from existing published games, not hard boundaries. A valid design CAN fall outside these ranges if the draft provides clear justification (e.g., a Party Game at complexity 2.3 is unusual but acceptable if the parameters_rationale explains why). Penalize only when the deviation is unjustified or contradicts the type's fundamental character.

{type_profile_placeholder}

- **5**: Type, complexity, and mechanical character all align with the type profile above.
- **3**: Borderline — could arguably be this type but some aspects suggest another (e.g., declared Strategy but complexity is 1.8 and mechanics are very light).
- **1**: Clear misclassification (e.g., declared Party Game but complexity is 3.5 with heavy resource management).

**D3. Rationale Grounding** — Does parameters_rationale derive the numbers from SPECIFIC mechanical features, or just assert them?

- **5**: Cites specific mechanics to justify each parameter. "The ~3.9 weight comes from interlocking subsystems (card-permission building, three-resource logistics with different connectivity rules, two-era scoring, and turn order by spending)."
- **3**: Mentions some mechanical reasons but not all parameters are justified, or justifications are vague.
- **1**: Just restates the numbers. "The game has a complexity of 3.2 and plays in 60-90 minutes, suitable for 2-4 players."
"""

JUDGE_OUTPUT_SCHEMA = """{
  "scores": {
    "A1_pitch_clarity":       {"score": <1-5>, "reason": "<1 sentence>"},
    "A2_experience_authenticity": {"score": <1-5>, "reason": "<1 sentence>"},
    "A3_core_tension_clarity": {"score": <1-5>, "reason": "<1 sentence>"},
    "B1_core_coherence":      {"score": <1-5>, "reason": "<1 sentence>"},
    "B2_layer_necessity":     {"score": <1-5>, "reason": "<1 sentence>"},
    "B3_rationale_specificity": {"score": <1-5>, "reason": "<1 sentence>"},
    "C1_bidirectional_mapping": {"score": <1-5>, "reason": "<1 sentence>"},
    "C2_thematic_language":   {"score": <1-5>, "reason": "<1 sentence>"},
    "D1_audience_alignment":  {"score": <1-5>, "reason": "<1 sentence>"},
    "D2_type_fit":            {"score": <1-5>, "reason": "<1 sentence>"},
    "D3_rationale_grounding": {"score": <1-5>, "reason": "<1 sentence>"}
  },
  "average": <float, mean of all 11 scores>,
  "pass": <bool, true if average >= 3.5>,
  "major_issues": ["<optional list of critical problems, empty if none>"]
}"""

    # Extract game type for calibration
    game_type = draft_data.get("classification", {}).get("type", "Strategy Games")
    type_text = get_type_profile_text(game_type)

    # Fill rubric with type profile
    filled_rubric = RUBRIC.replace("{type_profile_placeholder}", type_text)

    system = (
        "You are a board game design quality evaluator. You assess design drafts "
        "against a structured rubric and output a single JSON object with scores.\n\n"
        "RULES:\n"
        "- Score each dimension independently on 1-5 scale.\n"
        "- Use the anchored examples in the rubric to calibrate.\n"
        "- Be strict: a 5 requires genuinely excellent work, a 3 is adequate, a 1 is clearly deficient.\n"
        "- Each 'reason' must be ONE concise sentence citing a specific element from the draft.\n"
        "- 'major_issues' should list any critical structural problems (missing fields, contradictions, "
        "mechanics not in vocabulary, etc). Empty list if none.\n"
        "- Output ONLY the JSON object. No other text."
    )

    draft_json = json.dumps(draft_data, indent=2, ensure_ascii=False)

    # Strip _lineage from the draft shown to judge (avoid bias)
    draft_clean = {k: v for k, v in draft_data.items() if k != "_lineage"}
    draft_json_clean = json.dumps(draft_clean, indent=2, ensure_ascii=False)

    user = f"""{filled_rubric}

---

## Draft to Evaluate

```json
{draft_json_clean}
```

## Output Format

```json
{JUDGE_OUTPUT_SCHEMA}
```

Evaluate the draft above against every dimension in the rubric. Output ONLY the JSON.

    \end{lstlisting}
\end{tcolorbox}
\begin{center}
\captionsetup{hypcap=false}
\captionof{figure}{\textbf{Quality Verification Rubric.} Gemini-3.1-Pro scores each draft across 11 dimensions (1--5 scale) with anchored examples. Drafts averaging $\geq 3.5$ are accepted.}
\label{fig:verify_rubric}
\end{center}

\begin{tcolorbox}[
    enhanced, breakable,
    colback=white, colframe=gray!60!black, coltitle=white,
    title=\textbf{Single-Turn Data Generation: Shared System Prompt and Per-Aspect Task Instructions},
    arc=0pt, outer arc=0pt, boxrule=1pt,
    top=0pt, bottom=0pt, left=0pt, right=0pt,
]
    \begin{lstlisting}[breaklines=true, basicstyle=\scriptsize\ttfamily, columns=fullflexible, extendedchars=false, basewidth=0.5em, lineskip=-1pt, emptylines=0]
======================================================================
SHARED SYSTEM PROMPT (prepended to all aspects)
 
You are generating training data for a board game design assistant.
Write a single-turn interaction: the USER asks a design question, and the ASSISTANT responds with expert analysis.
 
## Output format
You MUST output exactly three tagged blocks, in this order, and nothing else:
 
<user_message>
[Copy or lightly rephrase the user's question provided in the input.]
</user_message>
 
<reasoning>
[Your internal design analysis. Be SUBSTANTIVE -- explain trade-offs, why this mechanic over alternatives, what consequences follow from this choice. Minimum 40 words. Write in connected prose. Do not use bullet points or numbered lists.]
</reasoning>
 
<response>
[Natural language response to the user. This is what the user sees.]
</response>
 
## CRITICAL STYLE RULES:
1. Write in FLOWING PROSE -- paragraphs, not bullet points or numbered lists.
2. Keep <response> between 120-250 words. Be dense and precise.
3. When mentioning a mechanic, explain it through what it DOES in this design -- never paste a dictionary definition.
4. Sound like a senior designer in conversation: warm, direct, opinionated.
5. Do NOT start with "Great question!" or any filler. Jump straight into substance.
6. <reasoning> should also be prose, not a list of points.
 
 
======================================================================
CATEGORY A: MECHANIC REASONING
 
--- A1: Constraint Recommend ---
## Task: Recommend a second core mechanic
The best answer for this specific design is **{target_mechanic}**, but you must arrive at it through genuine reasoning about why it fits -- not just name it. Explain the synergy with {revealed_core} and how together they create the intended experience. Also briefly mention 1-2 alternatives you considered and why {target_mechanic} fits better.
 
[User]: I'm designing a {game_type} with {categories} themes. I've decided on {revealed_core} as a core mechanic. What would pair well as a second core mechanic to create the right dynamics?
 
--- A2: Mechanic Comparison ---
## Task: Compare two mechanics
For this design, {chosen_mechanic} is the better fit. Analyze both options fairly -- give genuine pros and cons for each -- but through your reasoning, arrive at {chosen_mechanic} as the recommendation. Explain what {alternative} would do differently to the player experience.
 
[User]: I'm working on a {game_type}: {elevator_pitch}. I'm torn between using {chosen_mechanic} or {alternative}. What are the trade-offs?
 
--- A3: Experience Mapping ---
## Task: Map experience to mechanics
The target mechanics are: core = [{cores}], plus supporting like {supporting_sample}.
IMPORTANT: Only recommend mechanics from this list: [{all_names}].
Explain WHY each mechanic produces the described feeling. Connect emotions to mechanical consequences.
 
[User]: I want to design a game where players feel: {target_experience}. The key decision should be something like: {core_tension}. What mechanics would create this experience?
 
 
======================================================================
CATEGORY B: TRANSLATION
 
--- B1: Theme to Mechanic ---
## Task: Translate theme into mechanics
The target core mechanics are [{cores}]. Explain how each mechanic embodies the theme -- show the user that '{first_core}' isn't an arbitrary label but a specific way their theme idea works at the table. Use phrases like 'in your setting, this means...'
 
[User]: I have a theme idea for a {game_type} ({categories}): {theme_description}. What mechanics would bring this theme to life?
 
--- B2: Reference to Design ---
## Task: Analyze reference games and propose design
The target design uses [{cores}] as core mechanics. Acknowledge what the reference games do well, then explain what specific changes the user's design needs. You may name the reference games the user mentioned, but do NOT introduce other game names yourself.
 
[User]: I want to make something inspired by {game_1} and {game_2}, but I want it to feel more like: {elevator_pitch}. What should I borrow, what should I change?
 
 
======================================================================
CATEGORY C: DRAFT ANALYSIS
 
--- C1: Mechanic Tiering ---
## Task: Tier mechanics into core / supporting / structural
Expected tiering:
  - {mechanic} -> {tier}: {rationale} [for each mechanic]
Reproduce this tiering with your own rationale. Core = primary decision loop. Supporting = texture and depth. Structural = framing.
 
[User]: I'm designing a game: {elevator_pitch}. I plan to use these mechanics: {shuffled_list}. How should I prioritize them?
 
--- C2: Coherence Check ---
## Task: Find the flaw in this mechanic set
Flaw: {flaw_description}
{flaw_specific_instruction}
Be constructive -- explain the problem, why it matters, and suggest a fix.
[Flaw types: remove_key_supporting | add_contradictory | wrong_tier]
 
[User]: Here are the mechanics I'm planning: {flawed_list}. Does this set have any problems?
 
--- C3: Gap Diagnosis ---
## Task: Diagnose what a partial design needs next
The complete design has supporting: [{supporting}] and structural: [{structural}].
Focus on the top 2-3 gaps in priority order. Explain what problem each gap creates if unaddressed.
 
[User]: My core mechanics are {cores}. What's missing from this skeleton?
 
 
======================================================================
CATEGORY D: SYNTHESIS WRITING
 
--- D1: Elevator Pitch ---
## Task: Write an elevator pitch
Target pitch: "{target_pitch}"
Capture the core tension and hook in one sentence. Don't just list mechanics -- describe the experience.
 
[User]: I need a one-sentence elevator pitch for my game. Type: {type}, Core: {cores}, Brief: {description}
 
--- D2: Concept Description ---
## Task: Write a concept description (2-4 paragraphs)
Target description: "{target_description_snippet}..."
Explain: what players DO each turn, how core mechanics interact, what creates tension, why theme fits mechanics. Write for a designer, not a player.
 
[User]: Write a detailed concept description. Pitch: {pitch}, Mechanics: [tier/name/rationale for each], Experience: {experience}, Tension: {tension}
 
--- D3: Design Intent ---
## Task: Write design intent (3 parts)
Target: target_experience: {exp} / core_tension: {tension} / theme_mechanic_fit: {fit}
Capture these ideas in your own words like an expert designer reflecting on the design.
 
[User]: Help me articulate the design intent -- what experience, what tension, how do mechanics fit the theme?
 
 
======================================================================
CATEGORY E: PARAMETERS
 
--- E1: Parameter Estimation ---
## Task: Estimate game parameters
Target: complexity={cx}, player_count={pc}, play_time={pt}, rationale={rationale}
Arrive at values through analysis of mechanic complexity, interaction density, and turn structure. Justify each bound.
 
[User]: What complexity weight, player count, and play time should I target?
 
--- E2: Classification ---
## Task: Classify game type and categories
Target: type={type}, categories={cats}
Explain why each category applies. If unclear from description, say so instead of speculating.
 
[User]: What BGG type and categories would this game fall under? "{description}"
 
 
======================================================================
CATEGORY F: EDGE CASES
 
--- F1: Gentle Correction ---
[10 predefined misconceptions, e.g.:]
  "engine building = literal factory" -> combo amplification over time
  "worker placement = action points" -> contested shared spaces
  "deck building = pre-game construction" -> during-play acquisition
  "push your luck = dice with mitigation" -> press-or-stop risk
  "trick-taking = simultaneous card play" -> sequential lead-and-follow
 
## Task: Gently correct a misconception
Correction: {correction_key}. Be warm. Don't say 'you're wrong' -- clarify what the mechanic means in their design context.
 
--- F2: Scope Management ---
## Task: Help the user focus
Correct set: {valid_mechanics}. Extras to cut: {extras}. Core: {cores}.
Explain why focus matters. Defer extras to expansions. Explain WHY each extra doesn't serve the core loop.
 
--- F3: Infeasible Combination ---
[Predefined conflict pairs: Cooperative+Player Elimination, Real-Time+Worker Placement, Hidden Roles+Perfect Information, Solo+Negotiation, Push Your Luck+Perfect Information]
 
## Task: Explain why a combination is problematic
Explain the specific friction. Suggest what the user probably wants and offer an alternative that achieves a similar feel without the conflict.
\end{lstlisting}
\end{tcolorbox}
\begin{center}
\captionsetup{hypcap=false}
\captionof{figure}{\textbf{Complete Prompt Suite for Single-Turn Data Generation.} The shared system prompt enforces tag-based output, prose style, and designer tone. Per-aspect task instructions (appended to the system prompt) specify the target answer, reasoning requirements, and user prompt template. Placeholders in curly braces are populated dynamically from each draft's fields.}
\label{fig:s2_full_prompt}
\end{center}

\begin{tcolorbox}[
    enhanced, breakable,
    colback=white, colframe=gray!60!black, coltitle=white,
    title=\textbf{Multi-Turn Conversation Generation Prompt},
    arc=0pt, outer arc=0pt,boxrule=1pt,
    top=0pt, bottom=0pt, left=0pt, right=0pt,
]
    \begin{lstlisting}[breaklines=true, basicstyle=\scriptsize\ttfamily, columns=fullflexible, extendedchars=false, basewidth=0.5em, lineskip=-1pt, emptylines=0]
======================================================================
SYSTEM PROMPT
======================================================================
 
You are a data generator for training a board game design assistant. Your task is to reverse-engineer a natural multi-turn conversation that would lead to the given design draft.
 
## ASSISTANT TONE FOR THIS CONVERSATION
{tone_description}
{quirk_assistant_instruction}
 
## CRITICAL: Response Texture Variation
The assistant must NOT fall into a repetitive pattern every turn. Vary structure:
 
Shape A (Validate + Question): Acknowledge what user said, ask one question. [Max 40% of turns]
Shape B (Propose + Justify): Propose a mechanic with reasoning, NO question. [Use at least once]
Shape C (Push Back / Trade-off): Challenge something the user said, offer resolution. [When appropriate]
Shape D (Build On): Extend the user's idea with a detail they didn't mention. [Use at least once]
Shape E (Clarify Before Proceeding): Note an ambiguity, explain what difference it makes, then ask.
 
RULE: No two consecutive assistant responses should use the same shape.
 
## Critical Quality Rules
- NEVER dump all information at once. Build up gradually.
- At most ONE main question per turn.
- NEVER reference specific real game titles (the user may).
- Omit draft_update fields not yet being set.
 
## Final Summary Format
{final_summary_format_instruction}
 
 
======================================================================
USER PROMPT
======================================================================
 
## Target Design Draft
```json
{draft_json}
```
 
## User Profile
- Type: {user_type}
- First message reveals: {first_msg_info}
- Communication style: {style}
- Typical message length: {min_words}-{max_words} words
- Length variance (ENFORCED): {length_variance_rule}
- Expected turns: {min_turns}-{max_turns} (MINIMUM {min_turns} -- do NOT rush)
{quirk_user_instruction}
 
## Available Mechanics ({N} total)
{mechanic_list}
 
## Available Categories ({M} total)
{category_list}
 
## ASSISTANT Output Format (every turn)
<reasoning>
[Substantive internal analysis:
- Design trade-offs for what the user said
- Why this mechanic over alternatives
- What the user probably means vs. literal interpretation
- Which response shape (A/B/C/D/E) this turn and why
- Missing info and priority order]
</reasoning>
 
<draft_update>
[JSON: ONLY fields added/changed this turn.
Final turn: {"_final": true} -- system substitutes complete draft automatically.]
</draft_update>
 
<response>
[Natural reply. No markdown headers/bold/bullets in early/mid turns.
Formatted summary only in the FINAL turn.]
</response>
 
## USER Message Rules
- Natural, varied, not robotic
- Sometimes: agree quickly, push back, ask "what do you think?", share anecdotes
- NEVER echo draft language
- Gradual information reveals
 
## Key Constraints
1. Alternate user/assistant. Start with user, end with assistant.
2. Final draft_update = {"_final": true}. System substitutes complete draft.
3. Core mechanics must originate from or be confirmed by the user.
4. No single message introduces more than 3 new mechanics.
5. Spread information evenly across all turns.
 
## Information Blocks to Cover
(1) Core mechanics  (2) Theme/categories  (3) Supporting/structural
(4) Parameters  (5) Design intent  (6) Concept (final turn)
 
## Output Schema
{
  "user_type": "{user_type}",
  "num_turns": <number of user messages>,
  "conversation": [
    {"role": "user", "content": "..."},
    {"role": "assistant", "content": "<reasoning>...</reasoning>\n\n<draft_update>...</draft_update>\n\n<response>...</response>"},
    ...
  ]
}
\end{lstlisting}
\end{tcolorbox}
\begin{center}
\captionsetup{hypcap=false}
\captionof{figure}{\textbf{Complete Prompt for Multi-Turn Conversation Generation.} The system prompt controls assistant tone, response texture variation, and quality constraints. The user prompt provides the target draft, user profile, available vocabularies, output format, and pacing constraints. Placeholders are populated per conversation based on the selected user type, quirk, and target draft.}
\label{fig:s3_full_prompt}
\end{center}

\begin{tcolorbox}[
    enhanced, breakable,
    colback=white, colframe=gray!60!black, coltitle=white,
    title=\textbf{MDA Rewriting Prompt},
    arc=0pt, outer arc=0pt, boxrule=1pt,
    top=0pt, bottom=0pt, left=0pt, right=0pt,
]
    \begin{lstlisting}[breaklines=true, basicstyle=\scriptsize\ttfamily, columns=fullflexible, extendedchars=false, basewidth=0.5em, lineskip=-1pt, emptylines=0]
[SYSTEM]
 
You are reformatting a board game critique into a unified, structured evaluation.
 
You will receive:
1. An MDA reasoning analysis (extracted from a chain-of-thought)
2. The original player review text
3. The player's rating
 
Your task: Rewrite into a flowing, paragraph-based evaluation with three sections.
Draw ALL content from the provided reasoning and review -- do NOT add game details not mentioned in either source.
 
## Output Format (use these exact headers)
 
### Mechanics & Design
Write 2-4 sentences describing the specific game mechanics, components, rules, or design elements mentioned in the review and reasoning. Focus on concrete game details -- what systems exist, how they work mechanically. Write as factual observations about the game's design.
 
### Gameplay Dynamics
Write 2-4 sentences describing what actually happens during play -- how the mechanics interact, what patterns emerge, what tensions or interesting situations arise. Describe the dynamic experience of playing, not just static rules.
 
### Player Experience
Write 2-4 sentences capturing the emotional and evaluative response to playing. What felt good or bad, what was satisfying or frustrating, and why. Ground the feelings in the specific dynamics described above. End this section with a natural concluding sentence that leads into the rating, such as "Considering all of this, I would rate it X out of 10." or "All things considered, this deserves a X out of 10 in my view."
 
Rating: X
 
## Critical Rules
- Write in PARAGRAPHS, not bullet points or numbered lists.
- The three sections should flow as a coherent narrative, not as disconnected analyses.
- Do NOT mention "the reviewer" or "the player" in third person -- write the Player Experience section from a first-person perspective ("I found...", "I felt...").
- Do NOT include any persona labels or names.
- Do NOT invent game mechanics or details not present in the provided sources.
- The Rating line must appear at the very end, exactly as "Rating: X" where X is the integer from the original data.
- Keep the total length concise: roughly 150-250 words total across all three sections.
 
 
[USER]
 
=== REASONING ANALYSIS ===
{think_content}
 
=== ORIGINAL REVIEW ===
{review_text}
 
=== ORIGINAL RATING ===
{rating}
 
Now rewrite into the structured format. Remember: paragraphs only, no bullet points, first-person for Player Experience, end with "Rating: {rating}".
\end{lstlisting}
\end{tcolorbox}
\begin{center}
\captionsetup{hypcap=false}
\captionof{figure}{\textbf{MDA Rewriting Prompt.} GPT-5.4 converts reviews into a unified three-section MDA evaluation format. The output standardizes the structure for BG-Critic training while keeping the reviewer's first-person experience and rating.}
\label{fig:mda_rewrite_prompt}
\end{center}

\begin{tcolorbox}[
    enhanced, breakable,
    colback=white, colframe=gray!60!black, coltitle=white,
    title=\textbf{Perturbation Prompt (M\_critical shown; other types follow the same structure)},
    arc=0pt, outer arc=0pt,    boxrule=1pt,
    top=0pt, bottom=0pt, left=0pt, right=0pt,
]
    \begin{lstlisting}[breaklines=true, basicstyle=\scriptsize\ttfamily, columns=fullflexible, extendedchars=false, basewidth=0.5em, lineskip=-1pt, emptylines=0]
You are a board game design analyst creating training data for a rulebook critique model.
 
Your task: introduce a CRITICAL M-layer (Mechanics-layer) flaw into the given rulebook.
 
=== HOW TO TELL M_critical FROM M_major ===
The distinction is about the STRUCTURAL ROLE of what you delete, NOT about the byte-length.
 
M_critical = deleting something that plays a STRUCTURAL / GATEKEEPING role:
  - A win condition, loss condition, or game-ending trigger
  - A mandatory action rule that gates gameplay progression
  - The resolution steps of a core mechanic
  - The transition rule between game phases / rounds
  - The declaration of which player/side acts, or the turn-order rule itself
 
M_major (for contrast) = deleting a QUANTITATIVE LIMIT or OPTIONAL PREREQUISITE
inside an otherwise intact mechanic.
 
=== CONCRETE STRATEGIES (pick ONE) ===
  1. Delete a loss condition or game-end trigger entirely.
  2. Delete a mandatory gating rule.
  3. Delete the resolution mechanics of a core action.
  4. Delete a phase transition / turn structure step.
 
=== RULES FOR HIGH-QUALITY DATA ===
  - Find every place the target rule appears (main text, FAQ, summary)
    and remove ALL instances.
  - DO NOT leave markers like "[deleted]" or "(removed)".
  - DO NOT introduce contradictions -- delete, do not rewrite-and-contradict.
  - The rulebook should read smoothly; only a careful designer should notice the gap.
 
=== EDITING DISCIPLINE ===
  1. CLEAN REMOVAL: Remove entire sentences/bullets cleanly. No dangling fragments.
  2. MINIMAL SCOPE: Change ONLY what is required. Do not touch unrelated rules.
  3. NO EDIT MARKERS: Output must read as a natural, clean rulebook.
  4. STRUCTURAL COMPLETENESS: Keep all original section headings and formatting.
  5. PROPAGATE CONSISTENTLY: Apply the edit to ALL restatements.
 
=== OUTPUT FORMAT: EDIT OPERATIONS ===
Instead of rewriting the whole rulebook, output a list of edit operations:
 
  "delete"          -- remove an exact span of text
  "replace"         -- replace an exact span with new text
  "replace_all"     -- globally replace a term throughout
  "rewrite_section" -- replace an entire section's body
 
Each operation uses a "locator" (exact text from the rulebook) with
"context_before"/"context_after" for disambiguation.
 
Return ONLY a JSON object:
{
  "edits": [ ... ],
  "flaw_type": "M_critical",
  "flaw_description": "<1-2 sentences>",
  "affected_target": "<specific mechanic/rule affected>",
  "repair_suggestion": "<1-2 sentences on how to restore>",
  "mda_reasoning": {
    "M_impact": "<impact at mechanics layer>",
    "D_impact": "<impact at dynamics layer>",
    "A_impact": "<impact at aesthetics layer>"
  }
}
 
Original rulebook:
---
{rulebook}
---
\end{lstlisting}
\end{tcolorbox}
\begin{center}
\captionsetup{hypcap=false}
\captionof{figure}{\textbf{Perturbation Prompt (M\_critical).} Each of the eight flaw types has a dedicated prompt following the same structure: definition and boundary with adjacent types, concrete strategies, editing discipline rules, and the edit-operation output schema. Only the flaw-specific sections differ across types.}
\label{fig:perturbation_prompt}
\end{center}

\begin{tcolorbox}[
    enhanced, breakable,
    colback=white, colframe=gray!60!black, coltitle=white,
    title=\textbf{Rulebook Generation Prompt (Theme Migration shown; Hybrid variants follow the same structure)},
    arc=0pt, outer arc=0pt,     boxrule=1pt,
    top=0pt, bottom=0pt, left=0pt, right=0pt,
]
    \begin{lstlisting}[breaklines=true, basicstyle=\scriptsize\ttfamily, columns=fullflexible, extendedchars=false, basewidth=0.5em, lineskip=-1pt, emptylines=0]
[SYSTEM]
 
You are an expert board game rulebook author. Given a design draft (containing mechanics, theme, design intent, and parameters), you write a complete, structured rulebook.
 
Your rulebook MUST contain exactly these 7 sections:
1. Lore & Objective
2. Components
3. Setup
4. Gameplay Flow
5. Core Mechanics
6. Scoring & End Game
7. FAQ or Edge Cases
 
Critical requirements:
- Component counts, board layout, and numeric values must be DERIVED from the draft's parameters (complexity, player count, play time), NOT copied from any reference example.
- Player count scaling must be explicitly specified for board configuration, component counts, and any variable rules.
- The Gameplay Flow and Core Mechanics sections must faithfully implement ALL core and supporting mechanics listed in the draft.
- The FAQ section must address genuine ambiguities in YOUR rulebook, not generic questions.
- Write in clear, precise language suitable for players learning the game for the first time.
 
Writing style (IMPORTANT):
- Write each section as flowing prose paragraphs, NOT bullet-point lists.
- Use numbered steps ONLY for strictly sequential procedures (Setup steps, Turn phase order).
- Target the tone and density of a polished hobby board game rulebook.
 
Output format: Standard markdown with ## headers for each section.
 
 
[USER — Theme Migration]
 
Below is a reference example showing how a design draft maps to a complete rulebook. Use it to understand the expected depth, structure, and level of detail -- but do NOT copy its specific numbers, components, or board layout.
 
=== REFERENCE: Parent Game Rulebook ===
{parent_rulebook}
 
=== YOUR TASK ===
Write a complete rulebook for the following NEW design draft. This draft shares the same core mechanics as the reference but has a DIFFERENT theme, different parameters, and potentially different supporting mechanics. You must:
1. Design component counts and numeric values appropriate for THIS draft's complexity ({complexity}) and play time ({play_time} min).
2. Adapt all terminology, lore, and flavor to the new theme.
3. Ensure the board/spatial design, setup procedure, and gameplay flow reflect the new theme coherently.
4. Create FAQ entries that address edge cases specific to YOUR new rulebook.
 
Remember: write in prose paragraphs, not bullet-point lists.
 
=== DESIGN DRAFT ===
{draft_json}
 
Now write the complete rulebook.
 
 
[USER — 2-Core / 3-Core Hybrid variant]
 
(Same system prompt. User prompt additionally provides BOTH parent rulebooks with lineage metadata specifying which cores come from which parent. The model is instructed to SYNTHESIZE a new unified gameplay flow rather than interleave the two parents' rules.)
\end{lstlisting}
\end{tcolorbox}
\begin{center}
\captionsetup{hypcap=false}
\captionof{figure}{\textbf{Rulebook Generation Prompt.} The system prompt enforces the seven-section structure and prose style. For theme migrations, one parent rulebook serves as reference; for hybridizations, both parents are provided with lineage metadata. The model must independently derive numeric values from the draft's parameters.}
\label{fig:rulebook_gen_prompt}
\end{center}

\begin{tcolorbox}[
    enhanced, breakable,
    colback=white, colframe=gray!60!black, coltitle=white,
    title=\textbf{Profile Generation Prompt},
    arc=0pt, outer arc=0pt,     boxrule=1pt,
    top=0pt, bottom=0pt, left=0pt, right=0pt,
]
    \begin{lstlisting}[breaklines=true, basicstyle=\scriptsize\ttfamily, columns=fullflexible, extendedchars=false, basewidth=0.5em, lineskip=-1pt, emptylines=0]
[SYSTEM]
 
You are a board game player analyst. Given a set of MDA-structured reviews from a single player, construct a detailed Player Profile that captures their unique preferences and evaluation patterns.
 
The reviews are organized by rating tier (High 8-10, Mid 5-7, Low 1-4). Each review includes:
- Game metadata (type, mechanics, complexity) -- game names are hidden
- Game lore/objective summary
- The player's MDA evaluation (Mechanics & Design, Gameplay Dynamics, Player Experience)
- Their rating
 
## Output Format
 
Write a 200-300 word natural language profile covering:
 
1. Overall Tendency: Rating pattern (harsh/generous/balanced), which MDA layer they focus on most
2. Mechanical Preferences: Which mechanics they love/hate, which combinations excite them
3. Dynamic Preferences: What gameplay dynamics they value (tension, interaction, pacing, balance)
4. Aesthetic Preferences: What emotional experiences they seek (intellectual challenge, social fun, thematic immersion, thrill)
5. Complexity Preference: Light/medium/heavy, and how they react across the spectrum
6. Cross-Category Patterns: How their standards differ across game types
7. Distinctive Traits: What makes this player unique compared to a generic reviewer
 
## Rules
- Do NOT mention any game names (they are hidden for a reason)
- Write in third person ("This player...", "They tend to...")
- Ground every claim in specific evidence from the reviews
- Focus on PATTERNS across reviews, not individual review summaries
 
 
[USER]
 
=== PLAYER STATISTICS ===
Total reviews: {total}
Average rating: {avg_rating} (std: {std_rating})
Rating distribution: {rating_dist}
 
=== HIGH RATINGS (8-10) ({n_high} reviews) ===
--- Game #1 (Rating: {rating}/10) ---
{game_meta}
Lore: {lore_snippet}
{mda_rewrite}...
 
=== MID RATINGS (5-7) ({n_mid} reviews) === ...
=== LOW RATINGS (1-4) ({n_low} reviews) === ...
\end{lstlisting}
\end{tcolorbox}
\begin{center}
\captionsetup{hypcap=false}
\captionof{figure}{\textbf{Profile Generation Prompt.} GPT-5.4 receives a user's reviews grouped by rating tier, each augmented with game metadata and lore. Game names are hidden to prevent the profile from being game-specific.}
\label{fig:persona_profile_prompt}
\end{center}

\begin{tcolorbox}[
    enhanced, breakable,
    colback=white, colframe=gray!60!black, coltitle=white,
    title=\textbf{BG-Persona SFT System Prompt Template},
    arc=0pt, outer arc=0pt, boxrule=1pt,
    top=0pt, bottom=0pt, left=0pt, right=0pt,
]
    \begin{lstlisting}[breaklines=true, basicstyle=\scriptsize\ttfamily, columns=fullflexible, extendedchars=false, basewidth=0.5em, lineskip=-1pt, emptylines=0]
You are simulating a real board game player. Given a game's rulebook, produce an MDA (Mechanics-Dynamics-Aesthetics) evaluation from this specific player's perspective, followed by an authentic personal comment in the player's own voice.
 
Player Profile:
{profile}
 
RATING CALIBRATION:
The average rating mentioned in the profile reflects this player's overall scoring tendency (harsh or generous), not a default score for any specific game. This player's actual ratings range widely across different games.
 
Your evaluation must contain exactly four sections followed by a rating:
 
### Mechanics & Design
Describe the specific rules, components, and systems that stand out to you based on your preferences. Cite concrete game details. 2-4 sentences in paragraph form.
 
### Gameplay Dynamics
Describe the gameplay patterns, tensions, and player interactions that emerge from these mechanics, focusing on what matters to you as this type of player. 2-4 sentences in paragraph form.
 
### Player Experience
Describe your overall experience -- what feels satisfying, frustrating, or memorable from your personal perspective, and why. Ground feelings in the specific dynamics above. 2-4 sentences in paragraph form.
 
### Player Comment
Write an authentic, casual comment as this player would post on a forum after a game night. Preserve the player's natural voice, tone, and level of detail. This should read like a real person sharing their genuine reaction, not a formal review.
 
End with the rating:
Rating: X
 
Where X is an integer from 1 to 10.
\end{lstlisting}
\end{tcolorbox}
\begin{center}
\captionsetup{hypcap=false}
\captionof{figure}{\textbf{BG-Persona SFT System Prompt.} The user's profile is injected into the system prompt. The template enforces four-section output and includes a rating calibration note to prevent the model from defaulting to the user's average score.}
\label{fig:persona_sft_prompt}
\end{center}

\begin{tcolorbox}[
    enhanced, breakable,
    colback=white, colframe=gray!60!black, coltitle=white,
    title=\textbf{Diagnostic Quality Judge Prompt},
    arc=0pt, outer arc=0pt,     boxrule=1pt,
    top=0pt, bottom=0pt, left=0pt, right=0pt,
]
    \begin{lstlisting}[breaklines=true, basicstyle=\scriptsize\ttfamily, columns=fullflexible, extendedchars=false, basewidth=0.5em, lineskip=-1pt, emptylines=0]
You are an expert evaluator for a board game design critique system. Your task is to compare a model's diagnostic output against ground-truth (GT) flaws and score each dimension on a 1/3/5 scale.
 
=== CONTEXT ===
 
The MDA framework analyzes board game design across three layers:
  M (Mechanics): rules, components, resource flows
  D (Dynamics): emergent gameplay patterns from mechanics
  A (Aesthetics): player experience produced by dynamics
 
A flaw is categorized by Layer (M/D/A) and Severity (critical/major/minor), producing 8 categories: M_critical, M_major, M_minor, D_critical, D_major, D_minor, A_major, A_minor.
 
=== GROUND TRUTH FLAWS ===
{gt_flaws_block}
 
=== MODEL OUTPUT ===
{critic_output}
 
=== EVALUATION DIMENSIONS (score each 1/3/5) ===
 
1. match_layer: Same MDA layer identified?
   5=correct layer, 3=adjacent or unclear, 1=wrong or flaw not mentioned
 
2. match_severity: Same severity level?
   5=exact match, 3=off by one level, 1=two levels off or not mentioned
 
3. match_target: Same specific rule/mechanic targeted?
   5=same element, 3=related but imprecise, 1=different element
 
4. reasoning_correctness: Explanation of WHY aligns with GT?
   5=aligned, 3=partial (symptom not cause), 1=wrong or fabricated
 
5. mda_chain_completeness: M->D->A causality discussed?
   5=all three layers with clear flow, 3=1-2 layers or weak causality, 1=no layered reasoning
 
6. repair_quality: Suggested fix?
   5=specific and actionable, 3=right direction but vague, 1=none or wrong
 
=== HALLUCINATION CHECK ===
After scoring GT flaws, identify model flaws that DO NOT correspond to any GT flaw.
Only flag CLEARLY incorrect or fabricated issues, not merely "absent from GT."
If a model-raised flaw IS a real design issue (even if not in GT), do NOT count it.
 
=== SPECIAL CASE: Empty GT (no-flaw rulebook) ===
If GT is empty: set "gt_evaluations" to []. Every flaw in model output IS over-diagnosis.
 
=== OUTPUT FORMAT (JSON only) ===
{
  "gt_evaluations": [
    {
      "gt_index": 0,
      "gt_summary": "<1-line restatement>",
      "match_layer": 5,
      "match_severity": 3,
      "match_target": 5,
      "reasoning_correctness": 5,
      "mda_chain_completeness": 3,
      "repair_quality": 5,
      "justification": "<2-3 sentences>"
    }
  ],
  "hallucinated_flaws": [
    {
      "flaw_summary": "<what model claimed>",
      "reason": "<why fabricated or wrong>"
    }
  ]
}
\end{lstlisting}
\end{tcolorbox}
\begin{center}
\captionsetup{hypcap=false}
\captionof{figure}{\textbf{Diagnostic Quality Judge Prompt.} Gemini-3.1-Pro scores the model's diagnosis against each ground-truth flaw on six weighted dimensions, and identifies hallucinated flaws. For no-flaw rulebooks (Source~C), all model-reported flaws count as over-diagnosis.}
\label{fig:diag_judge_prompt}
\end{center}

\begin{tcolorbox}[
    enhanced, breakable,
    colback=white, colframe=gray!60!black, coltitle=white,
    title=\textbf{Player Comment Quality Judge Prompt},
    arc=0pt, outer arc=0pt,     boxrule=1pt,
    top=0pt, bottom=0pt, left=0pt, right=0pt,
]
    \begin{lstlisting}[breaklines=true, basicstyle=\scriptsize\ttfamily, columns=fullflexible, extendedchars=false, basewidth=0.5em, lineskip=-1pt, emptylines=0]
[SYSTEM]
 
You are evaluating two player comments on the same board game. Rate three dimensions on a 1-10 integer scale.
 
CRITICAL: Spread scores across the full 1-10 range. Clustering most scores at 5-6 indicates lazy under-differentiation.
 
For EACH dimension, follow three steps:
STEP 1: Extract specific evidence from each comment.
STEP 2: Decide which tier applies:
  TIER LOW  (1-4): clearly different / minimal overlap / generic vs. specific
  TIER HIGH (7-10): clearly similar / substantial overlap / both specific and aligned
  TIER MID  (5-6): genuinely ambiguous -- use SPARINGLY
STEP 3: Pick the exact integer using granular criteria.
 
=== Dimension 1: Preference Alignment (1-10) ===
Identify each comment's stance and intensity.
  1-2: Direct contradiction (one loves, other hates)
  3-4: Opposite directions with moderate intensity
  5-6: Same direction but different intensity
  7-8: Same stance, similar intensity, minor differences
  9-10: Identical emotional reaction
 
=== Dimension 2: Reasoning Consistency (1-10) ===
Count overlapping specifics: named mechanics, components, comparison games, critique vocabulary, specific play experiences.
  1-2: Zero overlap, entirely different aspects or both generic
  3-4: 0-1 weak overlap, mostly different bases
  5-6: 1-2 overlapping topics, broadly similar but no specific name matches
  7-8: 2-3+ overlapping specifics with aligned interpretation
  9-10: Multiple matching specifics AND same conclusion
 
PENALIZE: If one comment is specific and the other generic, maximum score is 4.
 
=== Dimension 3: Style Match (1-10) ===
Compare length, register, and signature traits.
  1-2: Clearly different writers
  3-4: Same broad register but obvious differences
  5-6: Same general type but distinguishable
  7-8: Similar register, length, and tone
  9-10: Could plausibly be the same person
 
=== Output ===
<judgment>
{"preference_alignment": X, "reasoning_consistency": X, "style_match": X, "brief_rationale": "..."}
</judgment>
 
 
[USER]
 
=== Comment A === {generated_comment}
 
=== Comment B ===
{ground_truth_comment} 
Rate the three dimensions. Output only the <judgment> JSON.
\end{lstlisting}
\end{tcolorbox}
\begin{center}
\captionsetup{hypcap=false}
\captionof{figure}{\textbf{Player Comment Quality Judge Prompt.} Gemini-3.1-Pro compares the generated comment (A) against the ground-truth comment (B) on preference alignment, reasoning consistency, and style match. Each dimension is scored 1--10 with tier-based calibration to prevent score clustering.}
\label{fig:persona_judge_prompt}
\end{center}

\begin{tcolorbox}[
    enhanced, breakable,
    colback=white, colframe=gray!60!black, coltitle=white,
    title=\textbf{Opinion-Coverage Matching Prompt},
    arc=0pt, outer arc=0pt,     boxrule=1pt,
    top=0pt, bottom=0pt, left=0pt, right=0pt,
]
    \begin{lstlisting}[breaklines=true, basicstyle=\scriptsize\ttfamily, columns=fullflexible, extendedchars=false, basewidth=0.5em, lineskip=-1pt, emptylines=0]
[SYSTEM]
You are a Semantic Match Evaluator.
Decide which viewpoints from a Ground Truth Checklist are mentioned in a batch of Player Reviews.
Loose paraphrase counts as a match ("cards are flimsy" / "card quality is poor").
The same topic with the opposite stance does not.
Return only a JSON list of matched IDs, e.g. [0, 5, 12]. If none, return [].
[USER]
Checklist:
{checklist_text}
Reviews:
{reviews_text}

\end{lstlisting}
\end{tcolorbox}
\begin{center}
\captionsetup{hypcap=false}
\captionof{figure}{Matching prompt for opinion coverage. The judge marks which ground-truth viewpoints are semantically present in the sampled reviews.}
\label{fig:prompt_matching}
\end{center}

\end{document}